\documentclass[12pt,aps,pre,referee,showpacs,showkeys]{revtex4}
\usepackage{amsmath}
\usepackage{bm}
\usepackage{multirow}
\usepackage{subfigure}
\usepackage{longtable}
\usepackage{setspace}
\usepackage{setspace,xspace}
\usepackage{pdflscape}
\usepackage{algorithm}
\usepackage{algpseudocode}
\usepackage{url}

\usepackage{ifpdf}
\usepackage{graphicx,natbib,amssymb,lineno}
\ifpdf
\usepackage[%
  pdftitle={Instructions for use of the document class
    elsart},%
  pdfauthor={Simon Pepping},%
  pdfsubject={The preprint document class elsart},%
  pdfkeywords={instructions for use, elsart, document class},%
  pdfstartview=FitH,%
  bookmarks=true,%
  bookmarksopen=true,%
  breaklinks=true,%
  colorlinks=true,%
  linkcolor=blue,anchorcolor=blue,%
  citecolor=blue,filecolor=blue,%
  menucolor=blue,pagecolor=blue,%
  urlcolor=blue]{hyperref}
\else
\usepackage[%
  breaklinks=true,%
  colorlinks=true,%
  linkcolor=blue,anchorcolor=blue,%
  citecolor=blue,filecolor=blue,%
  menucolor=blue,pagecolor=blue,%
  urlcolor=blue]{hyperref}
\fi

\makeatletter
\def\elsartstyle{%
    \def\normalsize{\@setfontsize\normalsize\@xiipt{14.5}}
    \def\small{\@setfontsize\small\@xipt{13.6}}
    \let\footnotesize=\small
    \def\large{\@setfontsize\large\@xivpt{18}}
    \def\Large{\@setfontsize\Large\@xviipt{22}}
    \skip\@mpfootins = 18\p@ \@plus 2\p@
    \normalsize
}
\@ifundefined{square}{}{}
\makeatother

\newcommand{\mpr}{\textcolor{magenta}{MPR}\xspace}
\newcommand{\Z}{\mathbb{Z}}
\newcommand{\R}{\mathbb{R}}

\newcommand{\bfr}{\mathbf{r}}

\begin{document}

\title{Gibbs Markov Random Fields with Continuous Values based on the Modified Planar Rotator Model}
\author{Milan \v{Z}ukovi\v{c}}
 \email{milan.zukovic@upjs.sk}
\affiliation{Institute of Physics, Faculty of Science, P. J. \v{S}af\'arik University, Park Angelinum 9, 041 54 Ko\v{s}ice, Slovakia}
\author{Dionissios T. Hristopulos}
 \email{dionisi@mred.tuc.gr}   
 \affiliation{School of Mineral Resources Engineering, Technical University of Crete,
Chania 73100, Greece}

\date{\today}

\begin{abstract}
We introduce a novel Gibbs Markov random field for spatial data on Cartesian grids based on the modified planar rotator (\mpr) model of statistical physics.
The \mpr captures spatial correlations using nearest-neighbor interactions of continuously-valued spins and does not rely on Gaussian assumptions. The only model parameter is the reduced temperature, which we estimate by means of an ergodic specific energy matching principle. We propose an efficient hybrid Monte Carlo simulation algorithm that leads to fast relaxation of the \mpr model and allows vectorization. Consequently, the \mpr computational time for inference and simulation scales approximately linearly with system size. This makes it more suitable for big data sets, such as satellite and radar images, than conventional geostatistical approaches. The performance (accuracy and computational speed) of the \mpr model is validated with conditional simulation of Gaussian synthetic and non-Gaussian real data (atmospheric heat release measurements and Walker-lake DEM-based concentrations) and comparisons with standard gap-filling methods.



\end{abstract}

\maketitle

\section{Introduction}
The steadily increasing volume of  Earth observation data collected by remote sensing techniques requires the development of new methods capable of efficient (often real time) and preferably automated processing. Such processing includes filling of gaps that may arise due to various reasons, such as instrument malfunctions and obstacles between the remote sensing device and the sensed object (clouds, snow, heavy precipitation, ground vegetation coverage, undersea topography, terrain blockage, etc.)~\citep{Kadlec17,Lehman04,Bechle13,Cole11,Sun17,Yoo10}. Filling gaps is desirable  to obtain continuous maps of  observed variables and to avoid the adverse missing-data impact on  statistical estimates of means and trends~\citep{Sickles07}. Traditional  kriging methods~\citep{wack03} have favorable statistical properties (optimality, linearity, and unbiasedness under ideal conditions) and can thus outperform other gap-filling methods in prediction accuracy~\citep{Sun09}. However, they are not suitable for large data sets due to high computational cost. In addition, they require several user-specified inputs (variogram model, parameter inference method, kriging neighborhood)~\citep{dig07,pebezma98}.

To alleviate the computational burden of kriging, several modifications~\citep{cres08,furr06,kauf08,zhong16,marco18,ingram08} and parallelized schemes~\citep{cheng13,guti14,hu15,pesq11} have been implemented. Recently an alternative approach to traditional geostatistical methods, inspired from statistical physics, has been proposed~\citep{dth03,dthsel07}. It employs Boltzmann-Gibbs random fields with
joint densities that model spatial correlations  by means of  short-range interactions instead of the empirical variogram used in geostatistics. These so-called \emph{Spartan spatial random field models} have been shown to be computationally efficient and applicable to both gridded and scattered Gaussian data. Furthermore, the concept of deriving correlations from local interactions was extended to non-Gaussian gridded data by means of classical spin models~\citep{mz-dth09a,mz-dth09b}. The latter are defined in terms of discrete-valued processes and thus require discretization for application to continuous processes.  The spin-based approach is non-parametric and captures the spatial correlations in terms of interactions between the ``spins''. The  predictions are determined by matching the energy of the entire (filled) grid with that of the sample data. In a similar spirit, non-parametric models that capture the spatial correlations via geometric constraints have also been proposed~\citep{mz-dth13a,mz-dth13b}.

Spatial data on regular grids are often modeled by means of Gaussian Markov random fields (GMRFs)~\citep{Rue05}. GMRFs are based on the principles of conditional independence and
the imposition of spatial correlations via local interactions. The local interaction structure translates into sparse precision matrices, which allow for computationally efficient representations. While there has been considerable activity in the development of GMRFs [for a review see Chapters 12-15 in~\citep{Gelfand10}], there is  considerably less progress on non-Gaussian Markov random fields (NGMRFs). The prototypical non-Gaussian Markov random field is the binary-valued Ising model, widely studied in statistical physics. The Ising model has been introduced in the statistical community by Julian Besag~\citep{Besag74} and its application to an image restoration problem, mostly within the spin-glass theory, has been proposed in a series of papers~\cite{nishi99,wong00,inoue01,ino-carl01,tadaki01}. The Ising model is most suitable for data with binary values, even though it is possible to apply it to multi-valued discretized data by means of successive thresholding operations~\citep{mz-dth09a,mz-dth09b}.

This paper presents a novel Gibbs Markov random field for spatial processes that take continuous values in a closed subset of the real numbers. The NGMRF is based on the parametric planar rotator spin model from statistical physics, which has successfully been applied to binary image restoration~\citep{saika02}. The  spatial prediction method proposed herein was prompted by our recent study which revealed that a suitably modified planar rotator model changes its low-temperature quasi-critical behavior (which is characterized by power-law decaying correlation function) to a regime characterized by a flexible short-range spatial correlation function~\citep{mz-dth15}.
In particular, the planar rotator model is modified to account for spatial correlations that are typical in geophysical and environmental data sets. In thermodynamic equilibrium, the \emph{modified planar rotator} (\mpr) model is shown to display flexible  short-range correlations controlled by the  temperature (which is the only model parameter).
A hybrid Monte Carlo algorithm  for  parameter estimation and  conditional simulation of the model on regular grids is presented.
 The  spatial prediction of missing data is based on the mean of the respective conditional distribution at the target site given the incomplete measurements.
The \mpr-based prediction is shown to be computationally efficient (due to sparse precision matrix structure and vectorization), and thus particularly suitable for remote-sensing data that are typically massive and collected in raster data format.



The remainder of the paper consists of five sections. Section~\ref{sec:mpr} has three goals. First, we present the \mpr Gibbs Markov random field model. Then, we propose a method for estimating the key model parameter (temperature) based on the matching of sample-based and expected (ensemble averaged) constraints. Finally, we develop an algorithm for the computationally efficient conditional simulation of \mpr realizations on regular grids.
The conditional mean of the simulation ensemble is proposed as the \mpr prediction of the missing grid data. In section~\ref{sec:vali-design} we present the design of the validation approach that employs comparisons between the \mpr predictions with those of commonly used spatial interpolators in terms of various statistical measures.
Section~\ref{sec:gap-fill} presents and analyzes the results of the validation studies based on both synthetic data (Gaussian random fields with Whittle-Mat\'{e}rn covariance function) and real data (non-Gaussian measurements of latent heat release and Walker lake data). Section~\ref{sec:discuss} further explores the proposed specific-energy-matching parameter inference method and comments on the computational efficiency of the \mpr method. Finally, Section~\ref{sec:conclusions} lists our conclusions and highlights certain topics for further research.

\section{Model Definition, Parameter Inference, and Simulation}
\label{sec:mpr}

Let $(\Omega,\mathcal{F},P)$ denote a probability space and $G\subseteq \Z^{2}$ a two-dimensional (2D) rectangular grid $G$ of size $N_{G}=L_{x} \times L_{y}$.
$L_{x}$ and $L_{y}$ represent the number of nodes in the horizontal and vertical directions, respectively. For simplicity, but without loss of generality, we will consider
square grids, i.e., $L_x=L_y \equiv L$. The grid sites are denoted by the vectors $\mathbf{r}_{i}=(x_i, y_i) \in {\mathbb{R}}^2$, where $i=1, \ldots, N_{G}$ and
 $\R$ is the set of real numbers.

We consider continuously-valued 2D lattice random fields $Z(\bfr;\omega)$ that represent mappings from  $\Omega\times\,\Z^{d}$, where here and in the following $d=2$, into $V=[v_{1}, v_{2}] \subset \R$. We assume that the data represent a realization of the random field $Z(\bfr;\omega)$ sampled on   $G_{s} \subset G$, where $G_{s}= \{\mathbf{r}_{i}\}_{i=1}^{N}$ and  $N<N_{G}$. The values of the data set are denoted by $Z_{s}=\{z_{i} \in {\mathbb{R}} \}_{i=1}^{N}$.
The set of prediction points is denoted by $G_{p}=\{\mathbf{r}_{p}\, \}_{p=1}^{P} $ such that $G_{s} \cup G_{p} = G$, $G_{s} \cap G_{p} = \emptyset$, and $P+N = N_{G}$.
The set of the random field values at the prediction sites will be denoted by $Z_{p}$.

The joint density of the lattice random field  is assumed to follow the Boltzmann-Gibbs functional form, i.e.,
\begin{equation}
 \label{eq:bg-pdf}
 f =\frac{1}{{\mathcal Z}} \exp(-{\mathcal H}/k_{B}T),
\end{equation}
 where the normalization constant ${{\mathcal Z}}$ is the partition function,
 $k_B$ is the Boltzmann constant, $T$ is the temperature parameter (higher temperature favors larger fluctuation variance), and ${\mathcal H}$ is an energy term that measures the ``cost'' of each configuration, so that higher cost
 configurations have a lower occurrence probability than lower cost ones. As we show below, the Boltzmann constant can be absorbed in the coupling parameter.

\subsection{Data transformation to spin space}
Let the lattice \emph{spin vector random field}  $\mathbf{S}(\bfr;\omega)=\left(S_{1}(\bfr;\omega), S_{2}(\bfr;\omega)\right)^\top$  denote a mapping   $\Omega\times\,\Z^{d} \mapsto \mathbf{S}_{2}$, where $\mathbf{S}_{2}$ denotes the set of all  unit vectors in the plane.
This field is uniquely determined by the scalar \emph{spin angle} field $\Phi(\bfr;\omega): \Omega\times\,\Z^{d} \mapsto [0, 2\pi]$ that represents the orientation of the unit spin vector in the plane.

Let a monotonic transformation $U: V \to [0, 2\pi]$ so that $Z(\bfr;\omega) \mapsto U[Z(\bfr;\omega)] = \Phi(\bfr;\omega)$ provide the mapping from the
original space $V$ to the spin angle space $[0, 2\pi]$.
Assuming ergodicity so that the data sample the entire space $V$,  the following linear transformation can be used

\begin{equation}
\label{map}
Z_{s} \mapsto \Phi_{s} = \frac{2\pi(Z_{s}- z_{s,\min})}{(z_{s,\max} - z_{s,\min})},
\end{equation}
where $z_{s,\min}$ and $z_{s,\max}$ are the minimum and maximum sample values and $\Phi_{s}=\{\phi_{i}\}_{i=1}^{N}$ and $\phi_{i} \in [0,2\pi]$, for $i=1, \ldots, N$.

%

\subsection{Definition of the \mpr Gibbs Markov random field}
The \mpr Gibbs Markov random field is defined by
 means of the Boltzmann-Gibbs distribution~\eqref{eq:bg-pdf} with energy ${\mathcal H}$ given by the following  expression
\begin{equation}
\label{Hamiltonian_mod}
{\mathcal H}=-J\sum_{\langle i,j \rangle}\cos[q(\phi_i-\phi_j)],
\end{equation}
where  $J>0$ is the \emph{exchange interaction parameter}, $\langle i,j \rangle$ denotes the sum over nearest neighbor spins on the grid, and $q \leq 1/2$ is the \emph{modification factor}.
The exponent of the joint density~\eqref{eq:bg-pdf}  contains the factor $J/k_{B}T$ which combines the temperature with the constants $k_B$ and $T$. Without loss of generality, we replace $k_{B}T/J$ with a \emph{``reduced temperature''} $T$ by setting $J=k_{B}=1$. We use open boundary conditions, so that the boundary nodes have a reduced number of nearest neighbors.

Equation~\eqref{Hamiltonian_mod} differs from the well known in physics planar rotator (or classical XY) spin model~\citep{Chaikin00} due to the modification factor $q$
($q=1$ corresponds to the standard planar rotator model.) Non-integer values of $q$ allow the emergence of correlations that are typical in geophysical and environmental applications~\citep{mz-dth15}. In particular, the
slowly (power-law) decaying correlation function that is characteristic of the Kosterlitz-Thouless phase in the standard XY model~\citep{kost73}, changes in the \mpr model to short-range dependence that is reasonably well modeled by Whittle-Mat\'{e}rn covariance functions (a more detailed study will be presented elsewhere).

\begin{figure}[t]
\centering
\includegraphics[scale=0.7,clip]{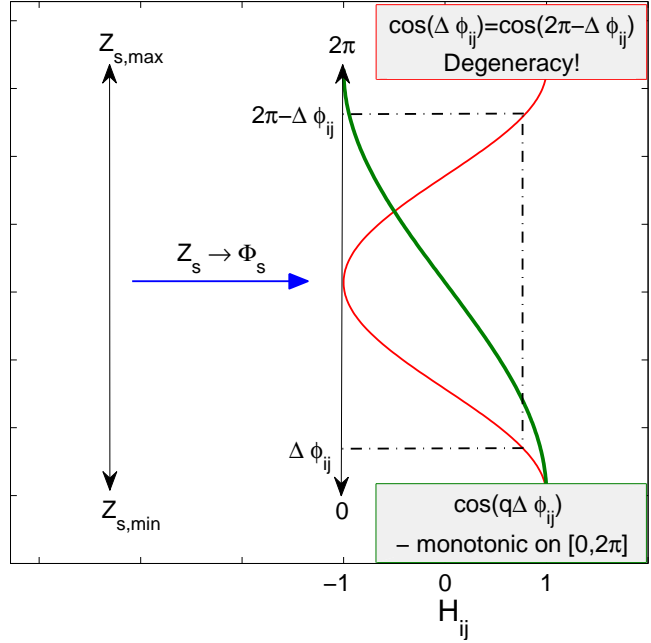}
\caption{Mapping of continuous data $Z_{s}=\{z_{i}\}_{i=1}^{N}$ where  $z_{s,\min} \le z_{i} \le z_{s,\max}$, for $i=1, \ldots, N$, to the spin angle set $\Phi_{s}=\{\phi_{i}\}_{i=1}^{N}$, where $0 \le \phi_{i} \le 2\pi.$ The \emph{modified pair interaction}  $H_{ij}=\cos(q\Delta\phi_{ij})$, with $q=1/2$ (thick green curve) removes the degeneracy of $q=1$ in the standard planar rotator energy $H_{ij}=\cos(\Delta\phi_{ij})$ (thin red curve) and provides a one-to-one mapping  $[z_{s,\min}, z_{s,\max}] \mapsto [0,2\pi]$.}\label{fig:degen}
\end{figure}

The choice $q \leq 1/2$ enables a \emph{monotonic mapping} between the spin values corresponding to the angles $\phi_i$ and the actual process values. In the standard planar rotator model, spin pairs with \emph{contrast angles} $\Delta\phi_{ij}=\phi_i-\phi_j$ and $2\pi-\Delta\phi_{ij}$ are degenerate (indistinguishable); i.e., if $q=1$ both terms contribute the same amount to the energy in~\eqref{Hamiltonian_mod}. However, such combinations  correspond to significantly different pair contrasts $z_{i} - z_{j}$ in terms of actual process values
according to~\eqref{map}.
This is  not satisfactory for geostatistical data, since neighbors with similar values (lower contrast) are more likely (i.e., have lower energy) than
neighbors with higher contrast. The undesirable degeneracy is lifted in the \mpr model with $q \leq 1/2$, which renders the energy~\eqref{Hamiltonian_mod}
 a monotonically increasing function of $\Delta\phi_{ij} \in [0,2\pi]$ as illustrated in Fig.~\ref{fig:degen}.
In the following, we arbitrarily set the value of the modification factor to $q=1/2$.

The \emph{sample \mpr specific energy} of the data is equal to the sample energy per spin pair and is estimated by means of the following sample average
\begin{equation}
\label{eq:mpr-sse}
e_s = - \frac{1}{N_{SP}}\sum_{i = 1}^{N}\sum_{j \in nn(i)}\cos[q(\phi_i-\phi_j)],
\end{equation}
where $j \in nn(i)$ denotes the \emph{sum over the non-missing nearest neighbors} of the point $i$, and $N_{SP}$ represents
the \emph{total number of the nearest-neighbor sample pairs} on $G_{s}$.

\subsection{Parameter estimation}
\label{ssec:temp-infer}
The two characteristic parameters of the \mpr  model are the grid size $L$, which is fixed, and the reduced temperature $T$. The latter needs to be estimated from
the gappy data in agreement with the sample constraints and the \mpr model. We propose a temperature estimation method that is based on matching the
\emph{sample \mpr specific energy} defined by~\eqref{eq:mpr-sse} with the respective \emph{equilibrium \mpr specific energy} defined by~\eqref{ene} below.
We will refer to this estimation method as \emph{specific energy matching (SEM)}.

The \emph{equilibrium \mpr specific energy} is given by
\begin{equation}
\label{ene}
e(T,L) = \frac{\langle {\mathcal H} \rangle}{N_{GP}},
\end{equation}
where $\langle {\mathcal H} \rangle$ is the expectation of the \mpr energy over all probable states,
and $N_{GP}=2L(L-1)$ is the number of nearest-neighbor pairs on the $L \times L$ grid with open boundary conditions.
The equilibrium \mpr specific energy varies as a function of $L$ and $T$.
The expectation  $\langle {\mathcal H} \rangle$ is numerically evaluated using unconditional simulation of the \mpr model
as described in Section~\ref{ssec:mc} below.

The principle of \emph{specific energy matching} is analogous to the method of moments: assuming ergodic conditions, it posits that  $e_s=e(\hat{T},L)$,
where $e_s$ is given by~\eqref{eq:mpr-sse}, $e(\hat{T},L)$ by~\eqref{ene}, and  $\hat{T}$ is the characteristic temperature of the gappy sample.
If  $e(T,L)$ is a known invertible function, so that $e(T,L)=x$, then $T = e^{-1}(x|L)$,
where $e^{-1}(\cdot\,|L)$ is the inverse specific energy  for fixed $L$.
Thus, we can  uniquely identify the  temperature of the gappy data configuration from $e_s$ and $e(T,L)$ by means of $\hat{T}=e^{-1}(e_s|L)$.

\begin{figure}[t]
\centering
\includegraphics[scale=0.7,clip]{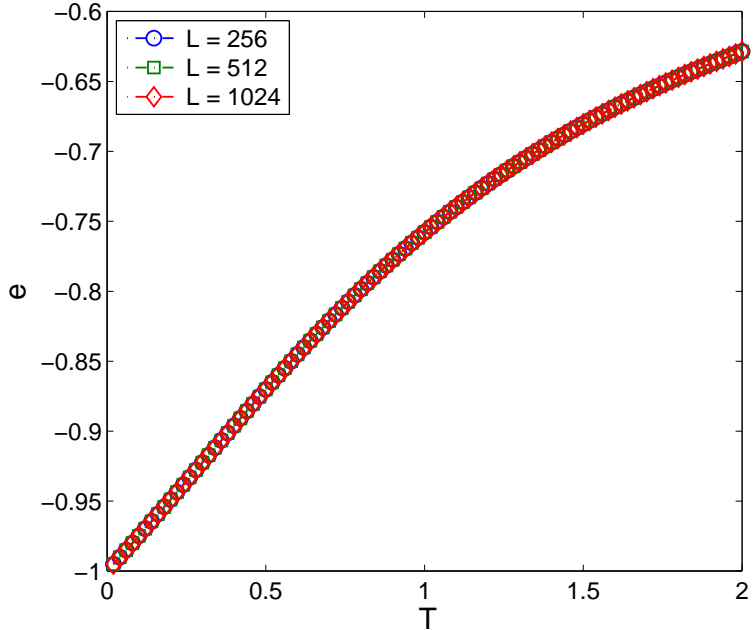}
\caption{Parametric curves of \mpr equilibrium specific energy $e(T,L)$ based on~\eqref{ene} as a function of temperature for three different grid sizes with length per side $L=256, 512, 1024$. The curves corresponding to the three values of $L$  collapse on each other.}\label{fig:ene}
\end{figure}

The function $e(T,L)$ is determined by calculating the equilibrium \mpr specific energy over any desired domain $[L_{\min}, L_{\max}]\times [T_{\min}, T_{\max}]$ of
 the $L$-$T$ parameter plane with some
fixed resolution $\Delta L, \Delta T$ that can be further refined by interpolation if needed.
As shown in Fig.~\ref{fig:ene}, the specific
energy varies smoothly with $T$ and is virtually independent of $L$.
Therefore, the curve $e(T,L)$ is calculated once, resulting in a look up table that can be used to estimate the temperature for all \mpr applications.

\subsection{Hybrid Monte Carlo update algorithm}
\label{ssec:mc}
The estimation of the energy expectation $\langle {\mathcal H} \rangle$  involves generating the ensemble of probable states by means of unconditional simulation of the \mpr model. The \mpr-based gap filling procedure involves the conditional simulation of the \mpr model at the estimated temperature $\hat{T}$. Both of these operations
require a method for the efficient exploration of the \mpr ensemble of states  (i.e., the configuration space).

We propose a \emph{hybrid Monte Carlo (MC) approach} that  combines the \emph{deterministic over-relaxation}~\citep{creutz87} and the \emph{stochastic Metropolis}~\citep{metro87} methods  (summarized in Algorithm~\ref{algo:mpr-relax}). This approach can be used for both   conditional and unconditional simulation. The main difference is that the sample set is empty and all the spins can vary in the latter case.
This implies significantly longer Monte Carlo sequences for unconditional simulation (typically $10^6$ MC sweeps for calculating equilibrium mean values at fixed temperature in addition to approximately $2\times10^5$ MC initial sweeps to reach equilibrium.)

In the \emph{initialization phase} the sampling locations are assigned the sample-derived values $\Phi_{s}$ that are kept fixed throughout the simulation. The remaining (prediction) locations are first initialized by spins with the set of spin angles $\hat{\Phi}_{p}^{(0)}$, where each angle is in the interval $[0,2\pi]$, and each spin angle value is updated according to the hybrid algorithm. The initial angle assignment is further discussed below.

In the \emph{over-relaxation update}, new spin angle values are obtained by a simple reflection of the spin about its local molecular field, generated by its nearest neighbors,
that conserves the energy.
This is accomplished by means of the following transformation
\begin{subequations}
\label{eq:over-relax}
\begin{align}
\phi'_{i}   = & \, (2\, \Phi_{i}-\phi_{i}) \mod {2\pi},
\\
\Phi_{i} = & \arctan2 \left( \sum_{j \in nn(i)} \sin{\phi_{j}}, \sum_{j \in nn(i)} \cos{\phi_{j}} \right),
\end{align}
\end{subequations}
where $nn(i)$ denotes the nearest neighbors of $\phi_{i}$, $i=1, \ldots, N$, and $\arctan2(\cdot)$ is the four-quadrant inverse tangent: for any  $x, y \in \mathbb{R}$
such that $|x| + |y| >0$, $\arctan2(y, x)$ is the angle (in radians) between the positive horizontal axis and the point  $(x, y)$. The over-relaxation transformation
reduces autocorrelations, and thus it can significantly speed up the \emph{relaxation process} (approach to equilibrium).
However, since it is energy-conserving and non-ergodic,  it has to be mixed with  Metropolis updates to achieve ergodicity and explore the probable energy states. In the standard $XY$ model such a hybrid update that uses an optimal ratio of Metropolis and over-relaxation sweeps achieves the correct dynamical critical exponent $z \approx 1.2$, in contrast with  $z \approx 2$ for the pure Metropolis algorithm~\citep{gupta88}.

The \emph{standard Metropolis} local update is rather inefficient, especially at low temperatures: At low $T$ most of the proposed local updates  get rejected, implying a very low \emph{acceptance ratio} $A$ (ratio of accepted updates over the total number of proposed updates). This regime is highly relevant to geostatistical simulations, because the presence of spatial correlations in the data implies a rather low \mpr temperature $T$. To increase the efficiency of the relaxation procedure we implement  a so-called \emph{restricted Metropolis algorithm} that generates a proposal spin-angle state according to the rule $\phi'=\phi+\alpha(r-0.5)$, where $r$ is a uniformly distributed random number $r \in [0,1)$ and $\alpha=2\pi/a \in (0,2\pi)$ is an adjustable scale factor (tunable parameter). The latter is automatically reset during the equilibration (typically reduced at lower $T$ and increased at higher $T$) to maintain the acceptance ratio $A$ close to a target value $A_{\textrm{targ}}$. Empirically, it is found that $A$ is controlled reasonably well by increasing the \emph{perturbation control factor} $a$ in linear proportion  to the simulation time, when $A$ drops below $A_{\textrm{targ}}$.

The proposed Metropolis state is then accepted or rejected with probability
\[
P(\Delta \mathcal{H}_i)=\min\{1,\exp(-\Delta \mathcal{H}_i/T)\},
\]
where $\Delta \mathcal{H}_i$ is the energy difference between the ``new state,'' generated by changing the value of the {\it i}-th spin angle and the old, i.e.,
\[
\Delta \mathcal{H}_i=\mathcal{H}_i^{\textrm{new}}-\mathcal{H}_i^{\textrm{old}}=-\sum_{j \in nn(i)}\{\cos[q(\phi'_i-\phi_j)]-\cos[q(\phi_i-\phi_j)]\}.
\]

\begin{figure}[t!]
\centering
\includegraphics[scale=0.7,clip]{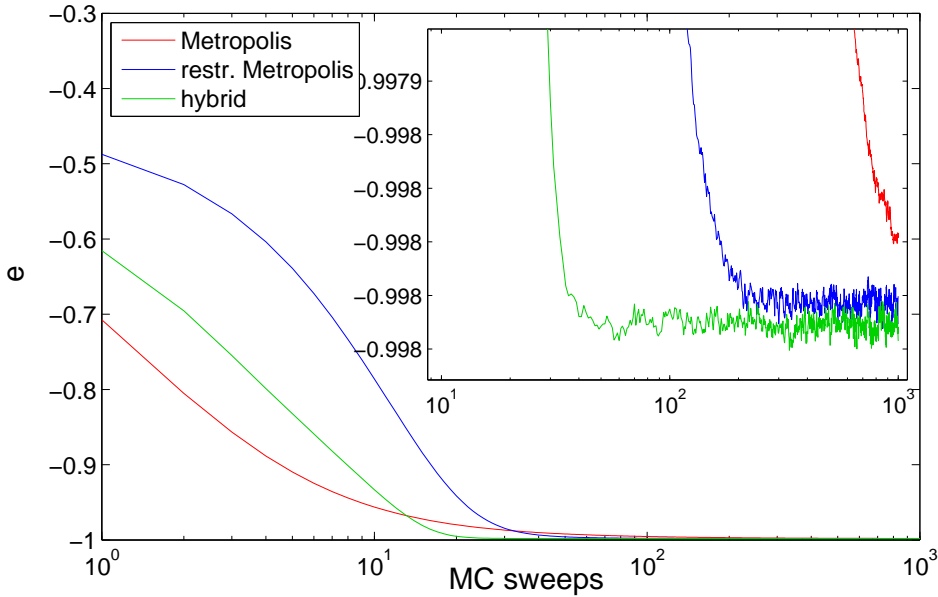}
\caption{Relaxation process demonstrated by the \mpr energy evolution under the application of the standard Metropolis,  restricted Metropolis and hybrid (restricted Metropolis combined with over-relaxation) algorithms. Synthetic data are generated by simulating a Gaussian random field with Whittle-Mat\'{e}rn covariance having parameters $\kappa=0.5$ and $\nu = 0.5$ on a square grid of size $L=512$ and by randomly removing  $p=90\%$ of the data.}\label{fig:equil_ene}
\end{figure}

The \emph{hybrid algorithm}  combining the \emph{restricted Metropolis} with  \emph{over-relaxation dynamics} can reduce the \mpr relaxation time by several orders of magnitude.  Fig.~\ref{fig:equil_ene} demonstrates the efficiency of the hybrid MC method with respect to standard and restricted Metropolis updates.
The synthetic data are simulated from a Gaussian random field  with Whittle-Mat\'{e}rn covariance  ($\kappa=0.5$ and $\nu = 0.5$), on a grid with $L=512$ followed by
random removal of $p=90\%$  of the data (see Section~\ref{ssec:synthetic} for details). After an initial phase (up to about $10^3$ MC sweeps) of fast relaxation, standard Metropolis significantly slows down due to extremely low acceptance ratio (true equilibrium is not reached even after $10^6$ MC sweeps), while hybrid dynamics drives the \mpr model to equilibrium after $\approx 50$ MC sweeps. Moreover, our numerical experiments show that the number of hybrid MC sweeps necessary to reach the \mpr equilibrium is  insensitive to grid size, requiring roughly the same number of MC sweeps for equilibration even for the largest $L$ considered.

In the initial \emph{non-equilibrium phase} the energy follows a decreasing trend.  To automatically detect the crossover to equilibrium (flat regime in the curves of Fig.~\ref{fig:equil_ene}), the energy is periodically evaluated every  $n_f$ MC sweeps and the variable-degree polynomial Savitzky-Golay (SG) filter is applied~\citep{savgol64}. The sampling of equilibrium  configurations for the evaluation of ensemble averages (unconditional simulation) or conditional probability distributions (conditional simulation) begins
at the point where the trend disappears and the energy shows only fluctuations around a stable level.

\begin{algorithm}[t!]
\caption{Hybrid updating algorithm that combines deterministic over-relaxation with the stochastic Metropolis step. $\mathbf{\hat{\Phi}}^{\mathrm{old}}$ is the
initial spin state, and $\mathbf{\hat{\Phi}}^{\mathrm{new}}$ is the new spin state. $\mathbf{\hat{\Phi}}^{\mathrm{old}}_{-p}$ is the initial spin
state excluding the point labeled by $p$. $U(0,1)$ denotes  the uniform probability distribution in $[0, 1]$.}
\label{algo:mpr-relax}
\begin{algorithmic}
\Procedure{Update}{$\mathbf{\hat{\Phi}}^{\mathrm{new}},\mathbf{\hat{\Phi}}^{\mathrm{old}},a,\hat{T}$}
\For{$p=1, \ldots, P$}  \Comment Loop over prediction sites
\State 1: ${\hat{\Phi}'}_p \gets \mathcal{R}\{\hat{\Phi}_p^{\mathrm{old}} \}$ \Comment Over-relaxation step according to~\eqref{eq:over-relax}
\State 2:   $u \gets U(0,1)$ \Comment Generate uniform random number
\State 3:   ${\hat{\Phi}''}_p \gets {\hat{\Phi}'}_p + 2\pi (u -0.5)/a \pmod{2\pi}$ \Comment Propose spin update
\State 4:   $ \Delta \mathcal{H} = \mathcal{H}({\hat{\Phi}''}_p, \mathbf{\hat{\Phi}}^{\mathrm{old}}_{-p}) - \mathcal{H}({\hat{\Phi}'}_p, \mathbf{\hat{\Phi}}^{\mathrm{old}}_{-p})$ \Comment Calculate energy change
\State 5:  $AP = \min\{1,\exp(-\Delta \mathcal{H}/\hat{T})\}$ \Comment Calculate acceptance probability
\State 6:  $\mathbf{\hat{\Phi}}^{\mathrm{new}}_{-p} \gets \mathbf{\hat{\Phi}}^{\mathrm{old}}_{-p}$ \Comment Perform Metropolis update
            \If{$AP > r \gets U(0,1)$}
            \State 6.1: {$\hat{\Phi}_p^{\mathrm{new}} \gets {\hat{\Phi}''}_p$}    \Comment Update the state
            \Else
            \State 6.2: {$\hat{\Phi}_p^{\mathrm{new}} \gets {\hat{\Phi}'}_p$} \Comment Keep the current state
\EndIf
\EndFor \Comment End of prediction loop
\State 7: \Return $\mathbf{\hat{\Phi}}^{\mathrm{new}}$  \Comment Return the ``updated'' state
\EndProcedure

\end{algorithmic}
\end{algorithm}

\subsection{Details of hybrid Monte Carlo \mpr simulation}

\begin{algorithm}[t!]
\caption{Simulation  of \mpr model. The algorithm involves the hybrid updating procedure \textsc{Update} described in Algorithm~\ref{algo:mpr-relax}.
$\mathbf{\Phi}_{s}$ is the vector of known spin values at the sample sites. $\mathbf{\hat{\Phi}}$  represents the vector of estimated spin values at the prediction sites.
$\hat{T}$ is the estimated reduced temperature. $G(\cdot)$ is the transformation from the original field to the spin field and $G^{-1}(\cdot)$ is its inverse. $\mathbf{\hat{Z}}(j)$, $j=1, \ldots, M$ is the $j$-th realization of the original field. $\mathbf{U}(0,2\pi)$ denotes a vector of random numbers from the uniform probability distribution in $[0, 2\pi]$. }
\label{algo:mpr-simul}
\begin{algorithmic}
\State 1: \emph{Initialize simulation parameters}
\State 1.1: Set $M$ \Comment $\#$ equilibrium configurations for statistics collection
\State  1.2:  Set $n_{f}$ \Comment verification frequency of equilibrium conditions
\State  1.3:  Set $n_{\textrm{fit}}$ \Comment $\#$ fitting points of energy evolution function
\State  1.4:  Set $A_{\textrm{targ}}$ \Comment target acceptance ratio of Metropolis update
\State  1.5:  Set $k_a$ \Comment defines variation rate of perturbation control factor $a$
\State  1.6:  Set $i_{\max}$ \Comment Monte Carlo relaxation-to-equilibrium steps (optional)
\State  1.7:  $ i \gets 0$      \Comment Initialize simulated state counter
\State  1.8:  $\mathbf{\hat{\Phi}}(0) \gets \mathbf{U}(0,2\pi)$  \Comment Initialize missing spin values from  uniform distribution
\State  1.9:  $ k(0) \gets - 1$    \Comment Initialize slope of energy evolution function
\State  1.10:  $ a(0) \gets 1$      \Comment Set spin angle perturbation control factor

\State 2: \emph{Data transformation}
\State  2.1: $\mathbf{\Phi}_{s} \gets G(\mathbf{Z}_{s})$  using~\eqref{map} \Comment Set data spin angles

\State 3: \emph{Parameter Inference}
\State 3.1: Estimate $e_s$ using~\eqref{eq:mpr-sse} \Comment Find sample specific energy
\State 3.2: $\hat{T} \gets e^{-1}(e_s|L)$ \Comment Estimate reduced temperature based on $e(\hat{T},L)=e_s$

\State 4: \emph{Non-equilibrium spin relaxation procedure}
\While{$[k(i)<0] \wedge [i \le i_{\max}]$ }    \Comment Spin updating with hybrid step
\State 4.1: \Call{Update}{$\mathbf{\hat{\Phi}}(i+1), \mathbf{\hat{\Phi}}(i), a(i),\hat{T}$}
\If{$A < A_{\textrm{targ}}$} \Comment Check if Metropolis acceptance ratio is low
\State 4.2: $a(i+1) \gets 1+(i+1)/k_a$ \Comment Update perturbation control factor
\EndIf
\State 4.3: Calculate $e(i+1) \gets \mathcal{H}(G)/N_{GP}$ \Comment Obtain current specific energy
\If{$[i \geq n_{\textrm{fit}}] \wedge [i \equiv 0 \pmod{n_f}]$} \Comment Check frequency for slope update of $e$
\State 4.4:  $k(i+1) \gets \mathrm{SG}$ \Comment Update slope of $e$ by SG filter using last $n_{\textrm{fit}}$ values
\EndIf
\State 4.5: $ i \gets i +1$ \Comment Update MC counter
\EndWhile

\State 5. \emph{Equilibrium state simulation}
\State 5.1: $\mathbf{\hat{\Phi}}^{\mathrm{eq}}(0) \gets \mathbf{\hat{\Phi}}(i)$ \Comment Initialize the equilibrium state
\For{$j=0, \ldots, M-1$}
\State 5.2: \Call{Update}{$\mathbf{\hat{\Phi}}^{\mathrm{eq}}(j+1), \mathbf{\hat{\Phi}}^{\mathrm{eq}}(j), 1,\hat{T}$} \Comment Generate equilibrium realizations
\State 5.3: $\mathbf{\hat{Z}}(j+1)  \gets G^{-1}\left[\mathbf{\hat{\Phi}}^{\mathrm{eq}}(j+1)\right]$ \Comment Back-transform spin states
\EndFor
\State 6: \Return Statistics of $M$ realizations $\mathbf{\hat{Z}}(j), \, j=1, \ldots, M$

\end{algorithmic}
\end{algorithm}

Algorithm~\ref{algo:mpr-simul} summarizes the main steps of the \mpr method for conditional simulation of gappy data.
 To avoid undesirable boundary effects, we add auxiliary nodes around the grid that are assigned the same values as their nearest grid neighbors. The augmented grid is used
 with open boundary conditions. Therefore, if ${\mathbf s}_{i,j}$ is a spin in the $i$-th row and $j$-th column of the grid, where $i,j=1,\dots, L$, then ${\mathbf s}_{i,L+1}={\mathbf s}_{i,L}$, ${\mathbf s}_{L+1,j}={\mathbf s}_{L,j}$, ${\mathbf s}_{i,0}={\mathbf s}_{i,1}$ and ${\mathbf s}_{0,j}={\mathbf s}_{1,j}$, where the indices $0$ and $L+1$ refer to auxiliary nodes.

Algorithm~\ref{algo:mpr-simul} involves several control factors that include: the number of equilibrium configurations for collecting statistics, $M$, the frequency of verification of equilibrium conditions, $n_f$, the number of points used for fitting the energy evolution function, $n_{\textrm{fit}}$,  the maximum number of Monte Carlo steps $i_{\max}$ (optional), and the parameters  $A_{\textrm{targ}}$ and $k_a$ used
in the restricted Metropolis update. Below, we comment on the selection of these factors and their impact on prediction performance.
\begin{itemize}
\item $M$ is set arbitrarily, depending on whether the main goal is computational efficiency or prediction performance. Lower (higher) values of $M$ increase (decrease) the computational speed and  decrease (increase) accuracy and precision. For the conditional  simulations we set $M=100$.

\item High-frequency checking of equilibrium conditions (small $n_f$) slightly slows down the simulation but can also lead to earlier onset of equilibrium calculations. A reasonable value of $n_f$ is determined based on the maximum total equilibration time. In conditional simulation the latter is about 50 MCS and $n_f=5$ in all cases.

\item The parameter $n_{\textrm{fit}}$ defines the memory length of the energy time series used to test  the onset of equilibrium. Close to equilibrium, where fluctuations can be considerable, $n_{\textrm{fit}}$ should be sufficiently large to ensure a robust fit (i.e., to distinguish the fluctuations from the trend). We found empirically that $n_{\textrm{fit}}=20$ is adequate for this purpose.

\item The factor  $i_{\max}$  prevents very long equilibration times,  if the convergence is very slow. Since the employed hybrid algorithm leads to very fast equilibration, the value of $i_{\max}$ is practically irrelevant.

\item We tested several initialization approaches for the spin angle state, including uniform and random assignments that correspond respectively to the ``ferromagnetic'' (cold start) and ``paramagnetic'' (hot start) initializations, typically used in spin system simulations. In conditional simulation we also tried configurations obtained by simple and fast interpolation of the sample data, e.g.,  nearest neighbors  and bilinear  methods. Since different initializations did not produce significant differences, we use the ``paramagnetic'' state as default with random values drawn from the uniform distribution in $[0,2\pi]$.

\item The adjustable scale parameters $A_{\textrm{targ}}$ and $k_a$ are introduced to avoid low-temperature inefficiency due to the  Metropolis acceptance ratio dropping to low values. Since their actual values appear to have little influence on the prediction performance, we arbitrarily set them to $A_{\textrm{targ}}=0.3$ and $k_a=3$.
\end{itemize}
In conclusion, the effect of the Monte Carlo simulation control factors on prediction performance is marginal. Thus,  the default values set above can be safely used in general. Combined with the fact that the temperature estimation is straightforward and does not require parameter tuning, this means that the \mpr conditional simulation method can be automatically applied without user intervention.

\section{Design of MPR-Prediction Validation and Comparison}
\label{sec:vali-design}
The \mpr model and Algorithms~\ref{algo:mpr-relax}-\ref{algo:mpr-simul} provide a framework for fast conditional simulation. The \mpr
predictions are based on the conditional mean as evaluated from the conditionally simulated reconstructions.
 We assess the \mpr performance as a gap-filling method by comparison with  established interpolation methods
using both synthetic and real data. We simulate missing values by setting aside a portion of the complete data to use as  \emph{validation set}.

The \mpr comparison with interpolation methods is implemented in the Matlab\circledR\ environment running on a desktop computer with 16.0 GB RAM and an Intel\textregistered Core\texttrademark2 i7-4790 CPU processor with an 3.60 GHz clock. The methods tested involve the triangulation-based nearest neighbor (NN), bilinear (BL) and bicubic (BC) interpolation using the built-in function \verb+griddata+, as well as the minimum curvature (MC) (or biharmonic spline) method~\citep{sand87}. We also include the deterministic inverse distance weighted (IDW)~\citep{shep68} interpolation, using the Matlab\circledR\, function \verb+fillnans+~\citep{Howat}, and the stochastic ordinary kriging (OK) method~\citep{wack03}, using the routines available in the Matlab\circledR\, library \verb+vebyk+~\citep{Sidler03}. We note that a number of functions useful in spatial and spatio-temporal geostatistical modelling can be also found in the freely distributed  R programming environment, such as the package \verb+gstat+~\citep{Pebe04,Pebe16}. The IDW, MC and OK methods are applied using the entire sample data set (without search neighborhoods). OK is applied to the Gaussian data using the ``true'' covariance parameters. Thus, it provides optimal predictions that serve as a standard for comparison with the \mpr estimates. The above spatial interpolation methods are commonly used in the environmental sciences~\citep{Li14}.

We employ several validation measures for performance comparison. Let $Z(\mathbf{r}_p)$ be the true value at $\mathbf{r}_p$ and $\hat{Z}(\mathbf{r}_p)$ its estimated value. The estimation error is defined as $\epsilon(\mathbf{r}_p)= Z(\mathbf{r}_p) - \hat{Z}(\mathbf{r}_p)$. The following validation measures are then defined:

\noindent Average absolute error
\begin{equation}
{\rm AAE} = (1/P)\sum_{\mathbf{r}_{p} \in G_{p}}|\epsilon(\mathbf{r}_p)|
\end{equation}
Average relative error
\begin{equation}
{\rm ARE}=(1/P)\sum_{\mathbf{r}_{p} \in G_{p}}\epsilon(\mathbf{r}_p)/Z(\mathbf{r}_p)
\end{equation}
Average absolute relative error
\begin{equation}
{\rm AARE} =(1/P)\sum_{\mathbf{r}_{p} \in G_{p}}|\epsilon(\mathbf{r}_p)|/Z(\mathbf{r}_p)
\end{equation}
Root average squared error
\begin{equation}
{\rm RASE} =\sqrt{\frac{1}{P}\sum_{\mathbf{r}_{p} \in G_{p}}\,\epsilon^2(\mathbf{r}_p)}.
\end{equation}
The above are complemented by the linear correlation coefficient $R$. Furthermore, for each method we record the  required CPU time, $t_{\mathrm{cpu}}$.
For each complete data set we generate $S$ different sample configurations with missing data and calculate the above validation measures. Global statistics,
denoted by MAAE, MARE, MAARE, MRASE, MR and $\langle t_{\mathrm{cpu}} \rangle$, are then calculated by averaging over all the sample configurations.

\section{Gap-filling Validation Results}
\label{sec:gap-fill}
\subsection{Synthetic data}
\label{ssec:synthetic}

Synthetic data are simulated on the square grid from the Gaussian random field $Z \sim N(m=50,\sigma=10)$ with Whittle-Mat\'{e}rn (WM) covariance given by
\begin{equation}
\label{mate} G_{\rm Z} (\|\mathbf{h}\|)=
\frac{{2}^{1-\nu}\, \sigma^{2}}{\Gamma(\nu)}(\kappa \,
\|\mathbf{h}\|)^{\nu}K_{\nu}(\kappa \, \|\mathbf{h}\|),
\end{equation}
where $\|\mathbf{h}\|$ is the Euclidean two-point distance, $\sigma^2$
is the variance, $\nu$ is the smoothness parameter, $\kappa$ is the
inverse autocorrelation length, and $K_{\nu}$ is the modified Bessel
function of index $\nu$. Hereafter, only the parameters $\kappa$ and $\nu$ will change. For such data we  use the abbreviation WM($\kappa,\nu)$.
The field is sampled on a square grid $G$ using the spectral method~\citep{drum87}. Incomplete samples  ${Z}(G_{s})$ of size $N=N_{G} - \lfloor
(p/100\%)\,N_{G} \rfloor$ are generated  by removing (i) randomly $P=\lfloor (p/100\%)\,N_{G} \rfloor$ points or (ii) a randomly selected solid square block of side length $L_B$.  For different degrees of thinning ($p=33\%$ and $66\%$) and block size ($L_B=5$ and $20$), we generate $S=100$ different sampling configurations. The predictions at the removed (validation) points are calculated and compared with the true values.

The WM family is flexible and includes several variogram models~\citep{Minasny05,pardo08,pardo09}. Small values of $\nu$, e.g., $\nu=1/2$, which is equivalent to the exponential model, imply that the spatial process is rough. On the other hand, large values, e.g., $\nu \to \infty$, which is equivalent to the Gaussian model, generate smooth processes.
In our simulations we use $\nu=0.25-0.5$, which is   appropriate for modeling rough spatial processes such as soil data~\citep{Minasny05}.

\begin{table}[t!]
\addtolength{\tabcolsep}{0pt} \caption{Interpolation validation
measures for the \mpr method and relative values, ${\rm XX^*=\mpr/XX}$, for all other methods. $S=100$ samples are generated from a Gaussian random field with mean equal to 50 on  a square grid with side length $L=16$. Two covariance models, WM($\kappa=0.5,\nu = 0.5$) and WM($\kappa=0.5,\nu = 0.25$) are used. Missing data are generated by (a) $p=33\%$ (b) $p=66\%$ random thinning and (c) random removal of square data block  with side length $L_B=5$. Boldfaced values denote that the respective method performs better than \mpr for the specific validation measure.} \vspace{3pt} \label{tab:synt_int_L16a}
\begin{scriptsize}
\resizebox{1\textwidth}{!}{
\begin{tabular}{|c|c|ccc|ccc|ccc|ccc|ccc|ccc|}
\hline
& & \multicolumn{3}{c|}{MAAE}  & \multicolumn{3}{c|}{MARE [\%]} &
 \multicolumn{3}{c|}{MAARE [\%]} & \multicolumn{3}{c|}{MRASE} &
 \multicolumn{3}{c|}{MR [\%]}  & \multicolumn{3}{c|}{$ \langle t_{\mathrm{cpu}} \rangle $}   \\
$\nu$ & & (a) & (b) & (c) & (a) & (b) & (c) & (a) & (b) & (c) & (a) & (b) & (c) & (a) & (b) & (c) & (a) & (b) & (c) \\
\hline
&\mpr &5.17&5.92&6.38&$-$1.89&$-$2.37&$-$0.10&11.11&12.75&12.70&6.56&7.47&7.89&73.01&63.90&37.62&0.01&0.02&0.01 \\
&${\rm NN^*}$ &0.75&0.81&0.80&{\bf 1.26}&{\bf 1.13}&$-$0.39&0.76&0.82&0.81&0.76&{\bf 1.16}&0.80&1.23&1.16&1.26&{\bf 10.63}&{\bf 15.69}&{\bf 7.89} \\
&${\rm BL^*}$ &0.94&0.98&0.88&{\bf 1.37}&{\bf 2.08}&$-$0.21&0.96&{\bf 1.01}&0.89&0.96&0.99&0.89&1.23&1.17&1.25&{\bf 8.00}&{\bf 14.08}&{\bf 5.45} \\
$0.5$ &${\rm BC^*}$ &0.95&0.97&0.86&{\bf 1.29}&{\bf 1.56}&$-$0.27&0.96&1.00&0.87&0.96&0.98&0.87&1.02&1.00&1.35&{\bf 12.99}&{\bf 21.40}&{\bf 8.83} \\
&${\rm MC^*}$ &0.96&0.97&0.93&{\bf 1.62}&{\bf 1.65}&0.08&0.96&0.98&0.93&0.97&0.96&0.93&{\bf 1.00}&{\bf 0.97}&1.01&{\bf 3.36}&{\bf 6.74}&{\bf 2.13} \\
&${\rm IDW^*}$ &0.99&{\bf 1.01}&0.99&0.82&0.93&0.26&0.99&{\bf 1.01}&0.99&0.99&1.00&0.99&1.01&{\bf 0.99}&{\bf 0.96}&{\bf 4.49}&{\bf 5.23}&{\bf 7.50} \\
&${\rm OK^*}$ &{\bf 1.04}&{\bf 1.05}&{\bf 1.03}&1.00&0.99&{\bf 1.06}&{\bf 1.04}&{\bf 1.04}&{\bf 1.03}&{\bf 1.04}&{\bf 1.05}&{\bf 1.03}&{\bf 0.97}&{\bf 0.94}&{\bf 0.89}&1e$-$4&5e$-$4&5e$-$5 \\
\hline
&\mpr &7.04&7.63&7.90&$-$3.19&$-$3.60&$-$3.90&15.21&16.62&17.12&8.75&9.44&9.59&47.87&36.69&29.77&0.01&0.02&0.01 \\
&${\rm NN^*}$ &0.78&0.81&0.78&{\bf 1.01}&{\bf 1.02}&{\bf 1.02}&0.79&0.82&0.80&0.78&0.81&0.79&1.37&1.17&1.43&{\bf 10.99}&{\bf 15.92}&{\bf 7.45} \\
&${\rm BL^*}$ &0.96&0.96&0.89&{\bf 1.16}&{\bf 1.11}&0.85&0.97&0.97&0.89&0.95&0.96&0.89&1.35&1.13&1.40&{\bf 8.29}&{\bf 14.04}&{\bf 5.13} \\
$0.25$ &${\rm BC^*}$ &0.94&0.94&0.87&{\bf 1.18}&{\bf 1.12}&0.88&0.95&0.95&0.87&0.93&0.93&0.86&1.04&{\bf 1.00}&1.58&{\bf 13.33}&{\bf 21.23}&{\bf 8.52} \\
&${\rm MC^*}$ &0.94&0.92&0.88&{\bf 1.20}&{\bf 1.07}&{\bf 1.18}&0.94&0.92&0.89&0.93&0.91&0.88&{\bf 0.99}&{\bf 0.97}&1.08&{\bf 3.44}&{\bf 6.66}&{\bf 2.01} \\
&${\rm IDW^*}$ &{\bf 1.01}&0.99&0.98&0.98&1.00&0.88&{\bf 1.01}&0.99&0.98&{\bf 1.01}&0.99& 0.98&{\bf 0.98}&{\bf 0.95}&{\bf 0.98}&{\bf 4.54}&{\bf 5.27}&{\bf 6.70} \\
&${\rm OK^*}$ &{\bf 1.01}&{\bf 1.02}&{\bf 1.01}&0.94&0.91&0.88&1.00&{\bf 1.01}&{\bf 1.01}&{\bf 1.02}&{\bf 1.03}&{\bf 1.02}&{\bf 0.97}&{\bf 0.92}&{\bf 0.91}&1e$-$4&5e$-$4&4e$-$5 \\
\hline
\end{tabular}
}
\end{scriptsize}
\end{table}

\subsubsection{Small grid size}

In Table~\ref{tab:synt_int_L16a} we present the \mpr interpolation validation measures for the smallest ($L=16$) grid size using an autocorrelation length $1/\kappa=2$. For this data size we  compare the performance of the \mpr model with all the methods presented above, including the optimal but computationally intensive  OK method. The actual values of the validation measures are shown only for the \mpr method. For  other methods XX (= NN, BL, BC, MC, IDW, OK) the validation measures are expressed relative to the \mpr method, i.e., ${\rm XX^*=\mpr/XX}$. Therefore, MAAE, MARE, MAARE, MRASE and $\langle t_{\mathrm{cpu}} \rangle$ less than one and correlation coefficient values higher than one indicate superior performance of \mpr. Boldfaced values denote that the respective method performs better than \mpr with respect to the specific validation measure or $\langle t_{\mathrm{cpu}} \rangle$.

The \mpr method performs better than the NN, BL and BC methods in terms of most measures, except for the MARE errors in the case of randomly thinned data and the CPU time.
The relative ``slowness'' of \mpr is due to the fact that the method performs conditional simulation (hence, it generates considerably more information than an ``optimal'' prediction).
The MC method returns better MARE while its MR is comparable with the \mpr method. Even better results are obtained with IDW, albeit still slightly worse than the \mpr  (except for the MR indicator and the CPU time). Comparing the \mpr and OK methods,  OK is optimal with respect to all the measures except for the MARE errors and the CPU time. The superior prediction performance of OK is not surprising, considering that it is the optimal model for Gaussian data with known covariance parameters. On the other hand, if a search neighborhood is not specified, the CPU time required by OK exceeds that of \mpr by about four orders of magnitude.

\subsubsection{Larger grids}

\begin{table}[t]
\addtolength{\tabcolsep}{0pt} \caption{Interpolation validation measures for the \mpr method and relative values, ${\rm XX^*=\mpr/XX}$, for the other methods except OK. $S=100$ samples are generated from a Gaussian random field with mean equal to 50 on  a square grid with side length $L=32, 64, 128$. The covariance model WM($\kappa=0.2,\nu = 0.5$) is used. Missing data are generated by (a) $p=33\%$ (b) $p=66\%$ random thinning and (c) random removal of square data block  with side length $L_B=20$. Boldfaced values denote that the respective method performs better than \mpr for the specific validation measure.} \vspace{3pt} \label{tab:synt_int_La}
\begin{scriptsize}
\resizebox{1\textwidth}{!}{
\begin{tabular}{|c|c|ccc|ccc|ccc|ccc|ccc|ccc|}
\hline
& & \multicolumn{3}{c|}{MAAE}  & \multicolumn{3}{c|}{MARE [\%]} &
 \multicolumn{3}{c|}{MAARE [\%]} & \multicolumn{3}{c|}{MRASE} &
 \multicolumn{3}{c|}{MR [\%]}  & \multicolumn{3}{c|}{$ \langle t_{\mathrm{cpu}} \rangle $}   \\
 $L$ & & (a) & (b) & (c) & (a) & (b) & (c) & (a) & (b) & (c) & (a) & (b) & (c) & (a) & (b) & (c) & (a) & (b) & (c) \\
\hline
&\mpr &3.48&4.03&6.95&$-$1.08&$-$1.46&$-$4.64&7.41&8.70&16.12&4.36&5.10&8.98&90.66&87.12&45.73&0.02&0.04&0.03 \\
&${\rm NN^*}$ &0.70&0.77&0.79&{\bf 1.48}&{\bf 1.82}&{\bf 1.02}&0.70&0.78&0.81&0.70&0.78&0.79&1.11&1.10&1.30&{\bf 9.47}&{\bf 17.67}&{\bf 7.72} \\
&${\rm BL^*}$ &0.95&0.97&0.85&{\bf 1.14}&{\bf 1.07}&0.86&0.95&0.98&0.85&0.95&0.97&0.87&1.11&1.09&1.27&{\bf 5.31}&{\bf 12.71}&{\bf 6.46} \\
$32$ &${\rm BC^*}$ &0.96&0.98&0.84&{\bf 1.57}&{\bf 1.45}&0.93&0.97&1.00&0.85&0.96&0.98&0.86&1.01&1.01&1.29&{\bf 10.05}&{\bf 23.06}&{\bf 12.20} \\
&${\rm MC^*}$ &{\bf 1.01}&{\bf 1.02}&0.90&{\bf 1.88}&{\bf 1.90}&{\bf 2.59}&{\bf 1.03}&{\bf 1.04}&0.94&{\bf 1.00}&{\bf 1.02}&0.91&{\bf 1.00}&{\bf 1.00}&{\bf 1.00}&0.59&{\bf 1.72}&0.74 \\
&${\rm IDW^*}$ &0.97&0.98&{\bf 1.00}&0.89&{\bf 1.17}&0.75&0.96&0.98&0.99&0.97&0.98&1.00&1.01&1.01&{\bf 0.94}&0.69&{\bf 1.11}&0.73 \\
\hline
&\mpr &3.38&3.86&6.69&$-$0.92&$-$1.28&$-$3.29&7.13&8.16&14.55&4.27&4.89&8.42&89.38&85.92&49.09&0.07&0.13&0.03 \\
&${\rm NN^*}$ &0.70&0.77&0.84&{\bf 1.43}&{\bf 1.62}&{\bf 1.18}&0.71&0.78&0.85&0.70&0.78&0.83&1.12&1.10&1.21 &{\bf 8.30}&{\bf 17.55}&{\bf 2.99} \\
&${\rm BL^*}$ &0.94&0.97&0.88&{\bf 1.09}&{\bf 1.17}&0.96&0.94&0.97&0.88&0.94&0.97&0.88&1.13&1.10&1.21&{\bf 4.14}&{\bf 11.39}&{\bf 1.45} \\
$64$ &${\rm BC^*}$ &0.94&0.97&0.86&{\bf 1.34}&{\bf 1.43}&{\bf 1.04}&0.95&0.98&0.87&0.95&0.97&0.87&1.01&1.01&1.36&{\bf 8.40}&{\bf 22.79}&{\bf 2.98} \\
&${\rm MC^*}$ &0.98&0.99&0.90&{\bf 1.61}&{\bf 1.70}&{\bf 1.47}&0.99&{\bf 1.00}&0.92&0.98&1.00&0.89&1.00&1.00&1.04&0.09&0.36&0.02 \\
&${\rm IDW^*}$ &0.97&0.97&0.98&0.88&{\bf 1.12}&0.94&0.97&0.98&0.98&0.97&0.98&0.98&1.01&1.01&1.08&0.20&0.25&0.27 \\
\hline
&\mpr &3.45&3.89&6.21&$-$1.05&$-$1.36&$-$3.89&7.37&8.34&13.75&4.34&4.90&7.86&90.50&87.71&55.40&0.27&0.49&0.05 \\
&${\rm NN^*}$ &0.71&0.77&0.82&{\bf 1.27}&{\bf 1.47}&{\bf 1.38}&0.71&0.78&0.85&0.71&0.77&0.81&1.10&1.09&1.24&{\bf 6.95}&{\bf 14.94}&{\bf 1.08} \\
&${\rm BL^*}$ &0.94&0.97&0.87&{\bf 1.08}&{\bf 1.15}&{\bf 1.15}&0.94&0.97&0.89&0.94&0.97&0.88&1.10&1.09&1.24&{\bf 3.43}&{\bf 9.54}&0.44 \\
$128$ &${\rm BC^*}$ &0.94&0.96&0.86&{\bf 1.28}&{\bf 1.37}&{\bf 1.16}&0.95&0.97&0.87&0.94&0.96&0.87&1.01&1.01&1.31&{\bf 6.94}&{\bf 18.79}&0.95 \\
&${\rm MC^*}$ &0.98&0.98&0.85&{\bf 1.51}&{\bf 1.65}&{\bf 1.06}&0.99&0.99&0.88&0.98&0.98&0.86&1.00&1.00&1.10&0.01&0.07&8e$-$4 \\
&${\rm IDW^*}$ &0.97&0.97&0.98&0.89&{\bf 1.08}&{\bf 1.12}&0.97&0.98&0.98&0.97&0.97&0.98&1.01&1.01&1.04&0.08&0.15&0.08 \\
\hline
\end{tabular}
}
\end{scriptsize}
\end{table}

\begin{table}[t!]
\addtolength{\tabcolsep}{0pt} \caption{Interpolation validation measures for the \mpr method and relative values, ${\rm XX^*=\mpr/XX}$, for the other methods except OK. $S=100$ samples are generated from a Gaussian random field with mean equal to 50 on  a square grid with side length $L=32, 64, 128$. The covariance model WM($\kappa=0.2,\nu = 0.25$) is used. Missing data are generated by (a) $p=33\%$ (b) $p=66\%$ random thinning and (c) random removal of square data block  with side length $L_B=20$. Boldfaced values denote that the respective method performs better than \mpr for the specific validation measure.} \vspace{3pt} \label{tab:synt_int_Lb}
\begin{scriptsize}
\resizebox{1\textwidth}{!}{
\begin{tabular}{|c|c|ccc|ccc|ccc|ccc|ccc|ccc|}
\hline
& & \multicolumn{3}{c|}{MAAE}  & \multicolumn{3}{c|}{MARE [\%]} &
 \multicolumn{3}{c|}{MAARE [\%]} & \multicolumn{3}{c|}{MRASE} &
 \multicolumn{3}{c|}{MR [\%]}  & \multicolumn{3}{c|}{$ \langle t_{\mathrm{cpu}} \rangle $}   \\
 $L$ & & (a) & (b) & (c) & (a) & (b) & (c) & (a) & (b) & (c) & (a) & (b) & (c) & (a) & (b) & (c) & (a) & (b) & (c) \\
\hline
&\mpr &5.50&5.81&6.62&$-$1.74&$-$1.90&$-$2.32&11.54&12.19&13.70&6.86&7.32&8.34&65.93&59.84&31.26&0.02&0.04&0.03 \\
&${\rm NN^*}$ &0.76&0.79&0.75&{\bf 1.21}&{\bf 1.35}&{\bf 1.84}&0.77&0.80&0.76&0.77&0.80&0.75&1.28&1.23&1.48&{\bf 9.59}&{\bf 17.94}&{\bf 7.60} \\
&${\rm BL^*}$ &0.93&0.94&0.82&0.88&0.95&{\bf 1.75}&0.93&0.94&0.83&0.93&0.94&0.82&1.28&1.22&1.43&{\bf 5.40}&{\bf 12.79}&{\bf 6.42} \\
$32$ &${\rm BC^*}$ &0.91&0.91&0.78&0.99&{\bf 1.09}&{\bf 1.68}&0.91&0.91&0.80&0.91&0.92&0.79&1.08&1.07&1.74&{\bf 10.29}&{\bf 23.45}&{\bf 12.19} \\
&${\rm MC^*}$ &0.92&0.90&0.66&{\bf 1.19}&{\bf 1.27}&{\bf 1.19}&0.93&0.91&0.67&0.92&0.91&0.68&1.05&1.06&1.29&0.60&{\bf 1.78}&0.75 \\
&${\rm IDW^*}$ &{\bf 1.00}&0.97&0.99&0.95&{\bf 1.07}&{\bf 2.45}&1.00&0.98&{\bf 1.01}&{\bf 1.00}&0.98&0.99&1.00&1.02&1.01&0.71&{\bf 1.13}&0.74 \\
\hline
&\mpr &5.53&5.88&7.18&$-$2.24&$-$2.51&$-$4.11&11.88&12.69&0.16&6.94&7.41&9.05&72.02&67.21&36.33&0.07&0.13&0.03 \\
&${\rm NN^*}$ &0.75&0.79&0.78&{\bf 1.17}&{\bf 1.23}&{\bf 1.06}&0.76&0.80&0.79&0.75&0.79&0.78&1.25&1.21&1.32&{\bf 8.29}&{\bf 17.64}&{\bf 3.01} \\
&${\rm BL^*}$ &0.93&0.94&0.87&{\bf 1.03}&{\bf 1.08}&{\bf 1.07}&0.93&0.95&0.88&0.93&0.94&0.87&1.25&1.21&1.32&{\bf 4.13}&{\bf 11.39}&{\bf 1.46} \\
$64$ &${\rm BC^*}$ &0.91&0.91&0.83&{\bf 1.12}&{\bf 1.16}&{\bf 1.14}&0.92&0.93&0.85&0.91&0.91&0.84&1.07&1.07&1.51&{\bf 8.38}&{\bf 22.89}&{\bf 2.99} \\
&${\rm MC^*}$ &0.93&0.91&0.73&{\bf 1.20}&{\bf 1.19}&{\bf 1.49}&0.94&0.92&0.75&0.92&0.91&0.74&1.04&1.05&1.32&0.09&0.36&0.02 \\
&${\rm IDW^*}$ &1.00&0.97&0.99&0.95&{\bf 1.06}&0.84&0.99&0.98&0.99&1.00&0.97&1.00&1.00&1.02&1.01&0.20&0.25&0.27 \\
\hline
&\mpr &5.53&5.84&7.32&$-$2.11&$-$2.43&$-$3.18&11.85&12.58&15.87&6.92&7.31&9.16&74.04&70.40&39.38&0.28&0.50&0.05 \\
&${\rm NN^*}$ &0.76&0.78&0.78&{\bf 1.15}&{\bf 1.21}&{\bf 1.26}&0.77&0.79&0.80&0.76&0.78&0.78&1.22&1.20&1.39&{\bf 7.29}&{\bf 15.39}&{\bf 1.10} \\
&${\rm BL^*}$ &0.93&0.94&0.86&{\bf 1.05}&{\bf 1.08}&{\bf 1.14}&0.93&0.94&0.87&0.93&0.94&0.87&1.22&1.20&1.39&{\bf 3.47}&{\bf 9.73}&0.45 \\
$128$ &${\rm BC^*}$ &0.90&0.91&0.83&{\bf 1.13}&{\bf 1.16}&{\bf 1.19}&0.91&0.92&0.84&0.90&0.91&0.83&1.06&1.07&1.62&{\bf 7.04}&{\bf 19.34}&{\bf 0.96} \\
&${\rm MC^*}$ &0.92&0.91&0.77&{\bf 1.20}&{\bf 1.23}&0.88&0.93&0.92&0.79&0.92&0.90&0.78&1.04&1.06&1.28&0.01&0.07&8e$-$4 \\
&${\rm IDW^*}$ &{\bf 1.00}&0.97&0.99&0.95&{\bf 1.04}&{\bf 1.06}&1.00&0.97&0.99&1.00&0.97&0.99&1.00&1.02&1.00&0.08&0.15&0.09 \\
\hline
\end{tabular}
}
\end{scriptsize}
\end{table}

\begin{table}[h!]
\addtolength{\tabcolsep}{0pt} \caption{Interpolation validation measures for the \mpr method. $S=100$ samples are generated from a Gaussian random field with mean equal to 50 on  a square grid with side length $L=2^n$, where $n=8, 9, 10, 11$. Two covariance models, WM($\kappa=0.2,\nu = 0.5$) and WM($\kappa=0.2,\nu = 0.25$) are used. Missing data are generated by (a) $p=33\%$ (b) $p=66\%$ random thinning and (c) random removal of square data block  with side length $L_B=20$.} \vspace{3pt} \label{tab:synt_int_MPR}
\begin{scriptsize}
\resizebox{1\textwidth}{!}{
\begin{tabular}{|c|c|ccc|ccc|ccc|ccc|ccc|ccc|}
\hline
& & \multicolumn{3}{c|}{MAAE}  & \multicolumn{3}{c|}{MARE [\%]} &
 \multicolumn{3}{c|}{MAARE [\%]} & \multicolumn{3}{c|}{MRASE} &
 \multicolumn{3}{c|}{MR [\%]}  & \multicolumn{3}{c|}{$ \langle t_{\mathrm{cpu}} \rangle $}   \\
 $\nu$ & $L$ & (a) & (b) & (c) & (a) & (b) & (c) & (a) & (b) & (c) & (a) & (b) & (c) & (a) & (b) & (c) & (a) & (b) & (c) \\
\hline
&$256$ &3.39&3.83&6.33&$-$0.98&$-$1.29&$-$3.34&7.19&8.16&14.02&4.26&4.84&7.95&90.47&87.62&52.90&0.89&1.58&0.10 \\
0.5 &$512$ &3.38&3.83&6.27&$-$0.98&$-$1.29&$-$2.41&7.16&8.16&13.29&4.24&4.83&7.91&90.49&87.55&54.36&3.47&6.71&0.51 \\
&$1024$ &3.38&3.83&6.34&$-$0.99&$-$1.30&$-$3.78&7.17&8.16&13.87&4.24&4.82&7.99&90.70&87.89&54.52&19.47&35.83&2.24 \\
&$2048$ &3.38&3.83&6.32&$-$0.99&$-$1.29&$-$2.86&7.17&8.15&13.71&4.24&4.82&7.98&90.59&87.74&52.41&85.83&149.11&9.12\\
\hline
.&$256$ &5.51&5.82&7.20&$-$2.10&$-$2.39&$-$4.19&11.79&12.51&15.90&6.92&7.31&9.01&72.24&68.23&36.05&0.88&1.55&0.11 \\
0.25 &$512$ &5.50&5.80&7.39&$-$2.09&$-$2.39&$-$3.56&11.77&12.45&16.18&6.90&7.27&9.28&72.69&68.87&36.32&3.59&6.42&0.51 \\
&$1024$ &5.48&5.79&7.25&$-$2.08&$-$2.38&$-$5.06&11.72&12.42&16.03&6.87&7.26&9.06&72.88&69.02&36.34&20.08&36.50&2.40 \\
&$2048$ &5.48&5.78&7.17&$-$2.08&$-$2.37&$-$3.76&11.71&12.40&15.50&6.87&7.25&8.97&72.70&68.83&36.08&85.04&148.77&9.28\\
\hline
\end{tabular}
}
\end{scriptsize}
\end{table}

For larger data sizes, we exclude the OK method from the comparison due to its  high computational cost. In Tables~\ref{tab:synt_int_La} and~\ref{tab:synt_int_Lb} we present similar comparisons as those in Table~\ref{tab:synt_int_L16a}, for square grids with side lengths $L=32,64$ and $128$. As $L$ increases, the relative performance of the \mpr method improves  (except for the MARE errors). Thus, for $L=64$ and $128$, the \mpr approach is superior to NN, BL, BC, MC and IDW methods  in terms of validation measures (except MARE).  \mpr  also has significantly shorter CPU times than MC and IDW. The lower MARE of the \mpr method is due to a less symmetric error distribution.

Finally, we study the performance of the \mpr method on increasing  grid sizes  $L=2^{n}$ where $n=8, 9, 10, 11$. The results are summarized in Table~\ref{tab:synt_int_MPR}. Increasing $L$ does not impact the validation measures. However, a closer look reveals a small but noticeable improving trend (more apparent for $\nu=0.25$),  in agreement with the trend  observed in Tables~\ref{tab:synt_int_La} and~\ref{tab:synt_int_Lb}. On the other hand, the CPU time increases drastically with $L$. Nevertheless, the scaling of the \mpr CPU time with $L$ is competitive with the alternative approaches, as we discuss in more detail below.

\subsubsection{Single sample statistics}

\begin{figure}[t!]
\begin{center}\hspace*{-7 mm}
    \subfigure[Original]{\label{fig:zo_L64}
    \includegraphics[scale=0.35]{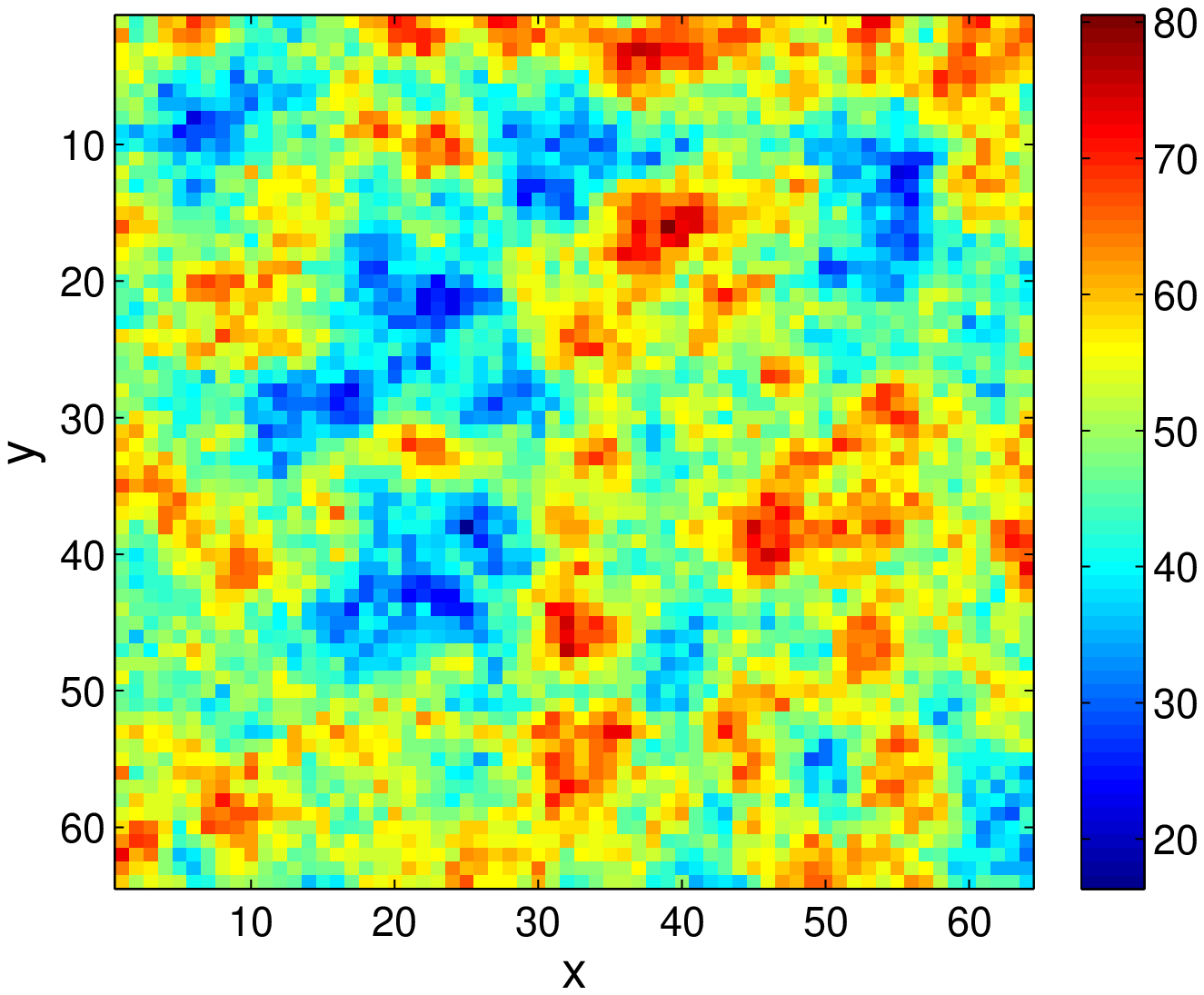}}\hspace*{-6 mm}
    \subfigure[$p=66\%$ missing]{\label{fig:zm_XY_L64_p66}
    \includegraphics[scale=0.35]{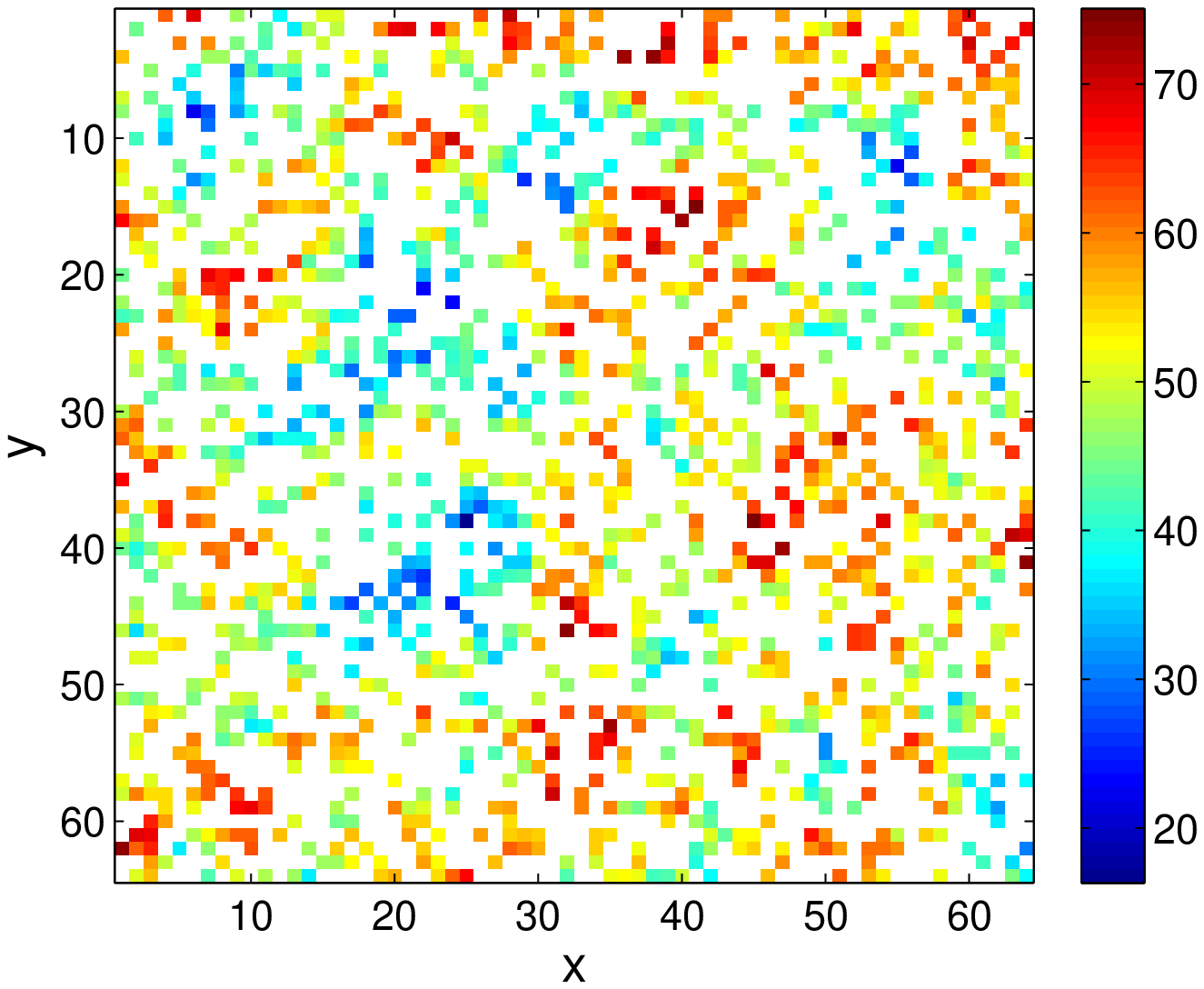}}\hspace*{-6 mm}
    \subfigure[Interpolated]{\label{fig:zr_XY_L64_p66}
    \includegraphics[scale=0.35]{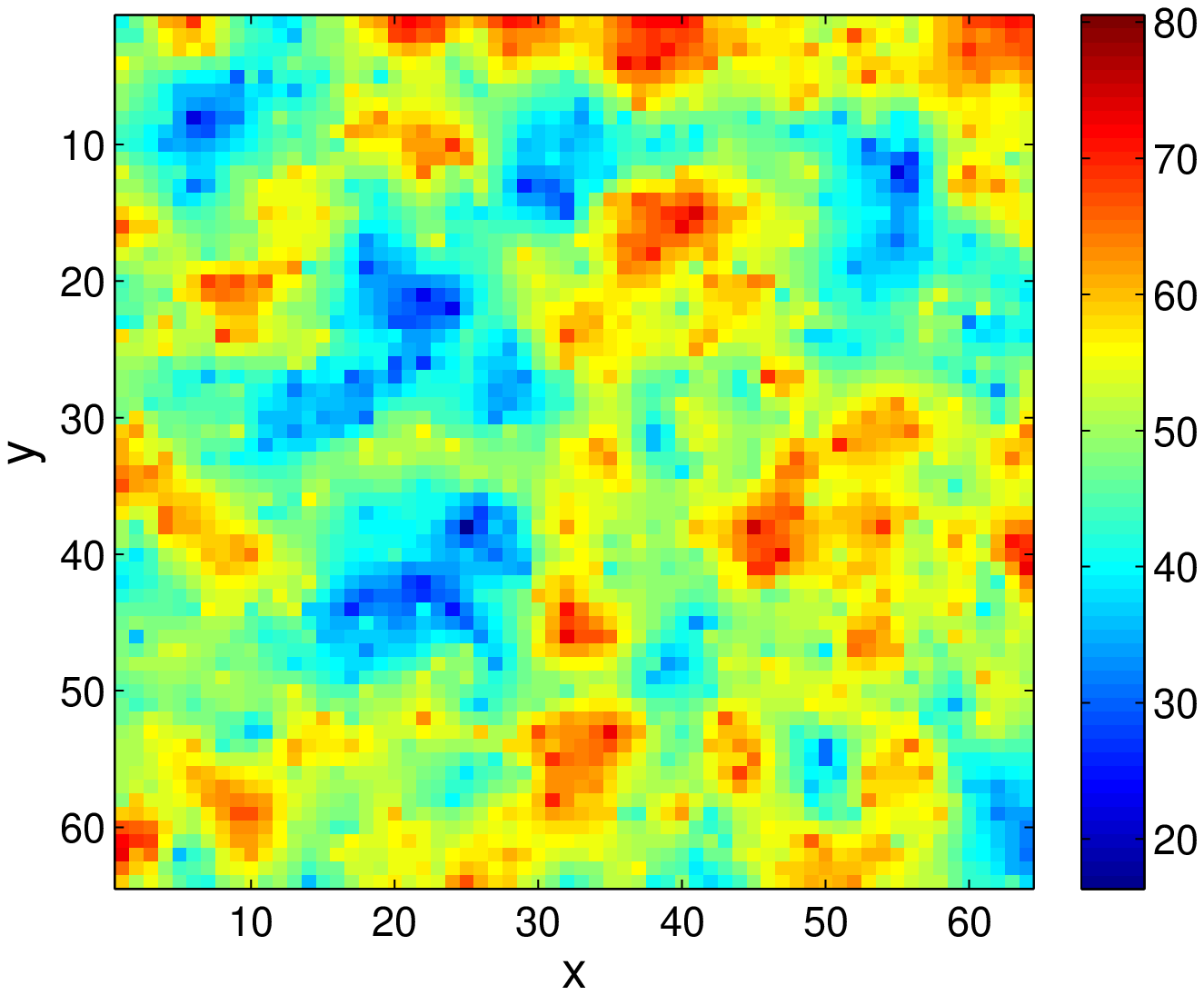}} \\ \hspace*{-6 mm}
    \subfigure[Standard deviation]{\label{fig:std_XY_L64_p66}
    \includegraphics[scale=0.35]{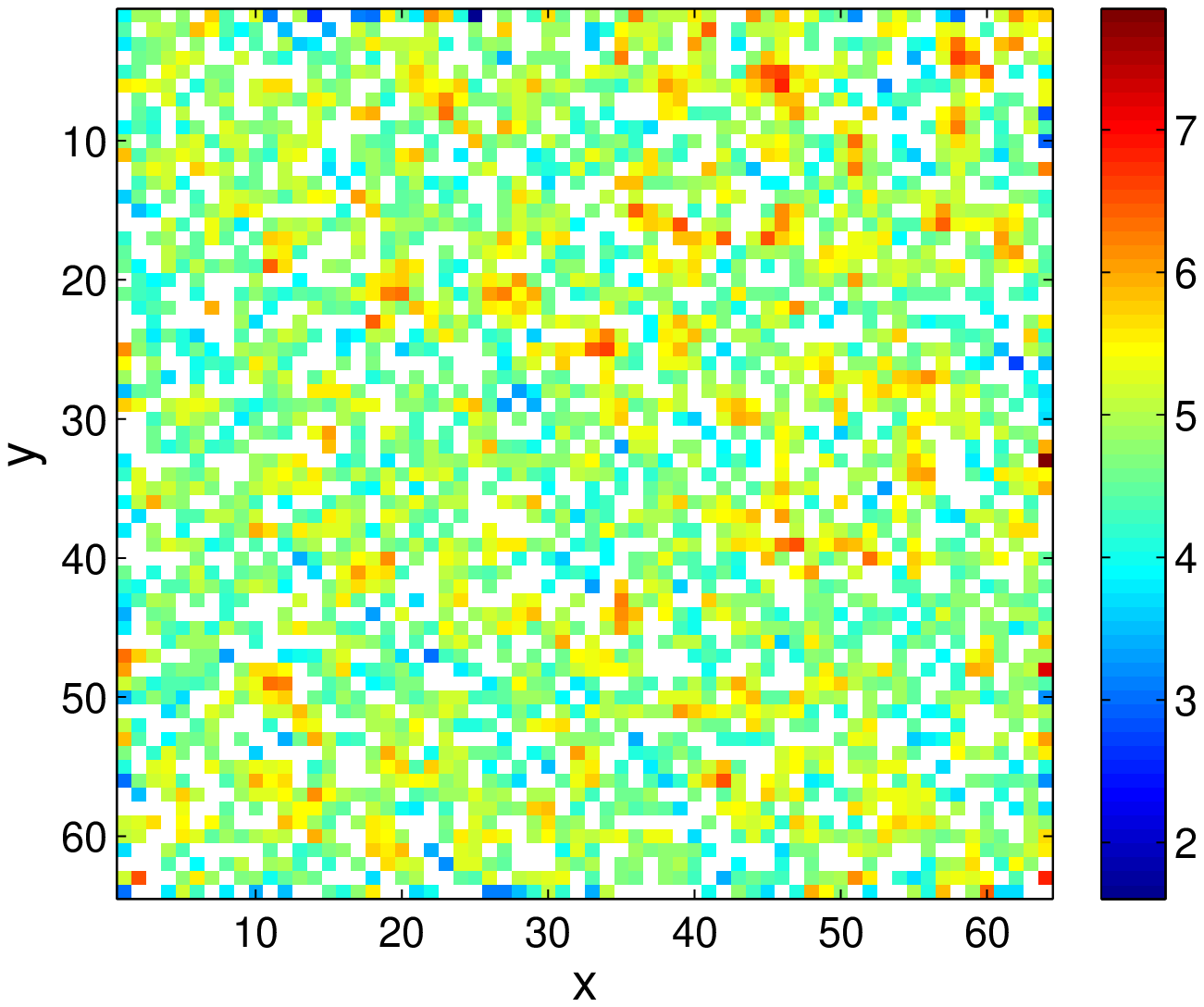}}\hspace*{-6 mm}
		\subfigure[Interpolation error]{\label{fig:err_XY_L64_p66}
    \includegraphics[scale=0.35]{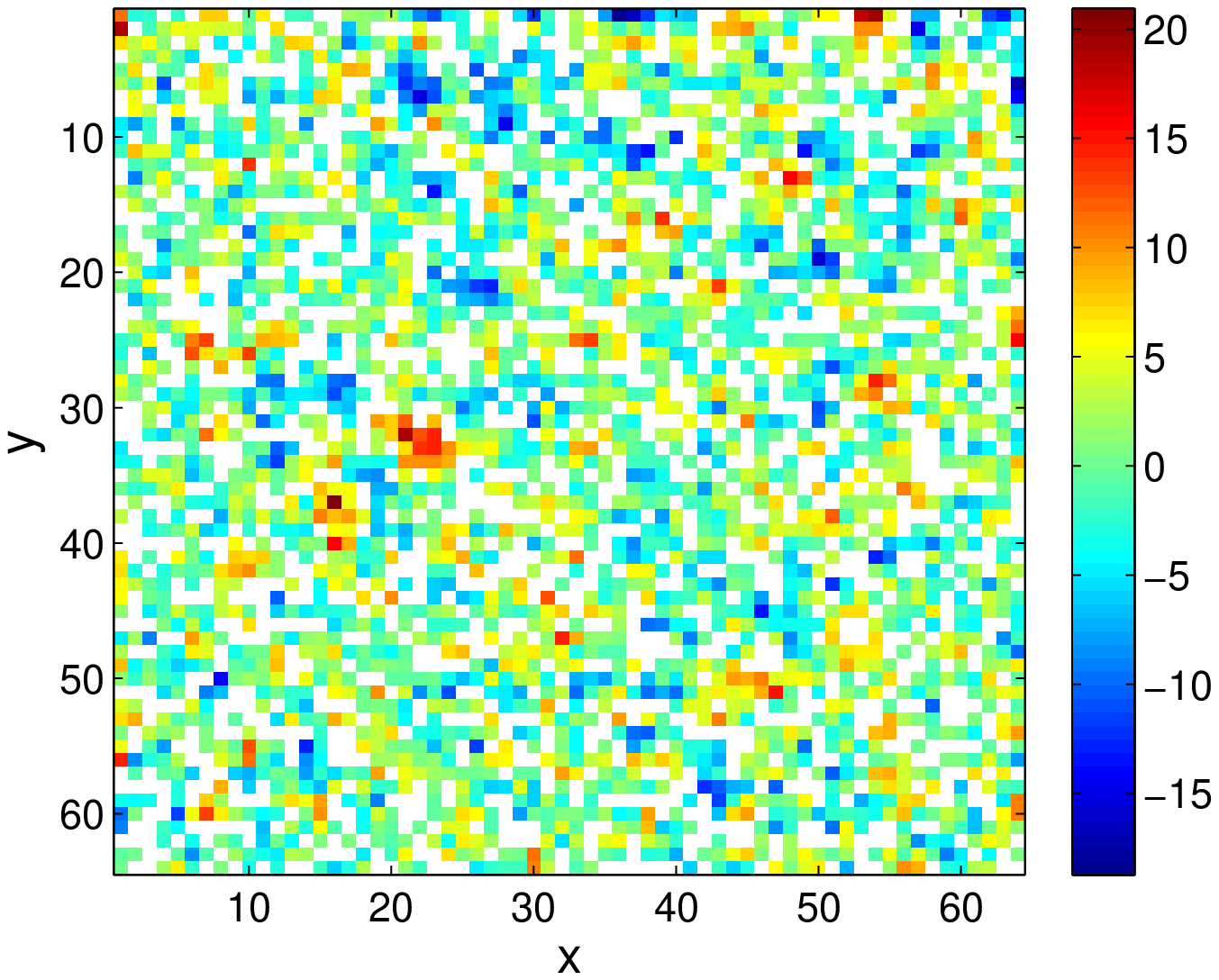}}\hspace*{-6 mm}
		\subfigure[Empirical cdf]{\label{fig:cdf_XY_L64_p66}
    \includegraphics[scale=0.35]{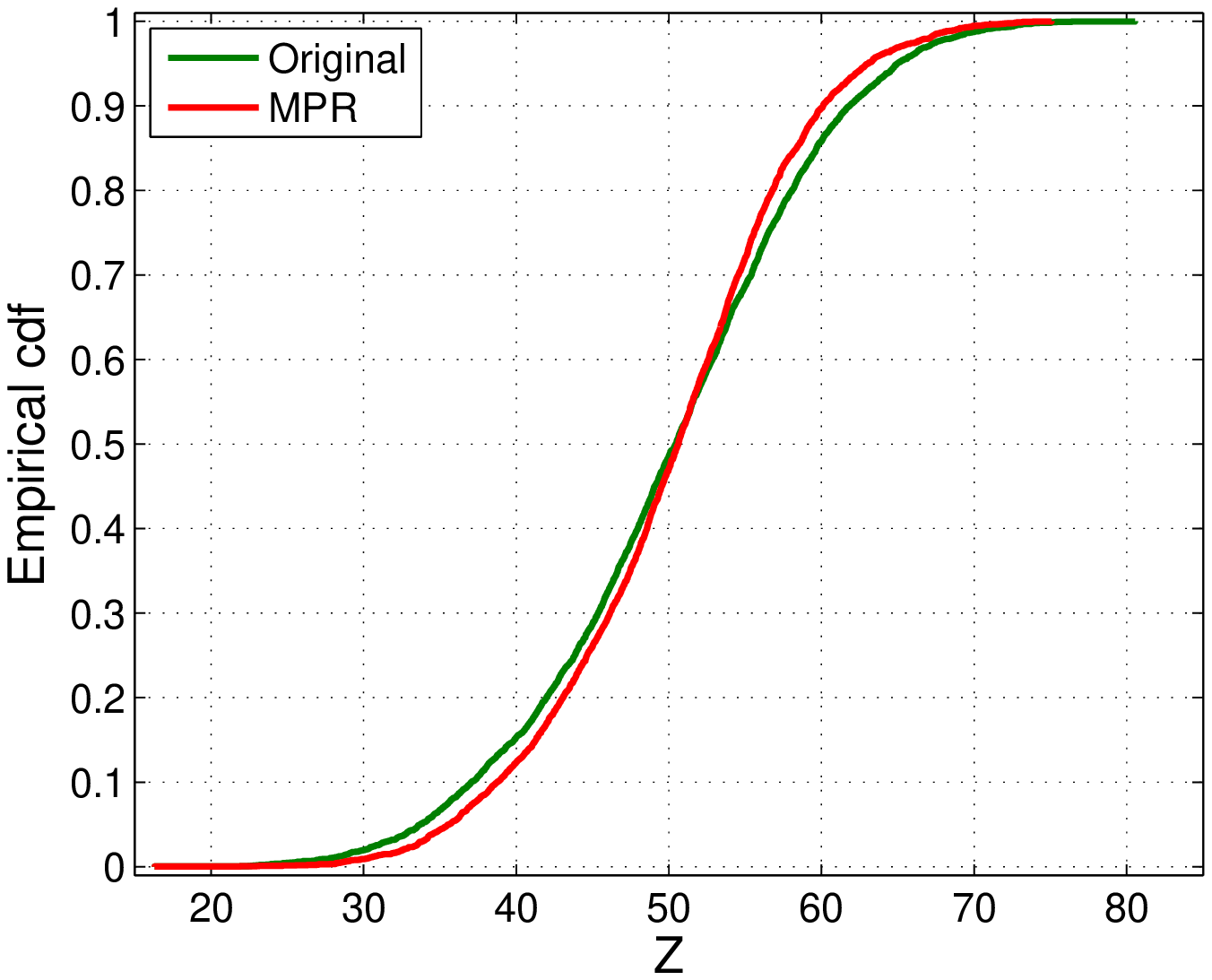}}
\end{center}
    \caption{\mpr interpolation results for WM($\kappa=0.2,\nu = 0.5$), with $L=64$, generated by $p=66\%$ random thinning. Subfigures include (a) original field, (b) thinned sample, (c) interpolated data,  (d) local standard deviation $\sigma(\mathbf{r}_p)$, (e) local estimation error $\epsilon(\mathbf{r}_p)$, and (f) comparison of the empirical cumulative distribution functions of the original and interpolated data.}
  \label{MPR}
\end{figure}

In Fig.~\ref{MPR} we investigate the prediction performance of the \mpr model based on a single synthetic sample. The sample is  generated by  simulating a random field with  WM($\kappa=0.2,\nu = 0.5$) covariance model on a square grid of size $L=64$.  Then, $p=66\%$ of the field values are randomly removed to generate the sample. Panels (a) and (b) show the simulated random field realization and the sample obtained after random thinning, respectively. The remaining panels illustrate the prediction performance in terms of (c) the interpolated data over the entire grid (d) the standard deviation and (e) the interpolation error $\epsilon(\mathbf{r}_p)$ at the missing points, as well as (f) the empirical cumulative distribution functions of the original and the interpolated data. It is evident that the \mpr predictor fairly accurately reconstructs basic statistical features of the original data.
On the other hand, panels (c) and (f) provide evidence that the interpolated field is overly smooth. The issue of over-smoothing and the ability of \mpr to capture the spatial data variability will be discussed more below in the context  of  block-missing real data.

\subsection{Real data}
\subsubsection{Data descriptive statistics}

\begin{figure}[t!]
\begin{center}
    \subfigure[Latent heat map]{\label{fig:zo_heat}
    \includegraphics[scale=0.46,clip]{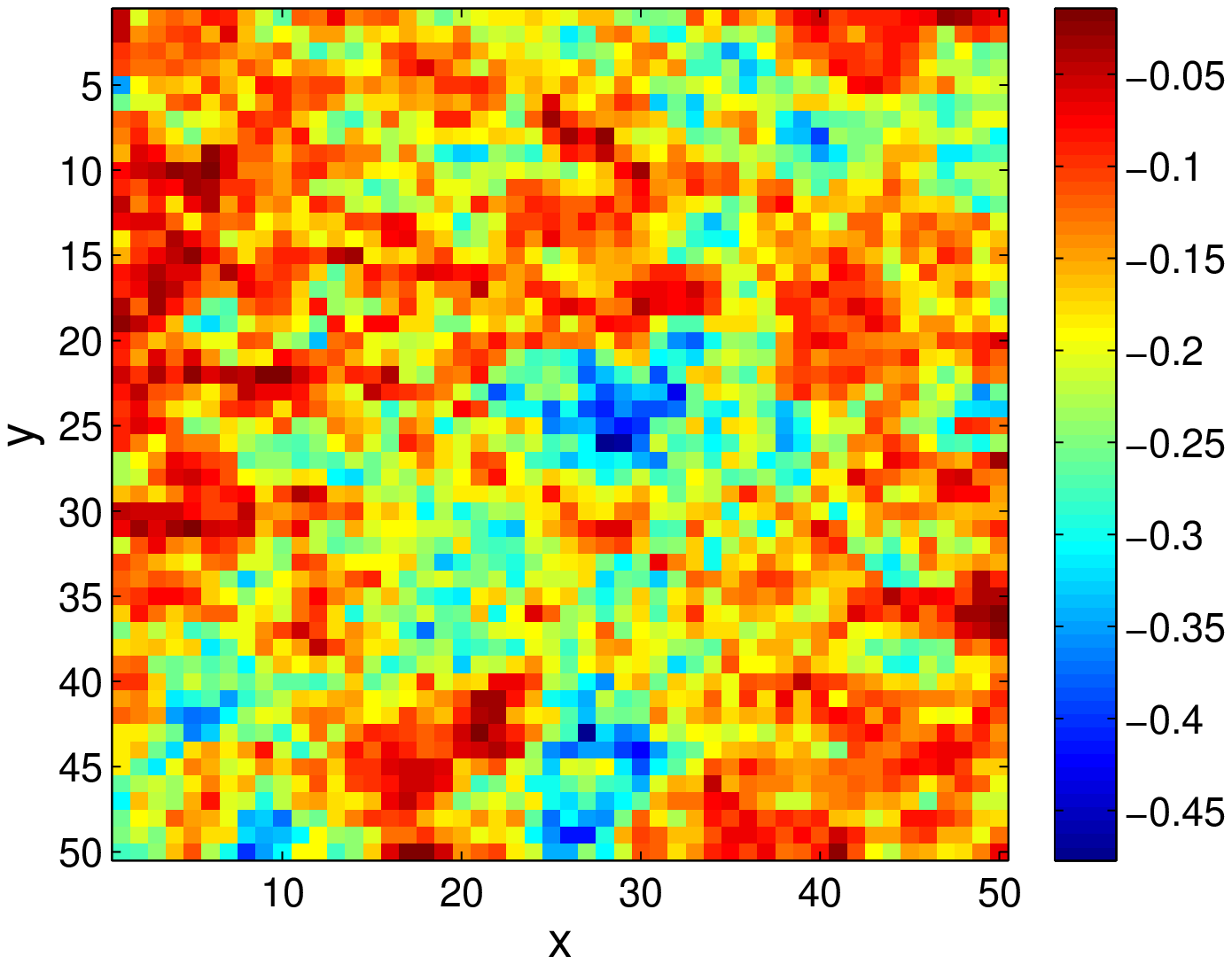}}
    \subfigure[Latent heat histogram]{\label{fig:hist_heat}
    \includegraphics[scale=0.46,clip]{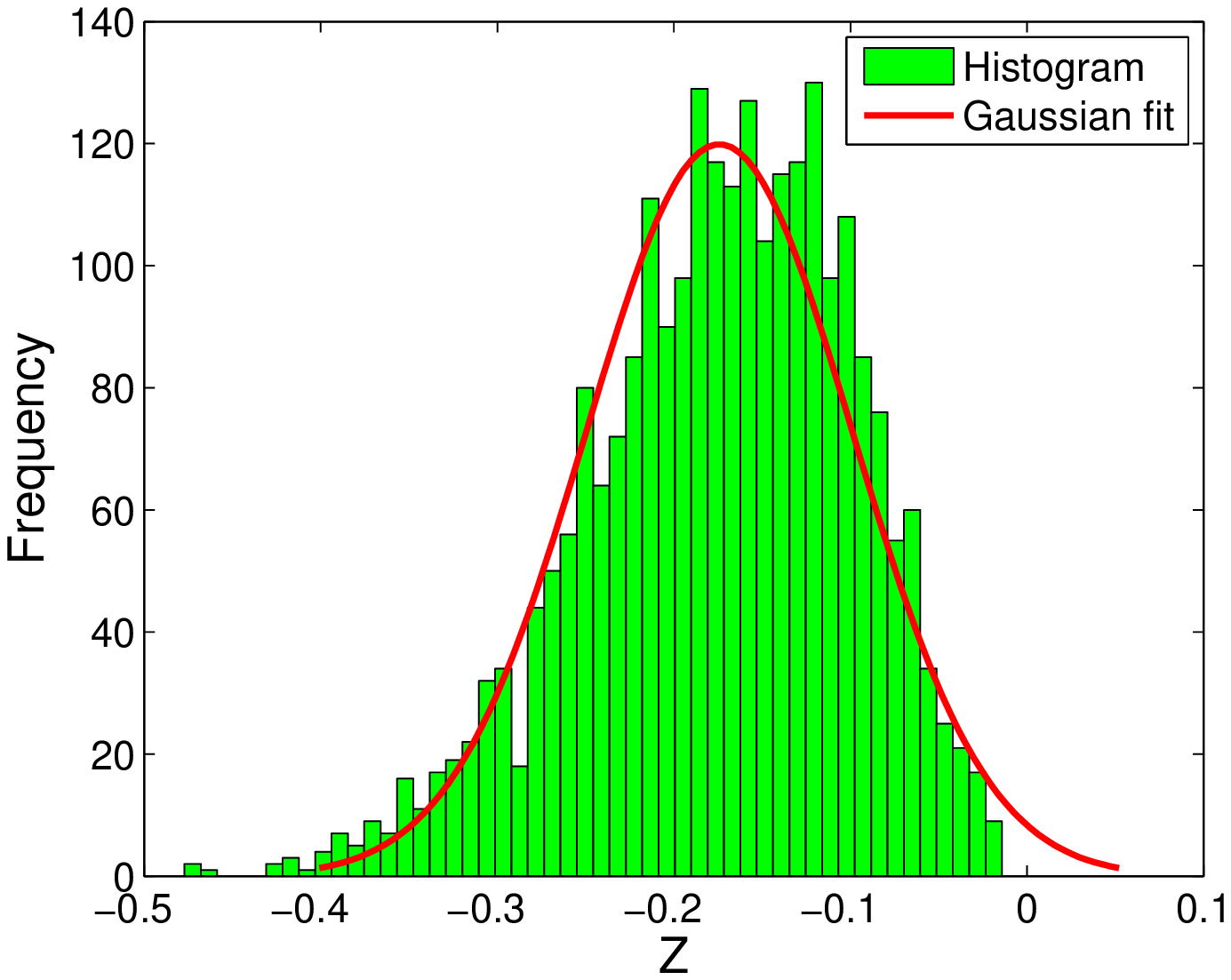}} \\
    \subfigure[Walker lake map]{\label{fig:zo_walker}
    \includegraphics[scale=0.46,clip]{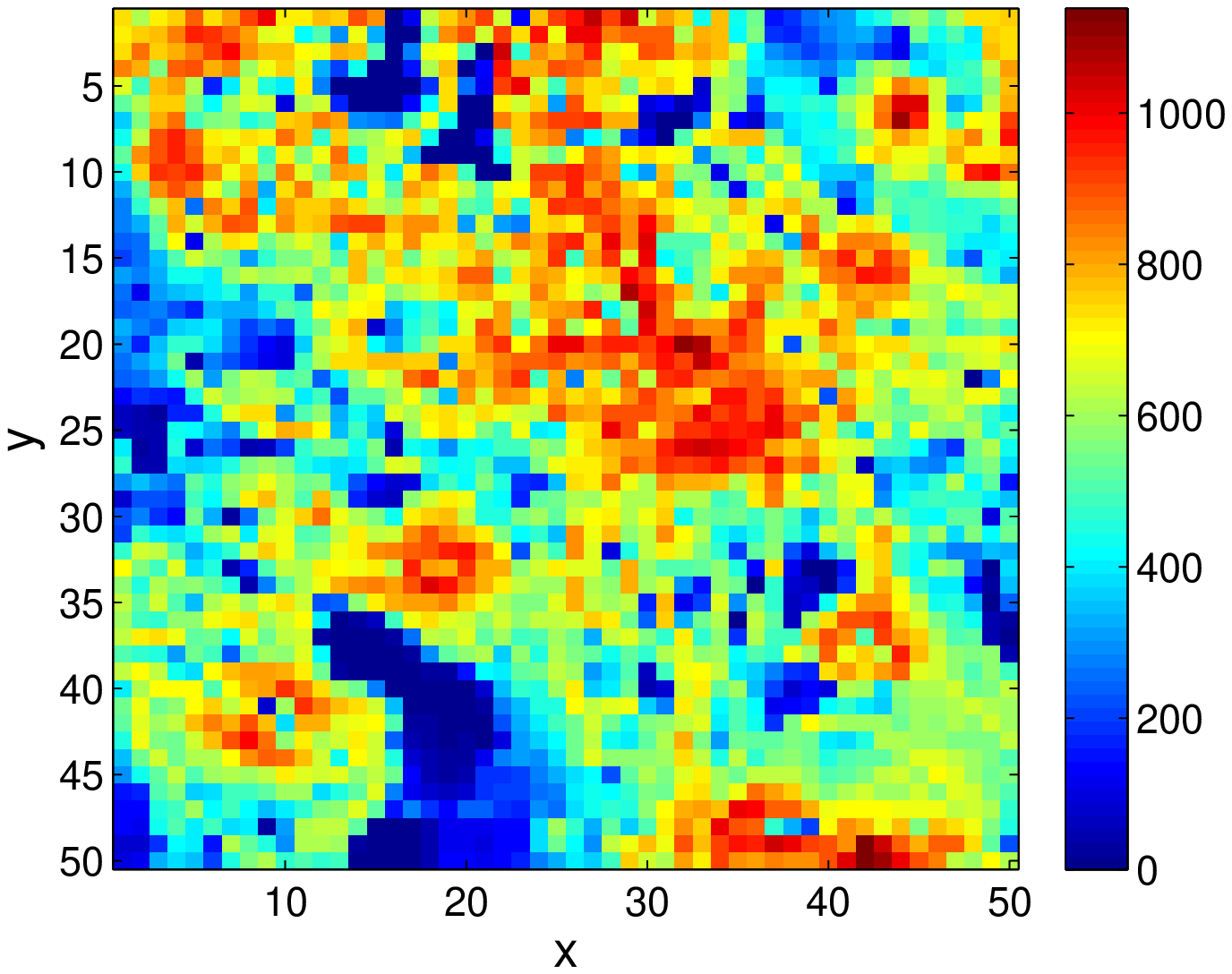}}
    \subfigure[Walker lake histogram]{\label{fig:hist_walker}
    \includegraphics[scale=0.46,clip]{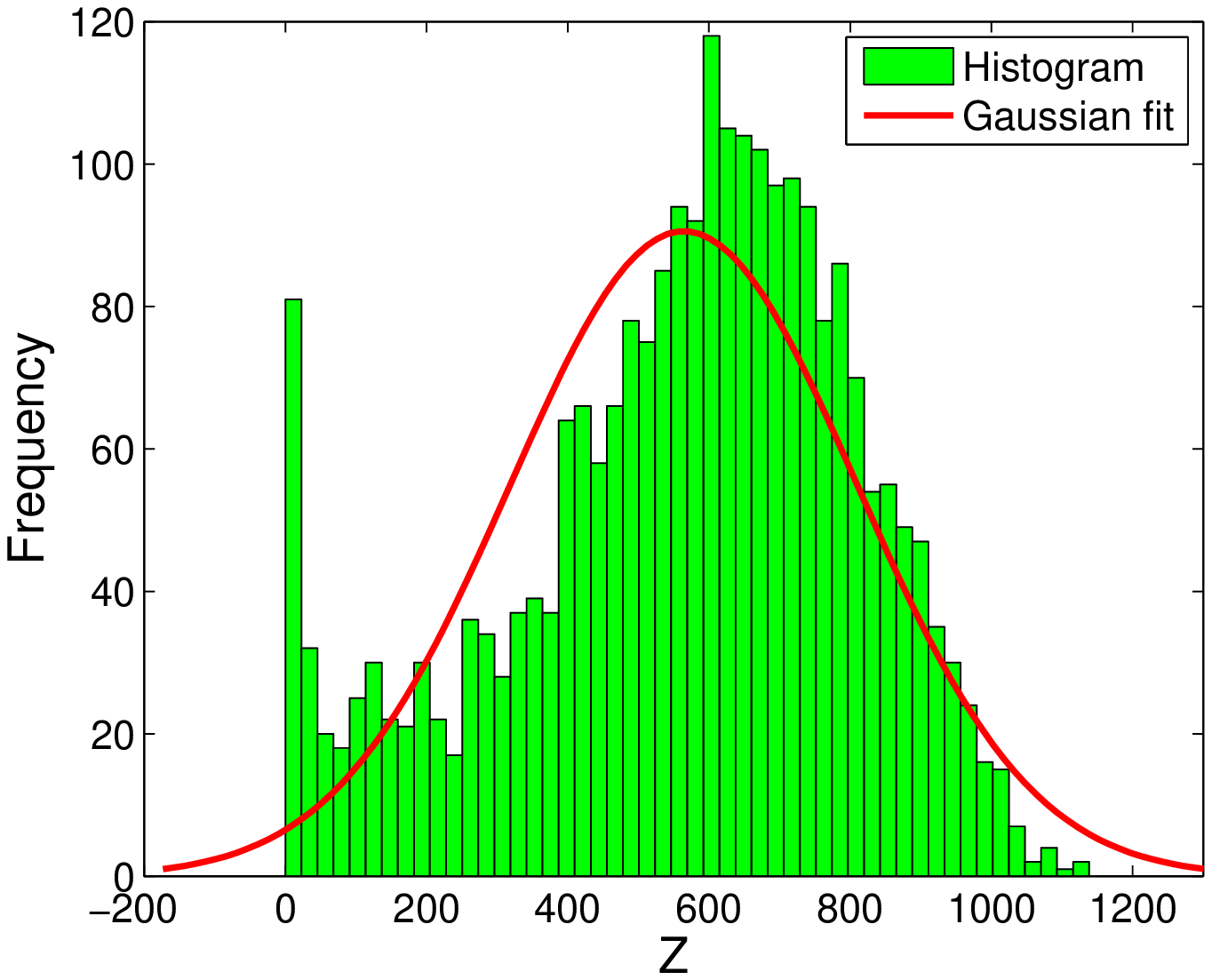}}
\end{center}
    \caption{(a,b) Latent heat and (c,d) Walker lake data maps and histograms.}
  \label{fig:real_data}
\end{figure}

We assess the performance of \mpr prediction by means of two real-world environmental data sets that follow non-Gaussian distributions.
The first set represents the monthly mean of vertically averaged atmospheric latent heat release measurements in January 2006~\citep{Tao06,{TRMM11}}. The data are on an $L=50$ grid with a $0.5^\circ \times 0.5^\circ$ cell size, extending in latitude from 16S to 8.5N and in longitude from 126.5E to 151E. The measurement units are C/hr (degrees Celsius per hour), and their  summary statistics  are as follows: $N=2500$, $z_{\min}=-0.4772$, $z_{\max}= -0.0141$, $\bar{z}=-0.1743$, $z_{0.50}= -0.1680$,
$\sigma_{z}=0.0755$, skewness coefficient equal to $-0.5153$, and
kurtosis coefficient equal to $3.1218$. Negative (positive) values correspond to latent heat absorption (release).

The second example is a subset of the DEM-based data from Walker lake area in Nevada~\citep{Isaak89}. The data denote chemical concentrations with units in parts per million (ppm). They are sampled on an $L=50$ grid, and they exhibit the following summary statistics: $N=2500$, $z_{\min}=0$, $z_{\max}= 1138.6$, $\bar{z}=564.1435$, $z_{0.50}= 601.0950$,
$\sigma_{z}=245.7724$, skewness coefficient equal to $-0.5295$, and
kurtosis coefficient equal to $2.7351$.

\begin{figure}[t!]
\begin{center}
    \subfigure[Latent heat variogram]{\label{fig:vario_heat}
    \includegraphics[scale=0.46,clip]{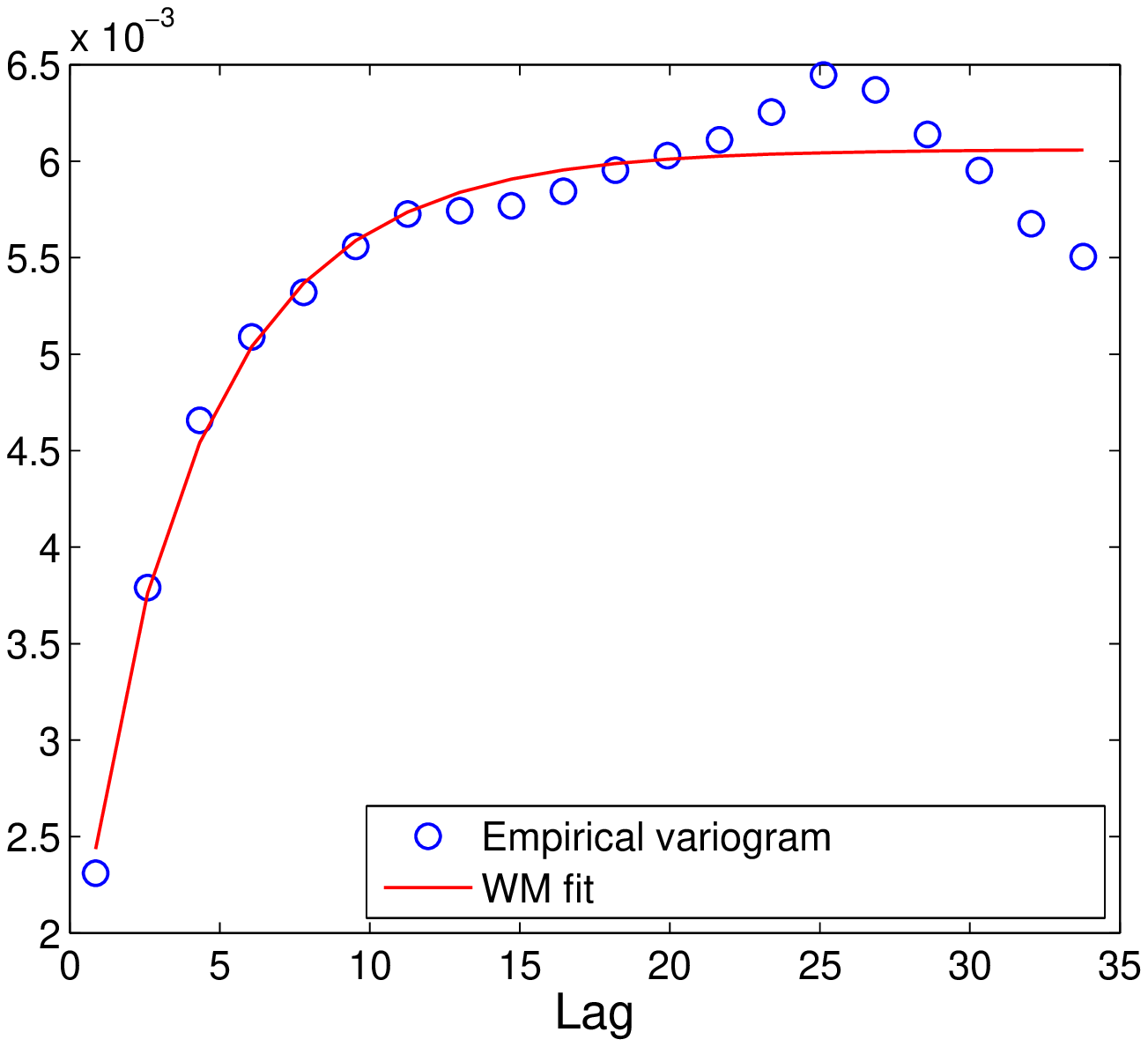}}
    \subfigure[Walker lake variogram]{\label{fig:vario_Walker}
    \includegraphics[scale=0.46,clip]{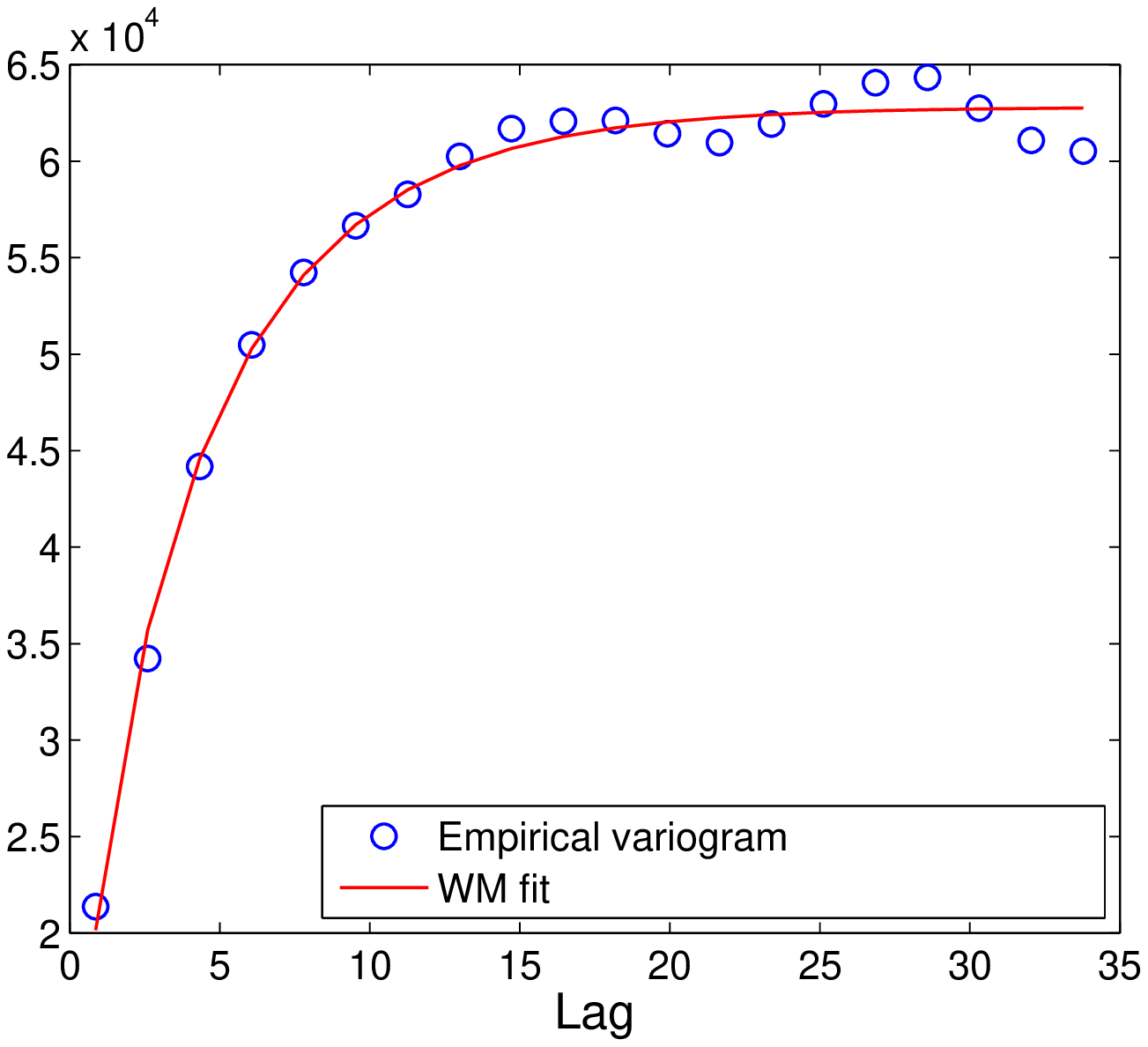}}
\end{center}
    \caption{Empirical (circles) and fitted (solid line) variograms of (a) latent heat and (b) Walker lake data. The estimated parameters of the WM variogram model are as follows:
    (a) $\sigma=0.07$, $\kappa=0.21$, and $\nu=0.34$; (b) $\sigma=250.62$, $\kappa=0.18$, and $\nu=0.29$.}
  \label{fig:vario_real_data}
\end{figure}

Surface and histogram plots for both data sets are shown in Fig.~\ref{fig:real_data}. The histograms clearly show the deviations  from the Gaussian distribution.
Fig.~\ref{fig:vario_real_data} presents the empirical variograms and their respective fits with the WM model using the weighted least squares method~\citep{cres85}. The estimated parameter values indicate that both data sets are examples of relatively rough spatial processes ($\nu=0.34$ and $\nu=0.29$ respectively), with  similar spatial variability to the synthetic data.

\subsubsection{Gap-filling performance}

The \mpr prediction performance evaluation is summarized for both data sets and different patterns of missing data in Table~\ref{tab:real}. Due to the presence of negative and zero values, the relative errors MARE and MAARE are excluded from the comparison. Table~\ref{tab:real} shows that in terms of prediction accuracy the \mpr method is superior to all other methods (NN, BL, BC, IDW, MC), except for a few cases where IDW  performs slightly better (cf. bold figures in the MAAE and MRASE columns). In terms of computational time, \mpr is more efficient than MC and IDW but less efficient than NN, BL and BC (cf. bold figures in the $\langle t_{\mathrm{cpu}} \rangle$ column).

\begin{figure}[t!]
\begin{center}\hspace*{-7 mm}
    \subfigure[\mpr, $p=33\%$]{\label{fig:zr_XY_heat_p33}
    \includegraphics[scale=0.35]{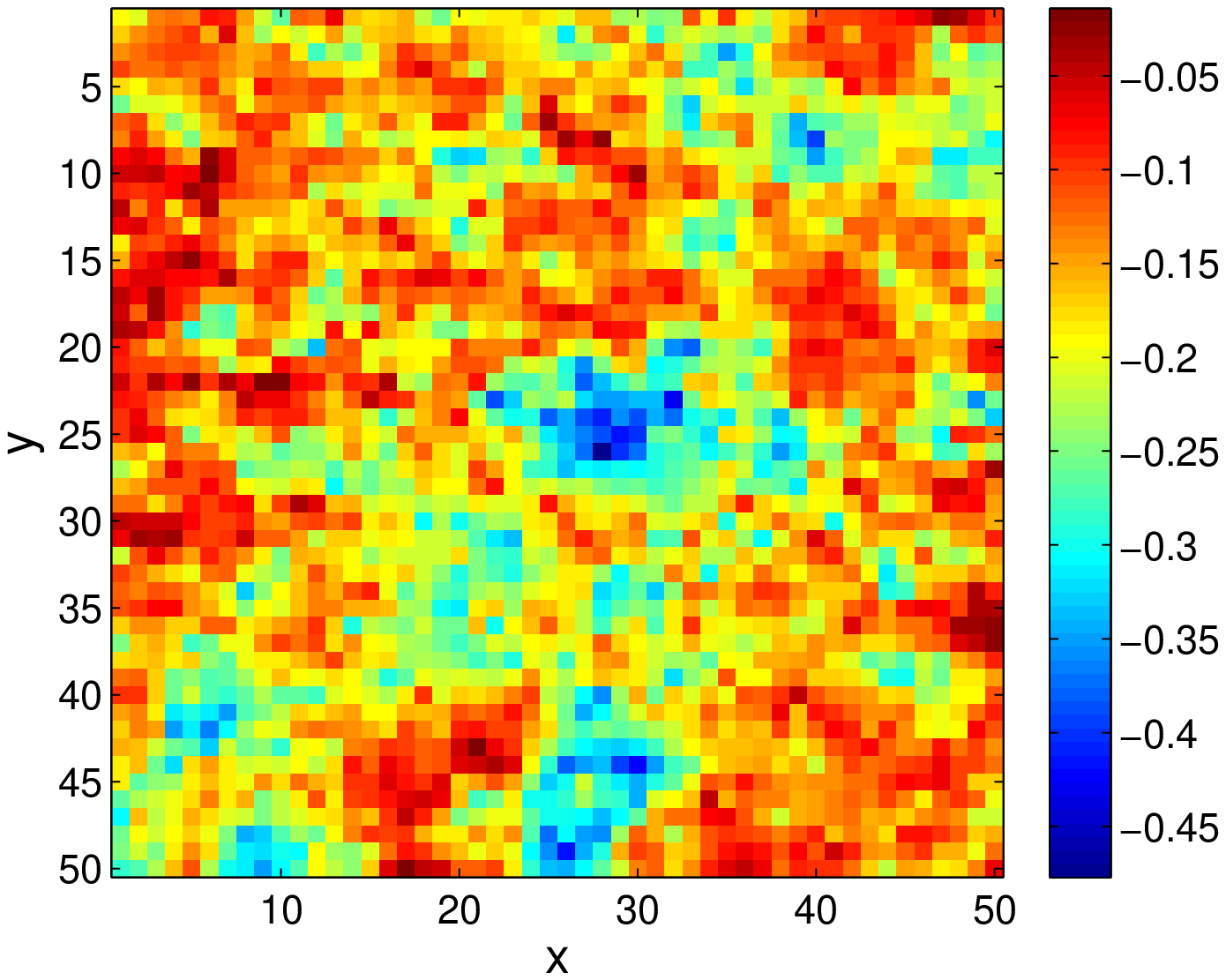}}\hspace*{-5 mm}
		\subfigure[\mpr, $p=66\%$]{\label{fig:zr_XY_heat_p66}
    \includegraphics[scale=0.35]{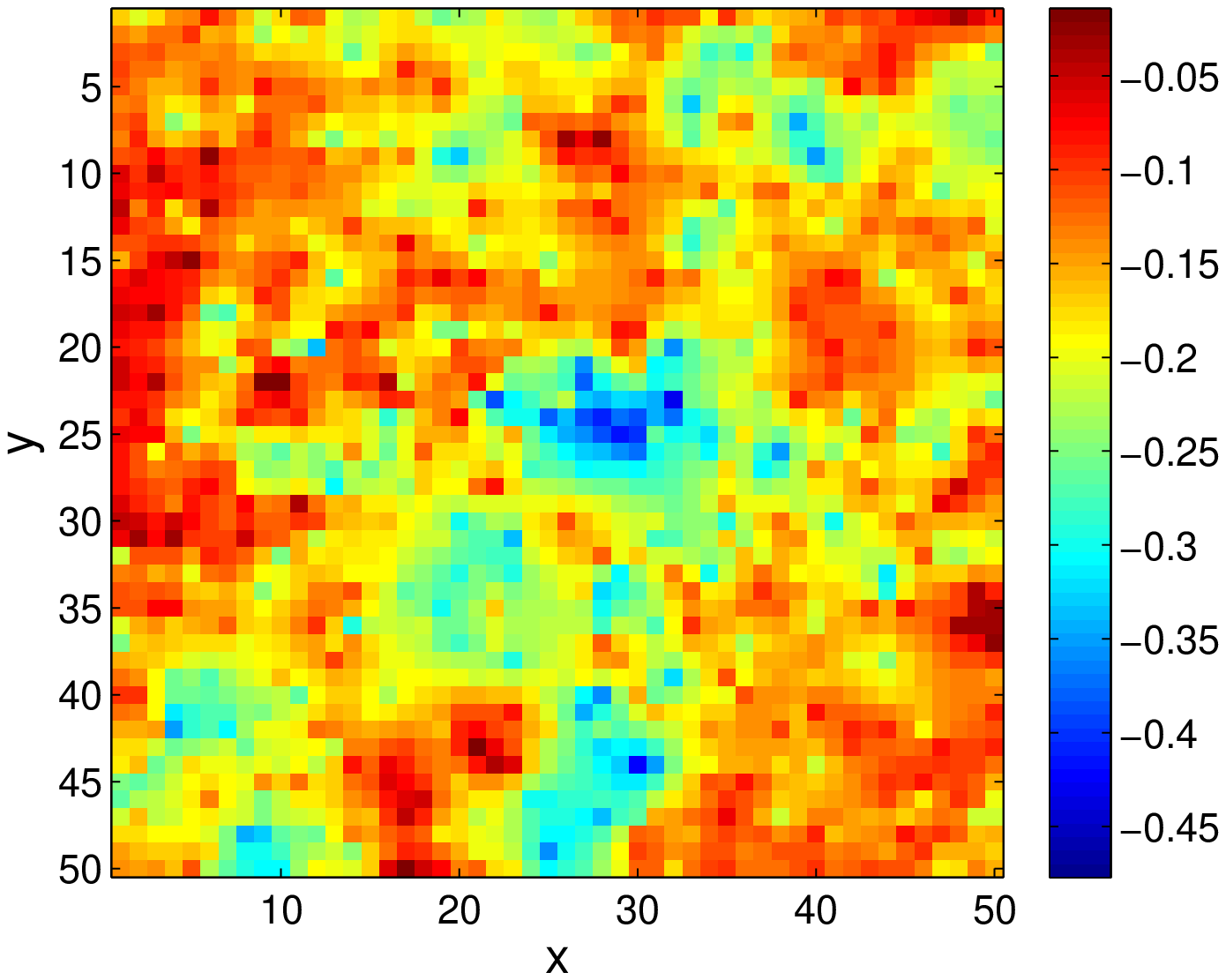}}\hspace*{-5 mm}
		\subfigure[\mpr, $L_B=20$]{\label{fig:zr_XY_heat_blk20}
    \includegraphics[scale=0.35]{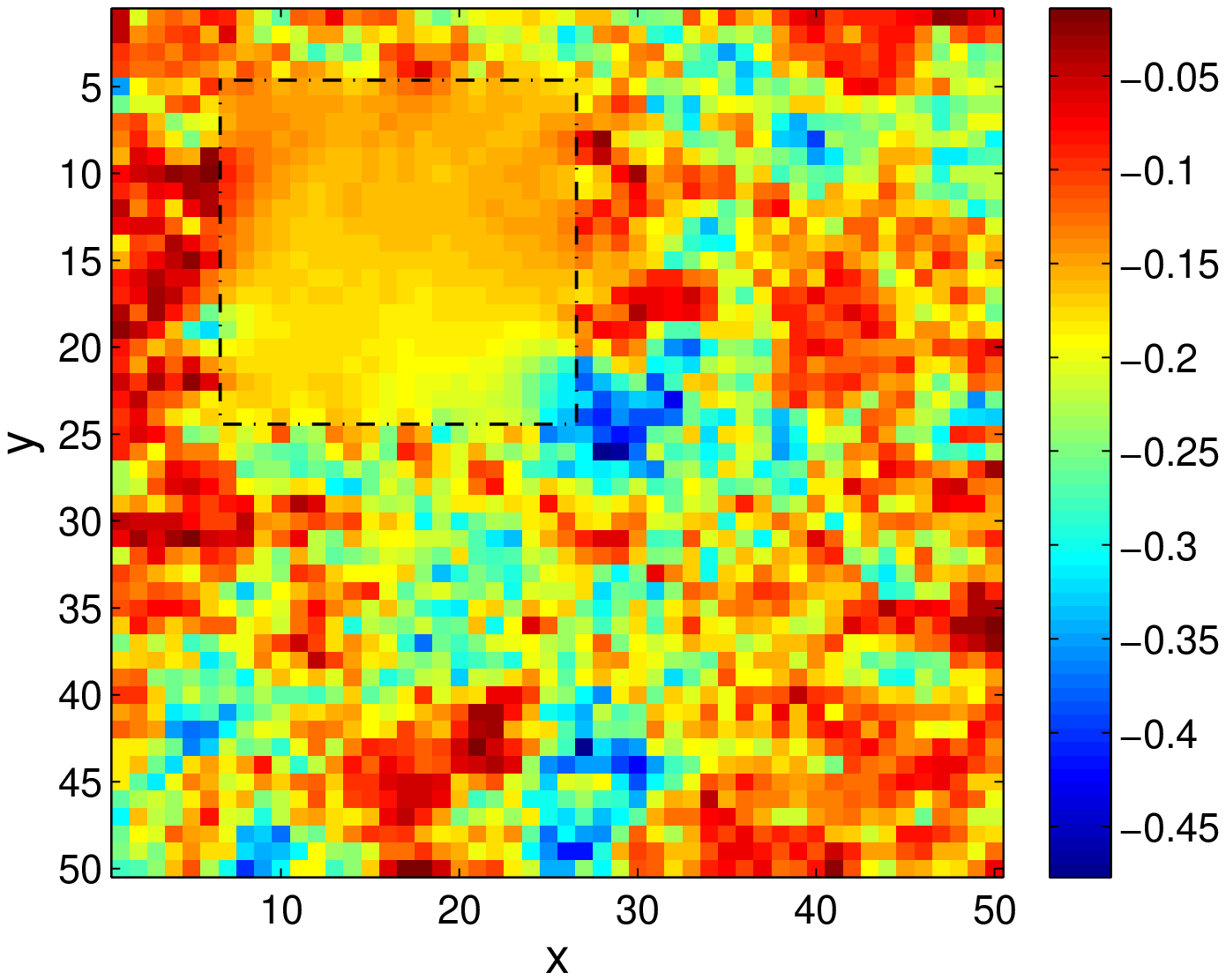}}\\\hspace*{-7 mm}
		\subfigure[IDW, $p=33\%$]{\label{fig:zr_ID_heat_p33}
    \includegraphics[scale=0.35]{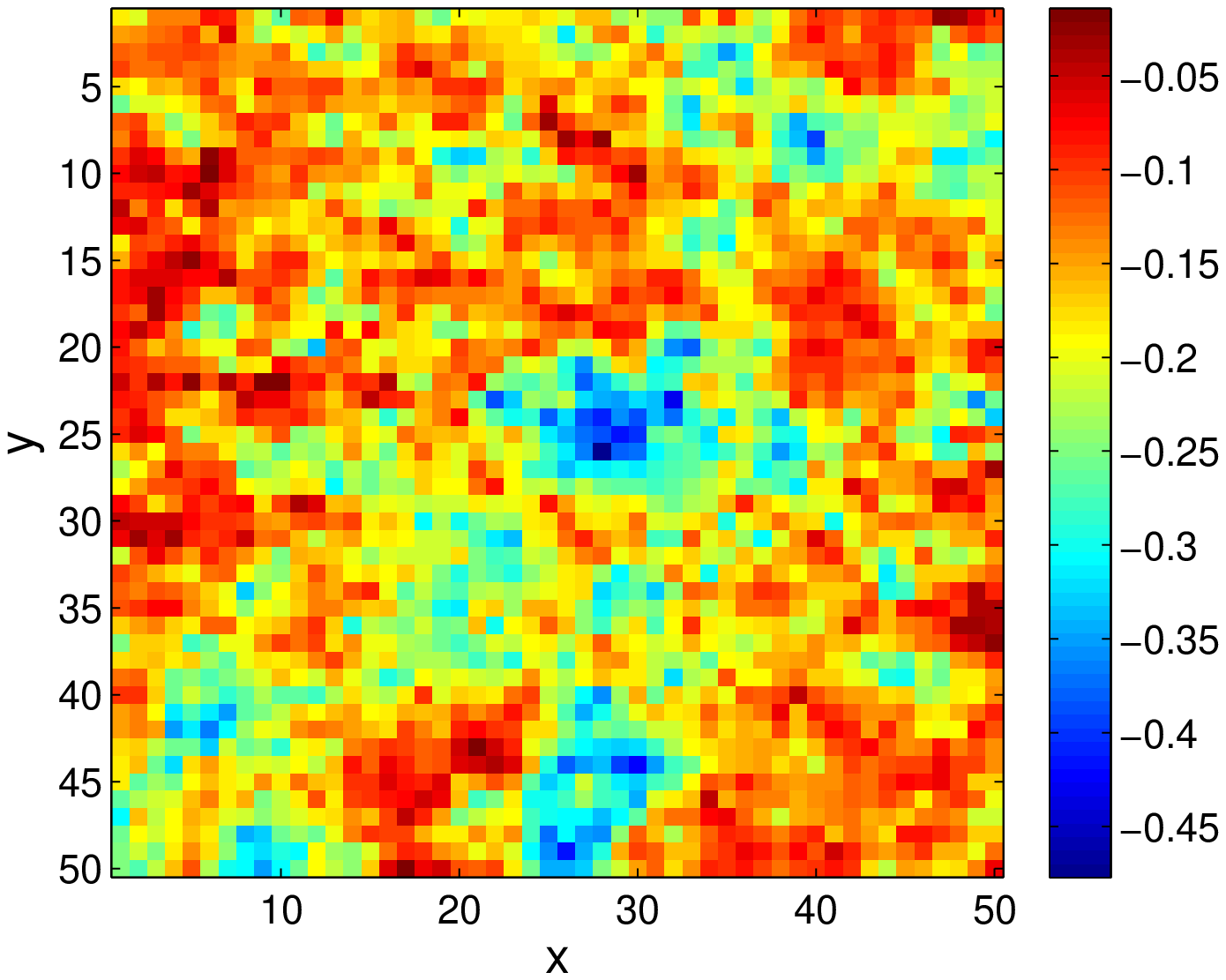}}\hspace*{-5 mm}
		\subfigure[IDW, $p=66\%$]{\label{fig:zr_ID_heat_p66}
    \includegraphics[scale=0.35]{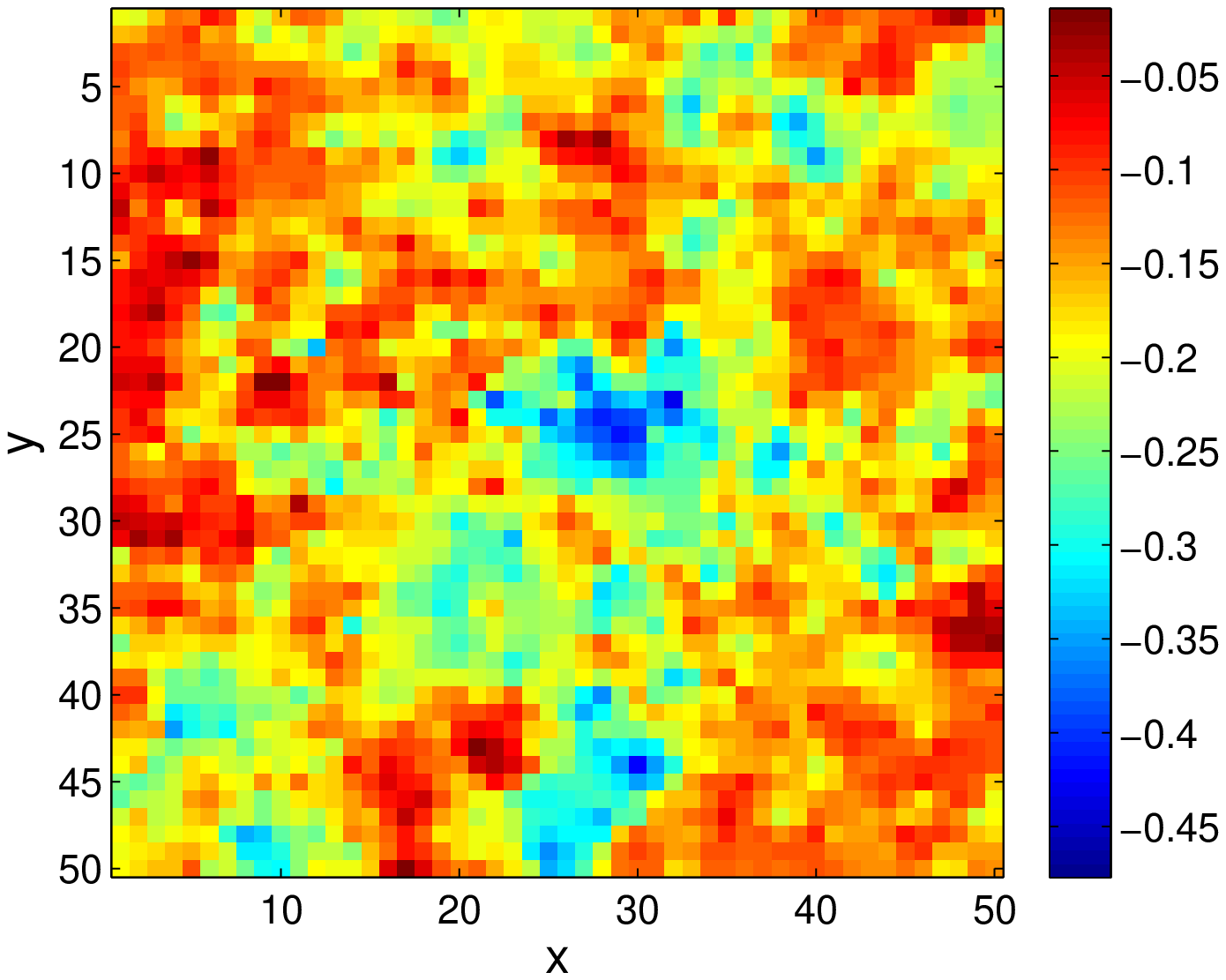}}\hspace*{-5 mm}
		\subfigure[IDW, $L_B=20$]{\label{fig:zr_ID_heat_blk20}
    \includegraphics[scale=0.35]{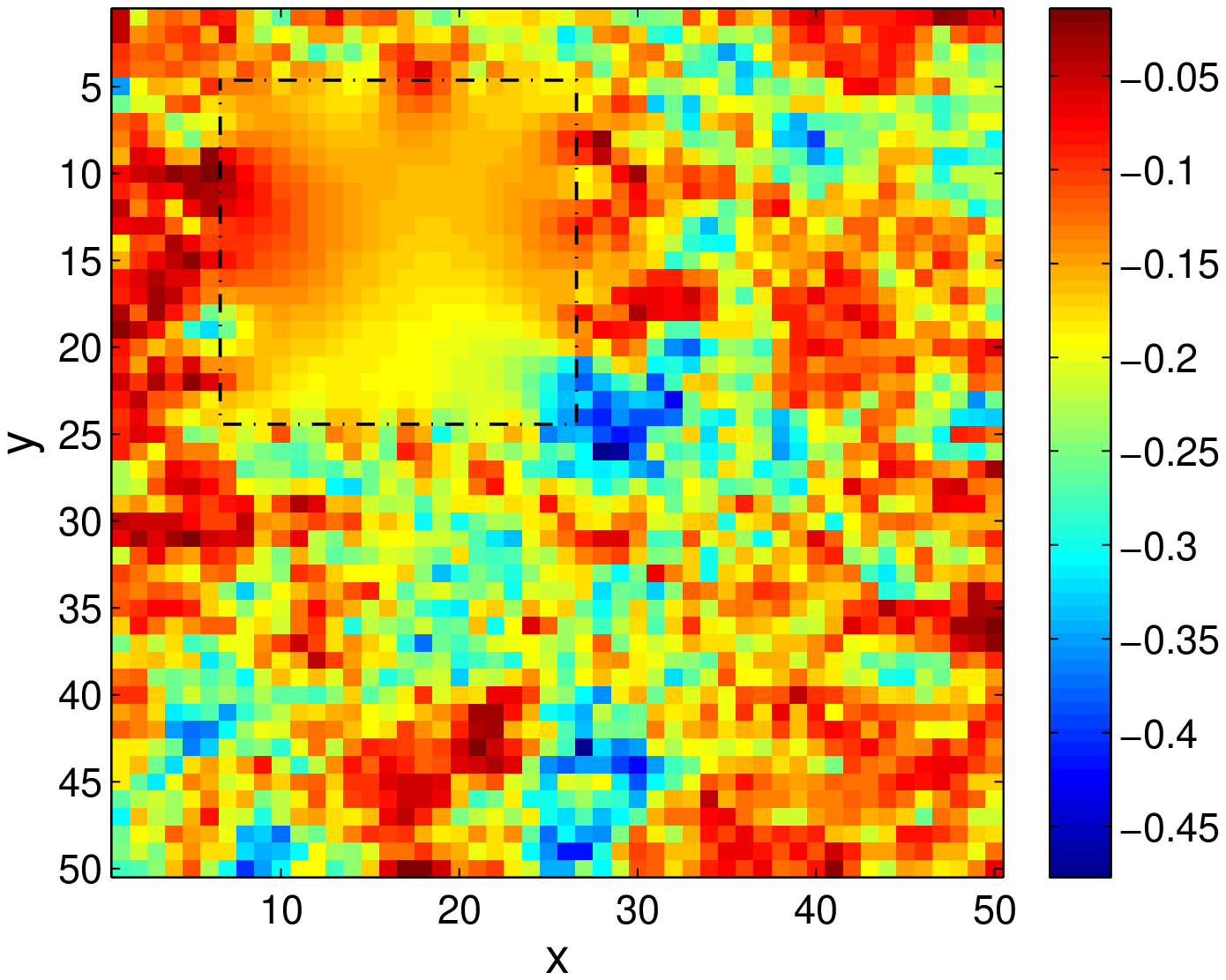}}\\\hspace*{-7 mm}
		\subfigure[MC, $p=33\%$]{\label{fig:zr_MC_heat_p33}
    \includegraphics[scale=0.35]{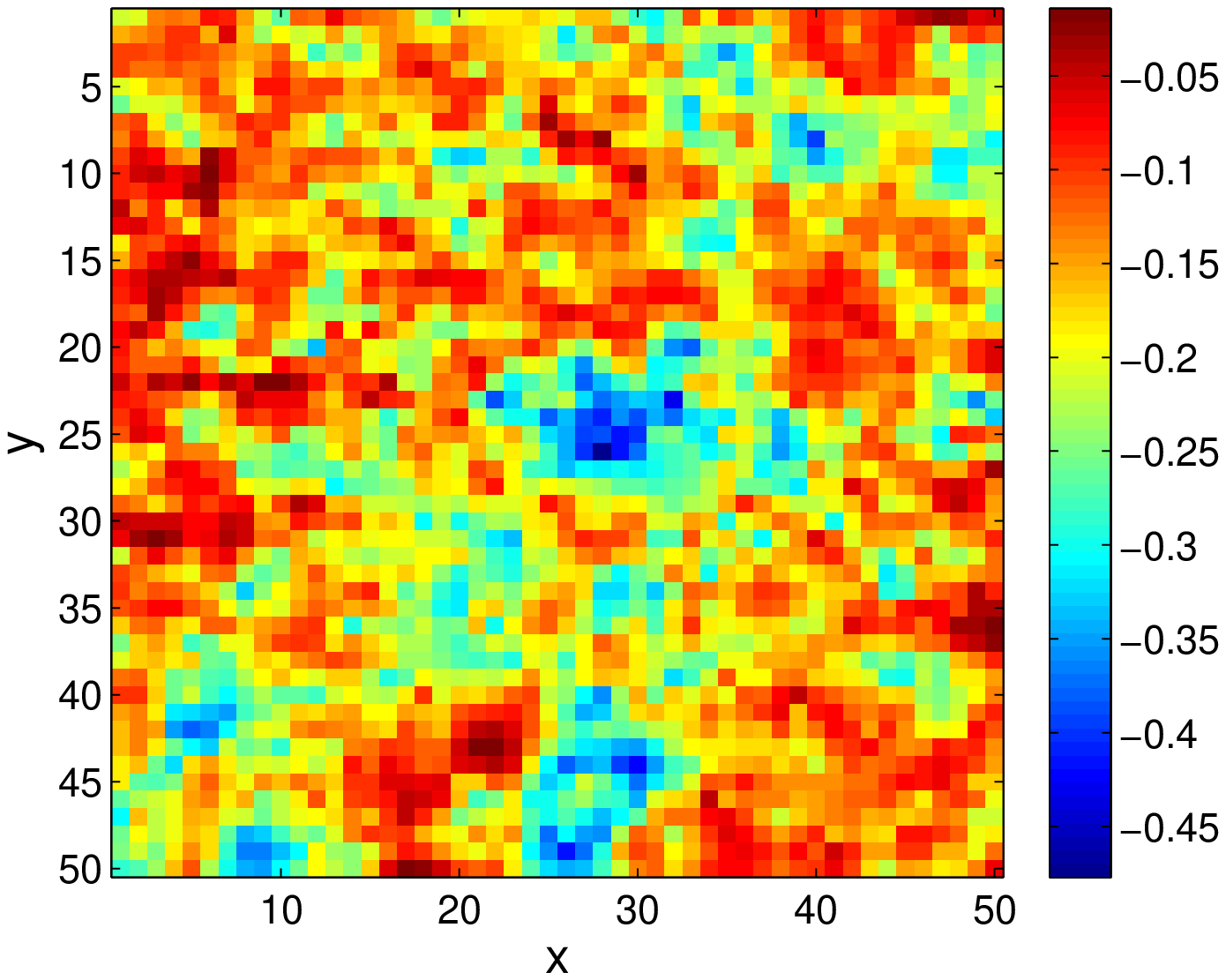}}\hspace*{-5 mm}
		\subfigure[MC, $p=66\%$]{\label{fig:zr_MC_heat_p66}
    \includegraphics[scale=0.35]{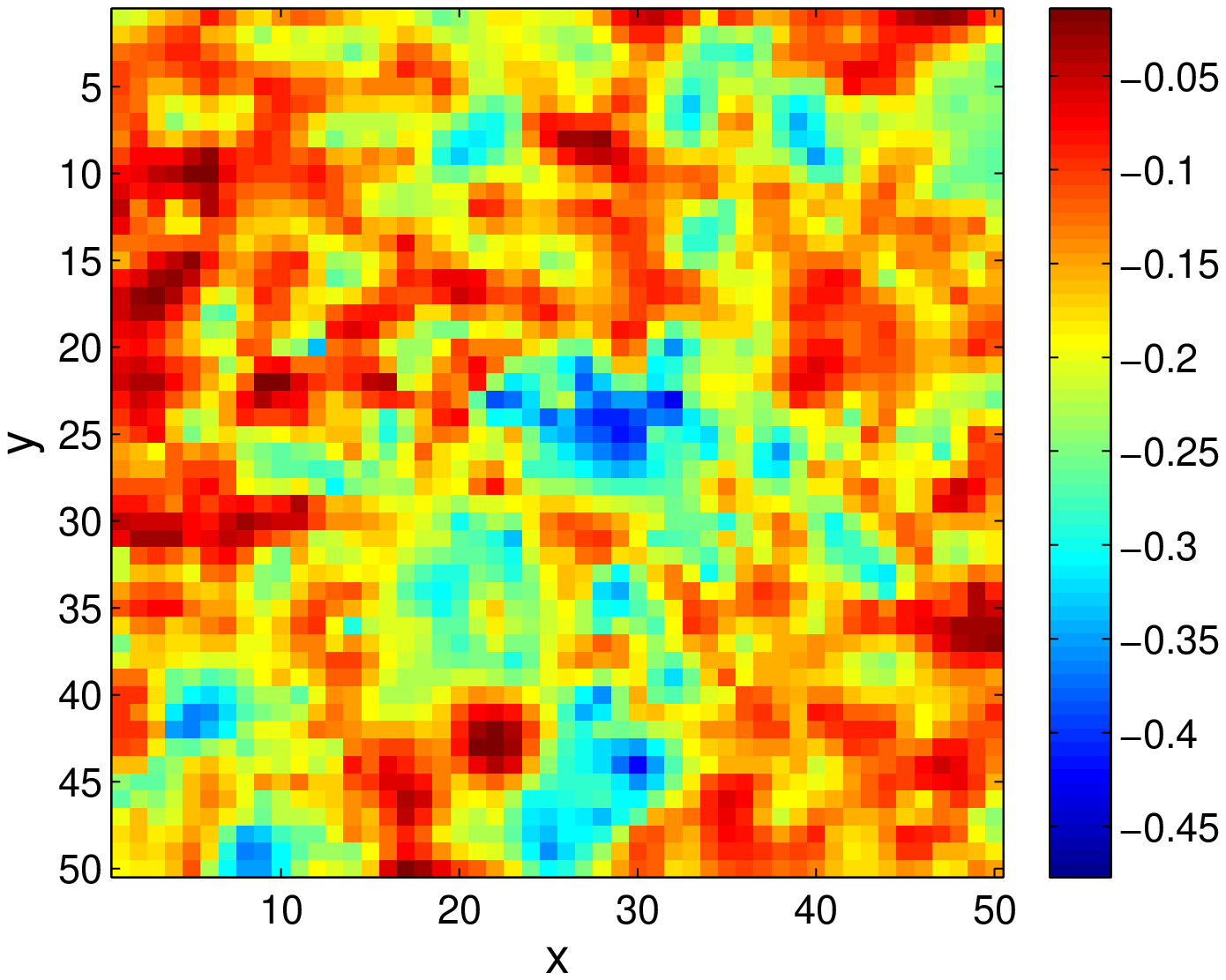}}\hspace*{-5 mm}
		\subfigure[MC, $L_B=20$]{\label{fig:zr_MC_heat_blk20}
    \includegraphics[scale=0.35]{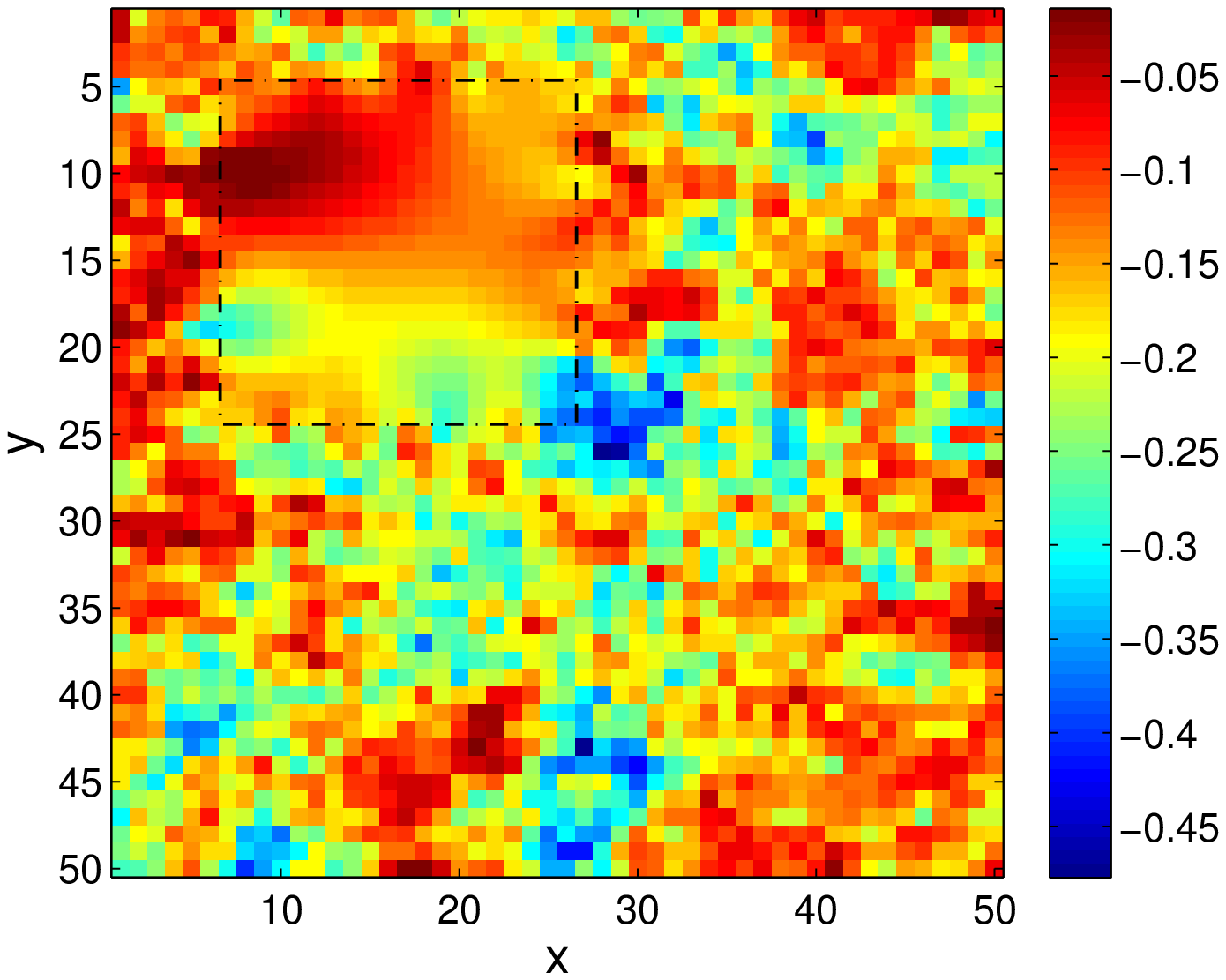}}\\\hspace*{-7 mm}
\end{center}
    \caption{Visual comparison of \mpr, IDW and MC interpolated maps for the latent heat data shown in Fig.~\ref{fig:zo_heat}. Three different  patterns of missing data are used: 33\% random thinning (left column), 66\% random thinning (middle column) and solid block removal (right column). The perimeter of the missing data block  is marked by the dashed line.}
  \label{fig:heat}
\end{figure}

\begin{figure}[t!]
\begin{center}\hspace*{-7 mm}
    \subfigure[\mpr, $p=33\%$]{\label{fig:zr_XY_walker_p33}
    \includegraphics[scale=0.35]{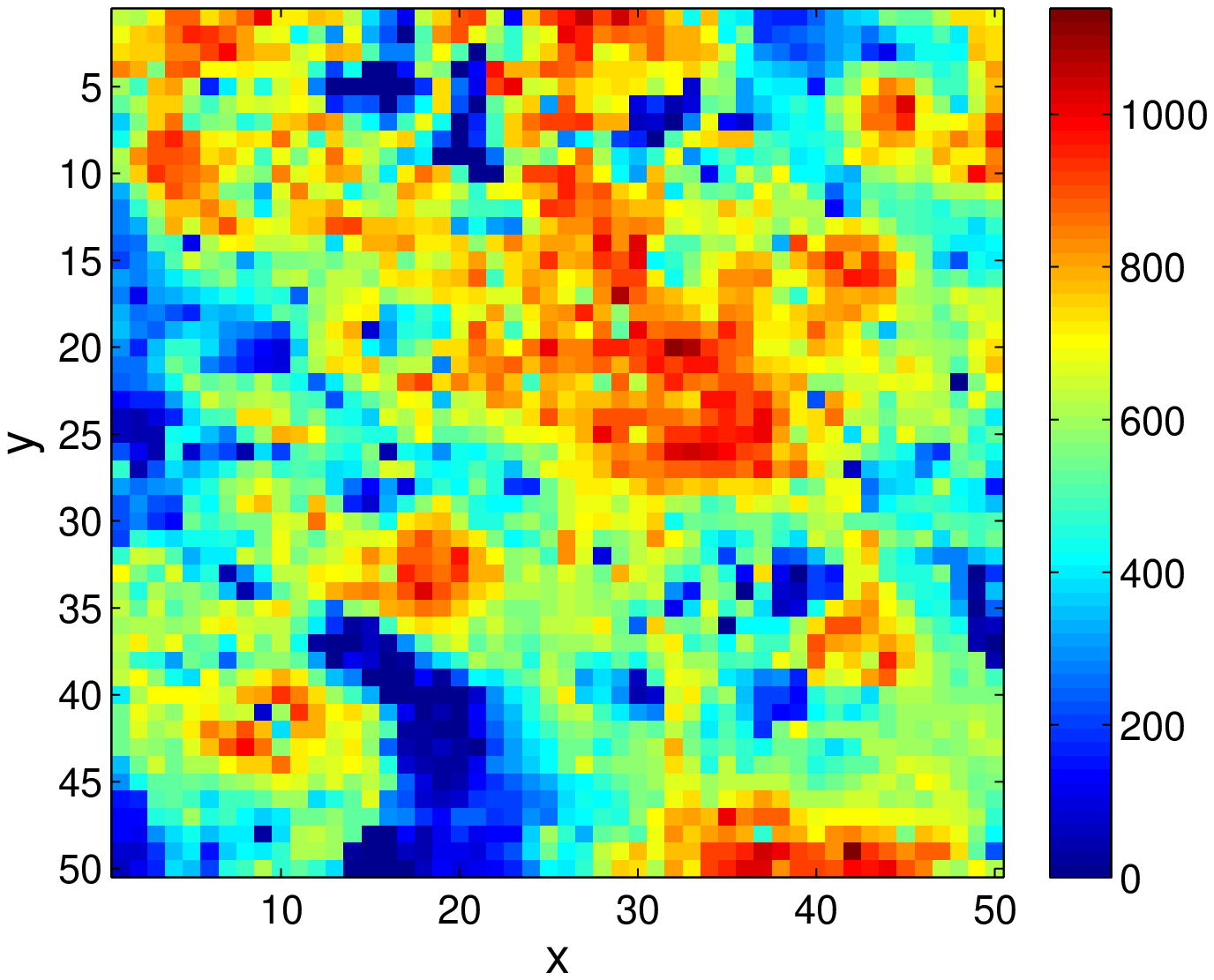}}\hspace*{-5 mm}
		\subfigure[\mpr, $p=66\%$]{\label{fig:zr_XY_walker_p66}
    \includegraphics[scale=0.35]{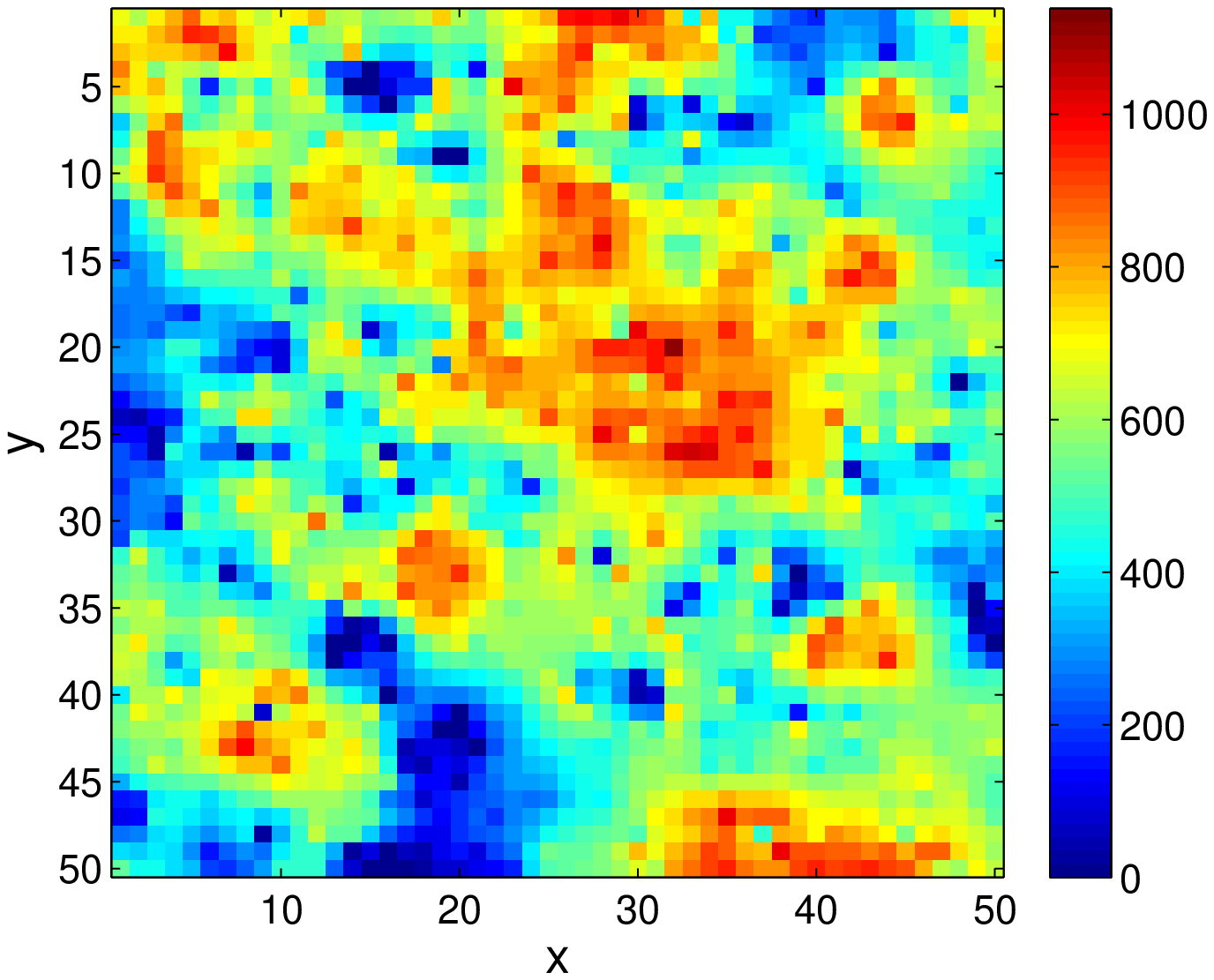}}\hspace*{-5 mm}
		\subfigure[\mpr, $L_B=20$]{\label{fig:zr_XY_walker_blk20}
    \includegraphics[scale=0.35]{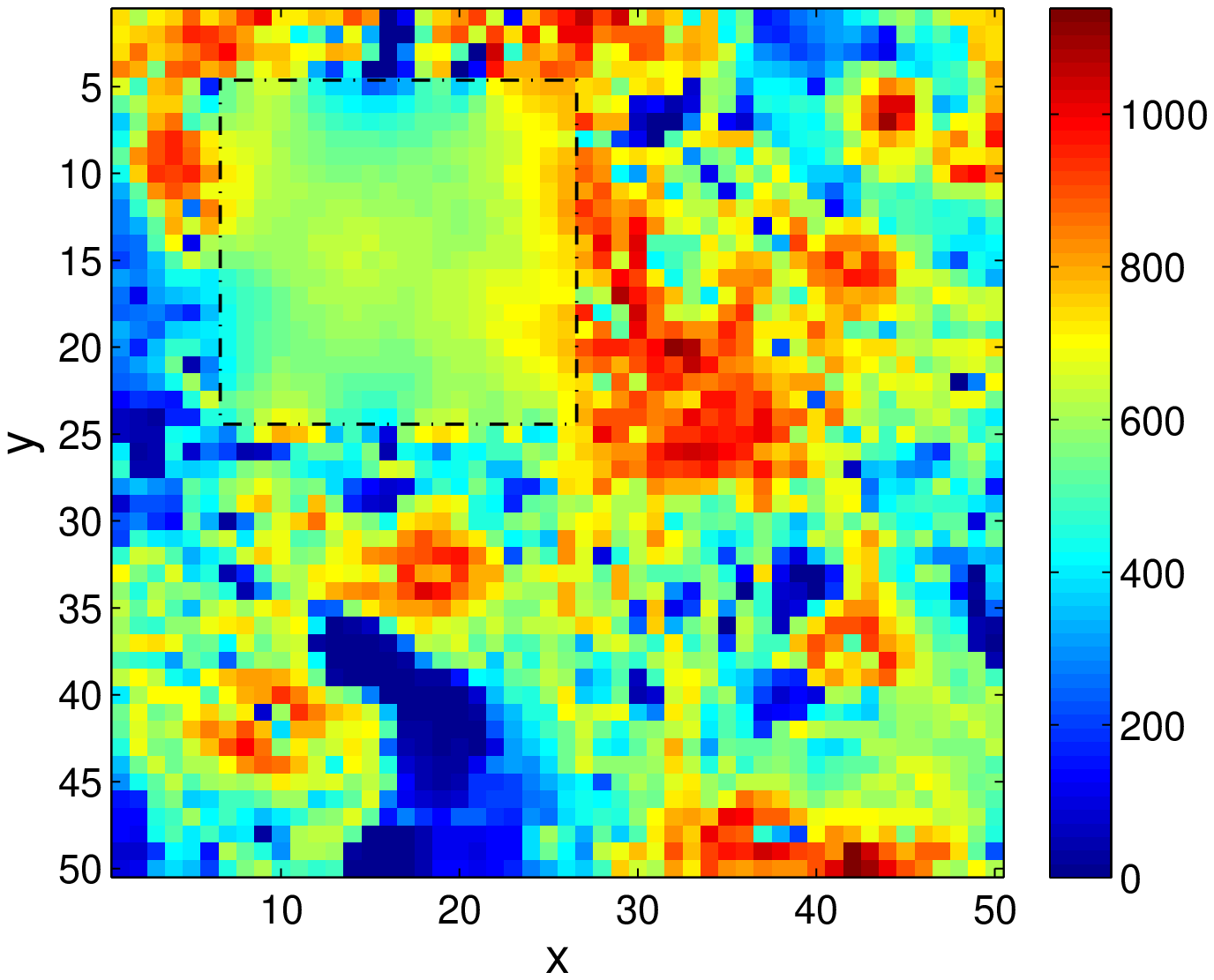}}\\\hspace*{-7 mm}
		\subfigure[IDW, $p=33\%$]{\label{fig:zr_ID_walker_p33}
    \includegraphics[scale=0.35]{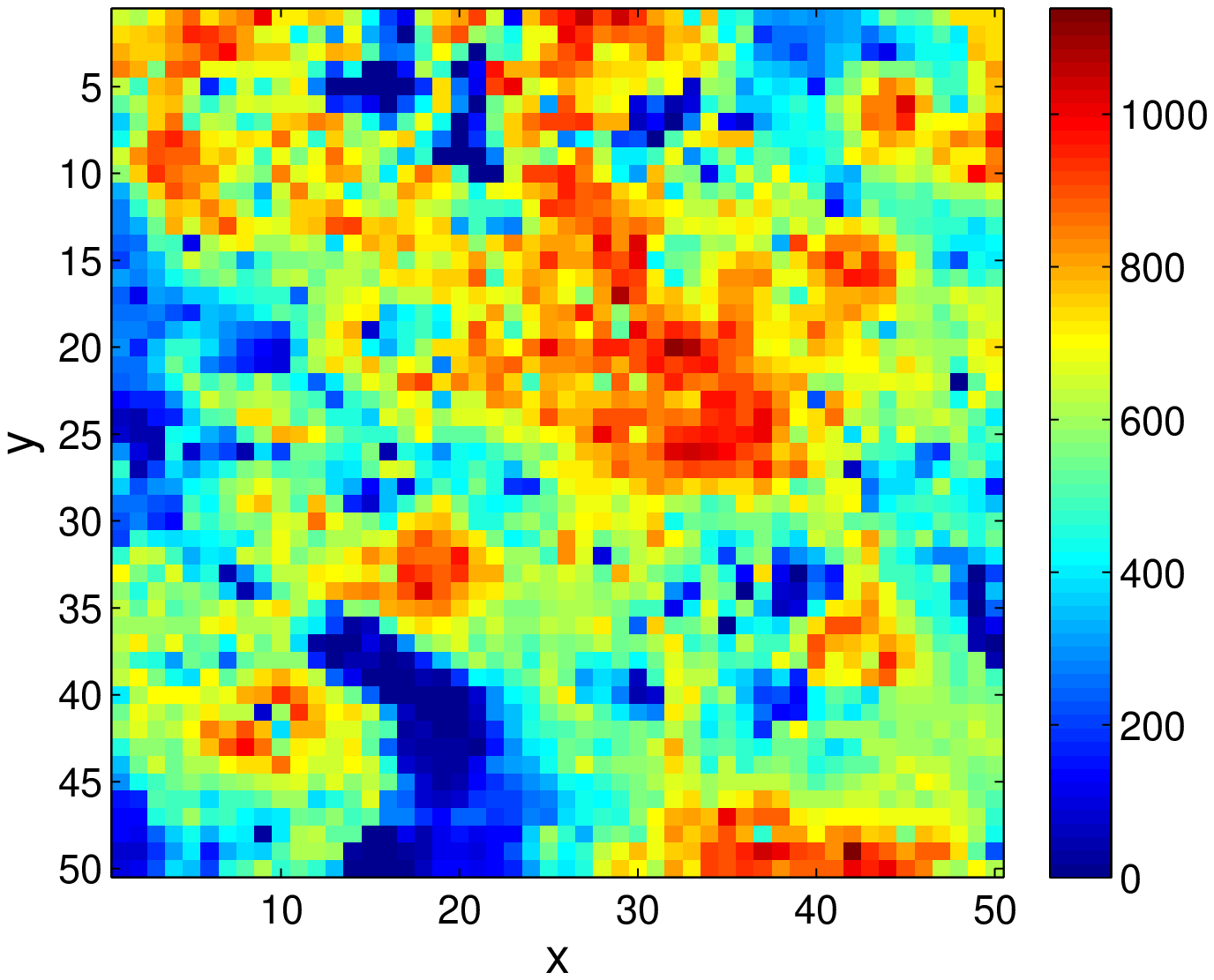}}\hspace*{-5 mm}
		\subfigure[IDW, $p=66\%$]{\label{fig:zr_ID_walker_p66}
    \includegraphics[scale=0.35]{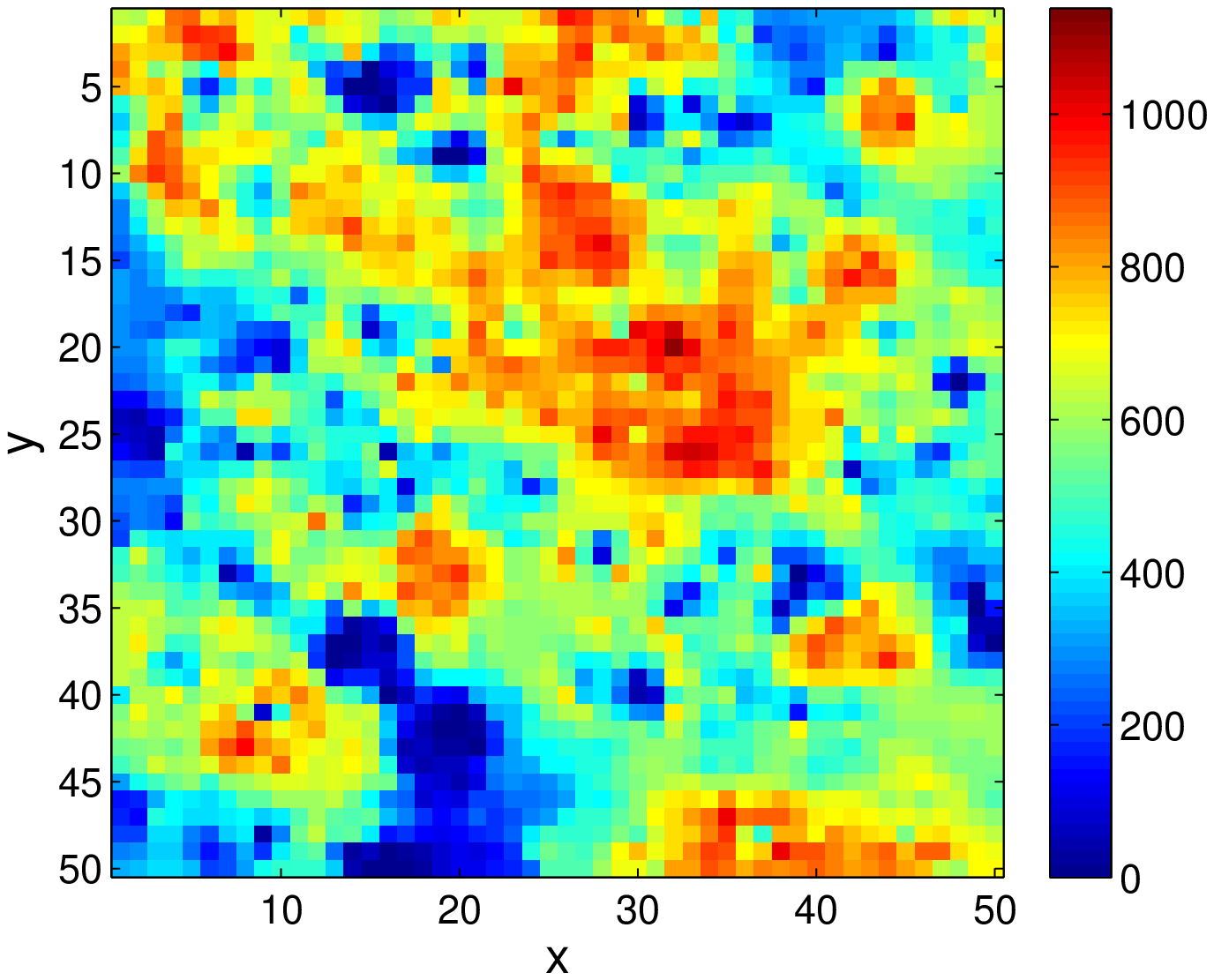}}\hspace*{-5 mm}
		\subfigure[IDW, $L_B=20$]{\label{fig:zr_ID_walker_blk20}
    \includegraphics[scale=0.35]{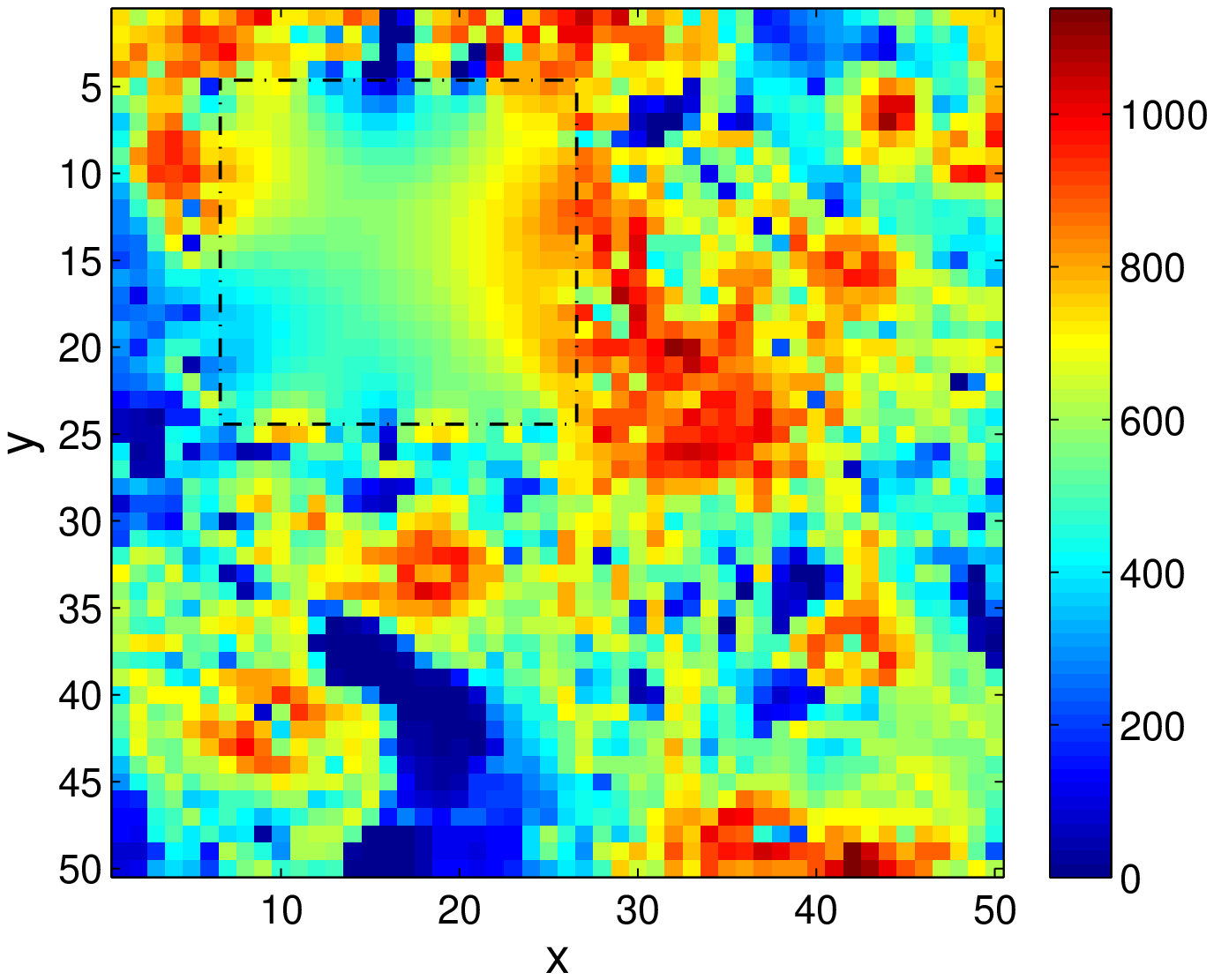}}\\\hspace*{-7 mm}
		\subfigure[MC, $p=33\%$]{\label{fig:zr_MC_walker_p33}
    \includegraphics[scale=0.35]{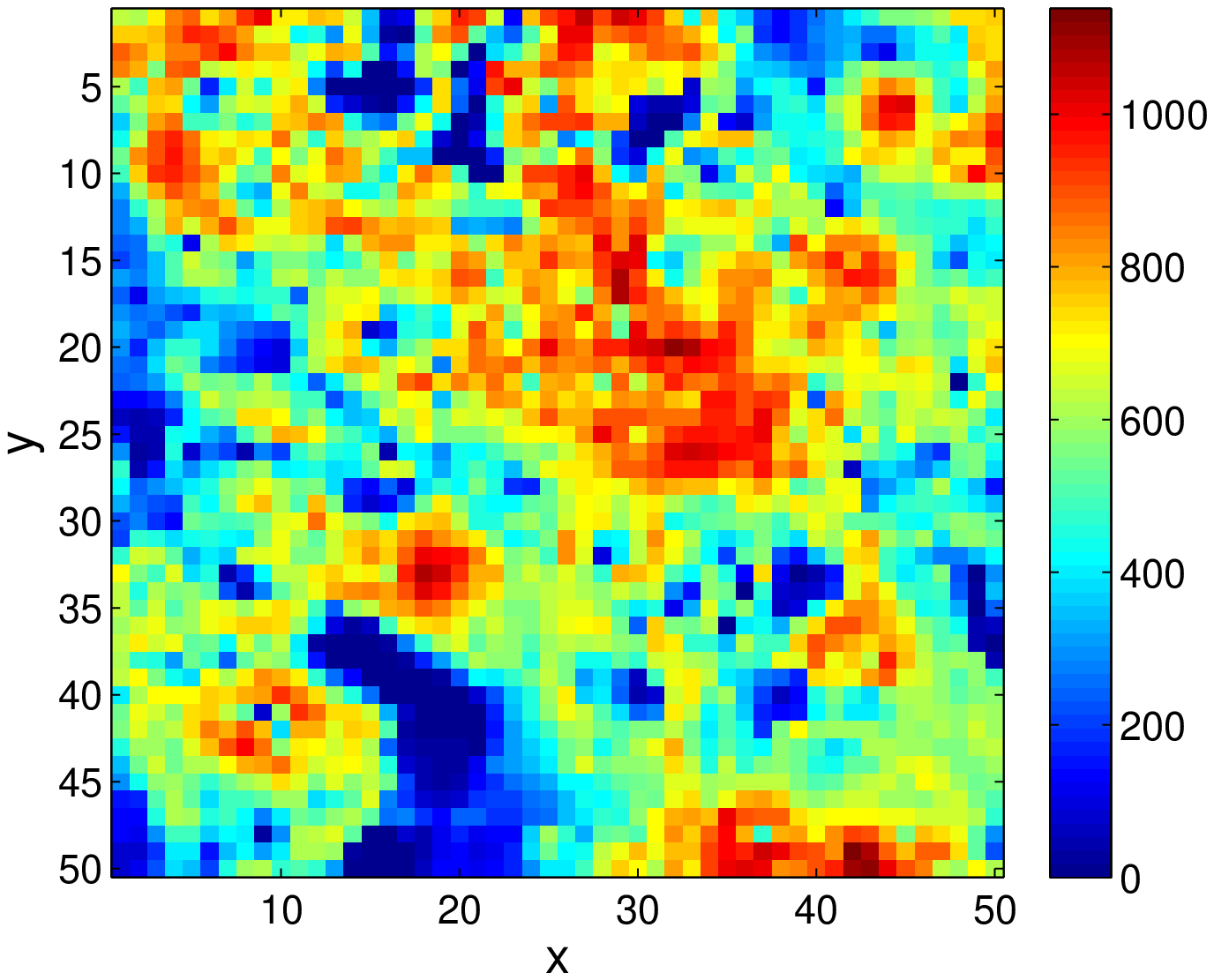}}\hspace*{-5 mm}
		\subfigure[MC, $p=66\%$]{\label{fig:zr_MC_walker_p66}
    \includegraphics[scale=0.35]{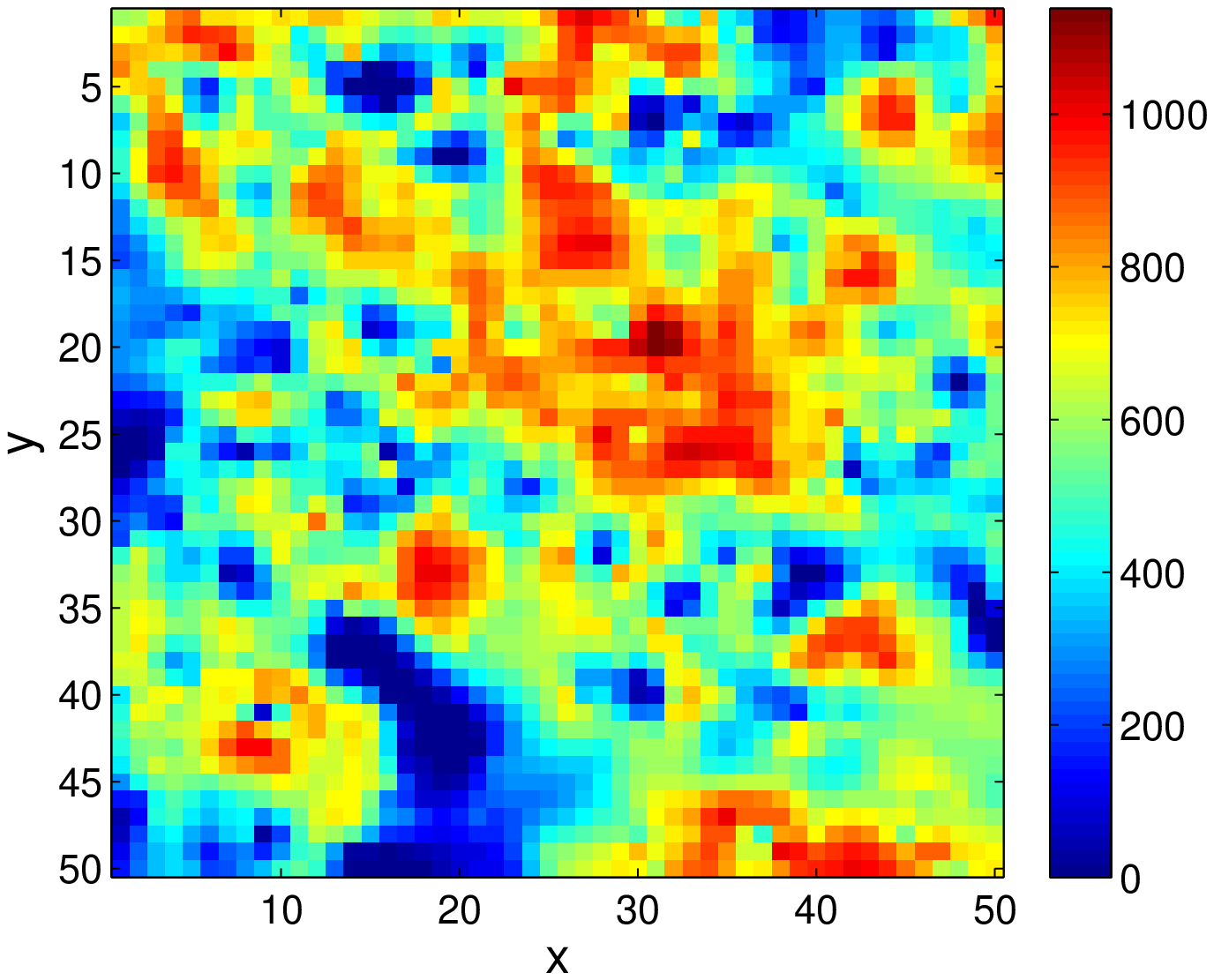}}\hspace*{-5 mm}
		\subfigure[MC, $L_B=20$]{\label{fig:zr_MC_walker_blk20}
    \includegraphics[scale=0.35]{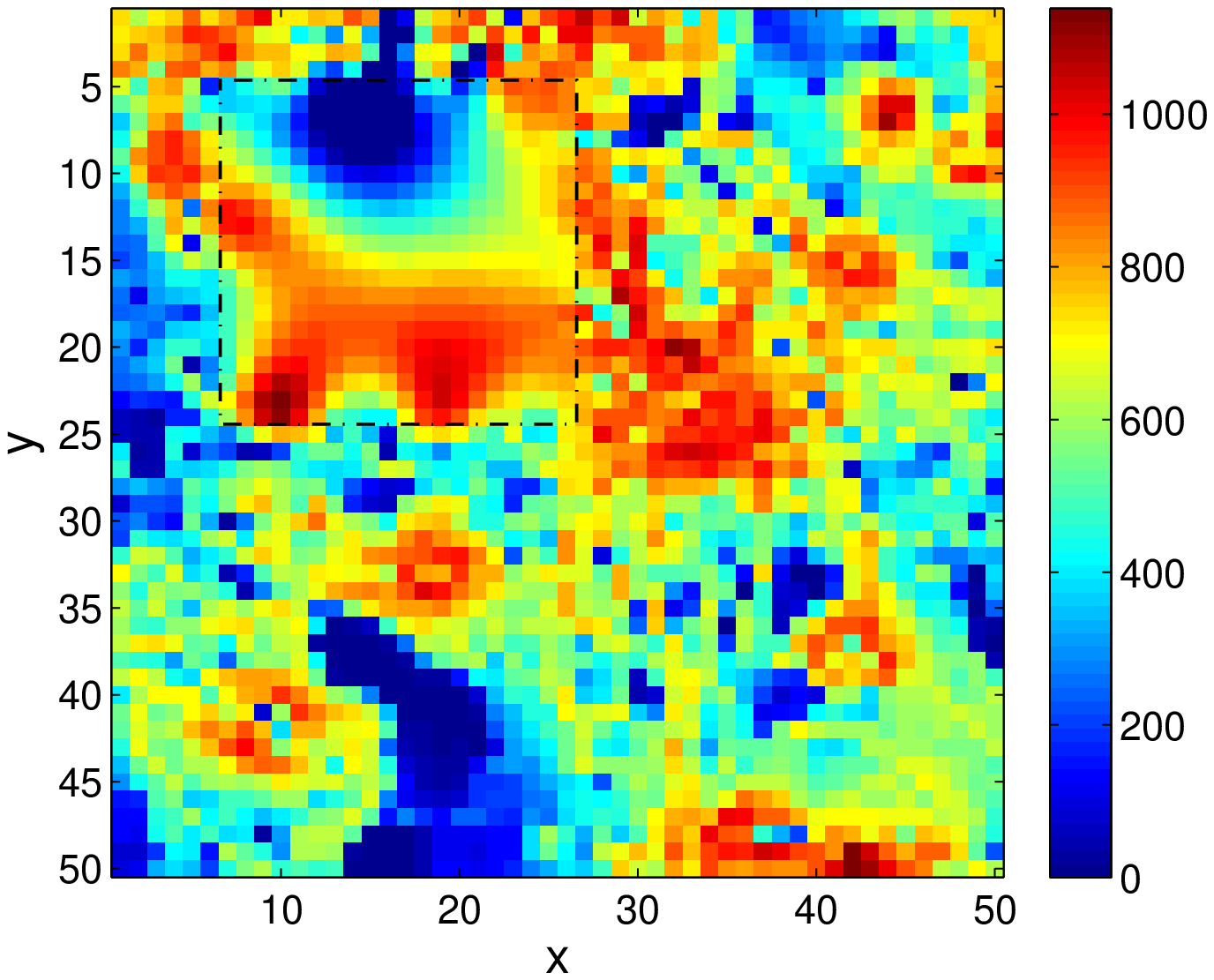}}\\\hspace*{-7 mm}
\end{center}
    \caption{Visual comparison of \mpr, IDW and MC interpolated maps for the Walker lake data shown in Fig.~\ref{fig:zo_walker}. Three different  patterns of missing data are used: 33\% random thinning (left column), 66\% random thinning (middle column) and solid block removal (right column). The perimeter of the missing data block  is marked by the dashed-line. }
  \label{fig:walker}
\end{figure}

In Figs.~\ref{fig:heat} and~\ref{fig:walker} we present examples of interpolated maps for different patterns of missing data, using three methods that give the best prediction performance, i.e., the \mpr, IDW and MC. For randomly thinned data, visual differences between predictions obtained by individual methods are not so obvious. All methods display smoothing that naturally increases with the fraction of missing data. The biggest differences between the methods appear in the case of missing solid data blocks.


\begin{table}[t!]
\addtolength{\tabcolsep}{0pt} \caption{Interpolation validation
measures for real data based on $S=100$ missing-data samples generated by (a) $p=33\%$, (b) $p=66\%$ random thinning and (c) random removal of square data block  with side length $L_B=20$.} \vspace{3pt} \label{tab:real}
\begin{scriptsize}
\resizebox{1\textwidth}{!}{
\begin{tabular}{|c|c|ccc|ccc|ccc|ccc|}
\hline
& & \multicolumn{3}{c|}{MAAE}  & \multicolumn{3}{c|}{MRASE} &
 \multicolumn{3}{c|}{MR [\%]}  & \multicolumn{3}{c|}{$ \langle t_{\mathrm{cpu}} \rangle $}   \\
 Data set & & (a) & (b) & (c) & (a) & (b) & (c) & (a) & (b) & (c) & (a) & (b) & (c) \\
\hline
\parbox[t]{2mm}{\multirow{6}{*}{\rotatebox[origin=c]{90}{Latent heat}}} &\mpr&0.04&0.04&0.06&0.05&0.05&0.07&79.25&72.40&38.14&0.05&0.09&0.03 \\
 &${\rm NN^*}$&0.72&0.80&0.80&0.71&0.79&0.78&1.25&1.17&1.35&{\bf 9.01}&{\bf 18.82}&{\bf 4.31} \\
&${\rm BL^*}$&0.95&0.96&0.87&0.94&0.95&0.87&1.25&1.16&1.36&{\bf 4.59}&{\bf 12.35}&{\bf 2.33} \\
 &${\rm BC^*}$&0.95&0.95&0.85&0.94&0.94&0.84&1.03&1.04&1.57&{\bf 9.28}&{\bf 24.85}&{\bf 4.79} \\
 &${\rm MC^*}$&1.00&0.97&0.81&0.99&0.96&0.80&1.00&1.01&1.21&0.18&0.65&0.08 \\
 &${\rm IDW^*}$&0.98&1.00&{\bf 1.02}&0.98&0.99&{\bf 1.01}&1.01&1.01&1.01&0.25&0.43&0.40 \\
\hline
\parbox[t]{2mm}{\multirow{6}{*}{\rotatebox[origin=c]{90}{Walker lake}}} &\mpr&102.02&117.52&167.93&138.97&156.57&212.55&82.79&77.51&45.32&0.05&0.09&0.03 \\
 &${\rm NN^*}$&0.76&0.85&0.83&0.73&0.80&0.80&1.17&1.14&1.27&{\bf 9.00}&{\bf 18.79}&{\bf 4.32} \\
 &${\rm BL^*}$&0.96&0.99&0.91&0.93&0.96&0.88&1.17&1.14&1.27&{\bf 4.59}&{\bf 12.43}&{\bf 2.35} \\
 &${\rm BC^*}$&0.96&0.98&0.89&0.93&0.94&0.86&1.03&1.04&1.33&{\bf 9.24}&{\bf 24.71}&{\bf 4.82} \\
 &${\rm MC^*}$&0.99&0.98&0.76&0.97&0.95&0.75&1.01&1.02&1.24&0.18&0.63&0.08 \\
 &${\rm IDW^*}$&0.98&{\bf 1.01}&1.00&0.98&0.99&0.99&1.01&1.01&1.02&0.25&0.43&0.41 \\
\hline
\end{tabular}
}
\end{scriptsize}
\end{table}

\begin{figure}[t!]
\begin{center}
    \subfigure[Standard deviation]{\label{fig:std_XY_walker_blk20}
    \includegraphics[scale=0.46,clip]{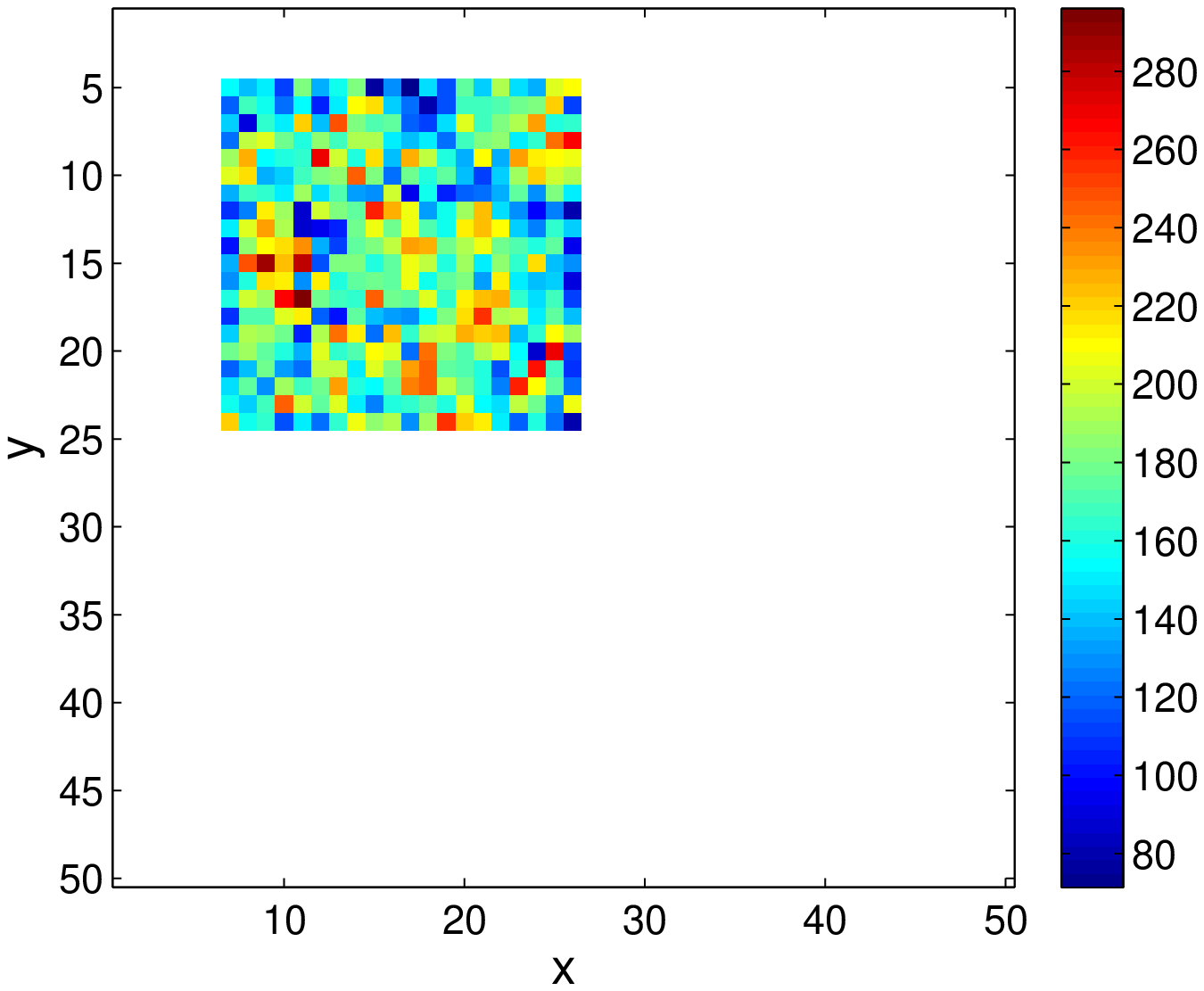}}
    \subfigure[Realization 1]{\label{fig:zr_XY_walker_blk20_real1}
    \includegraphics[scale=0.46,clip]{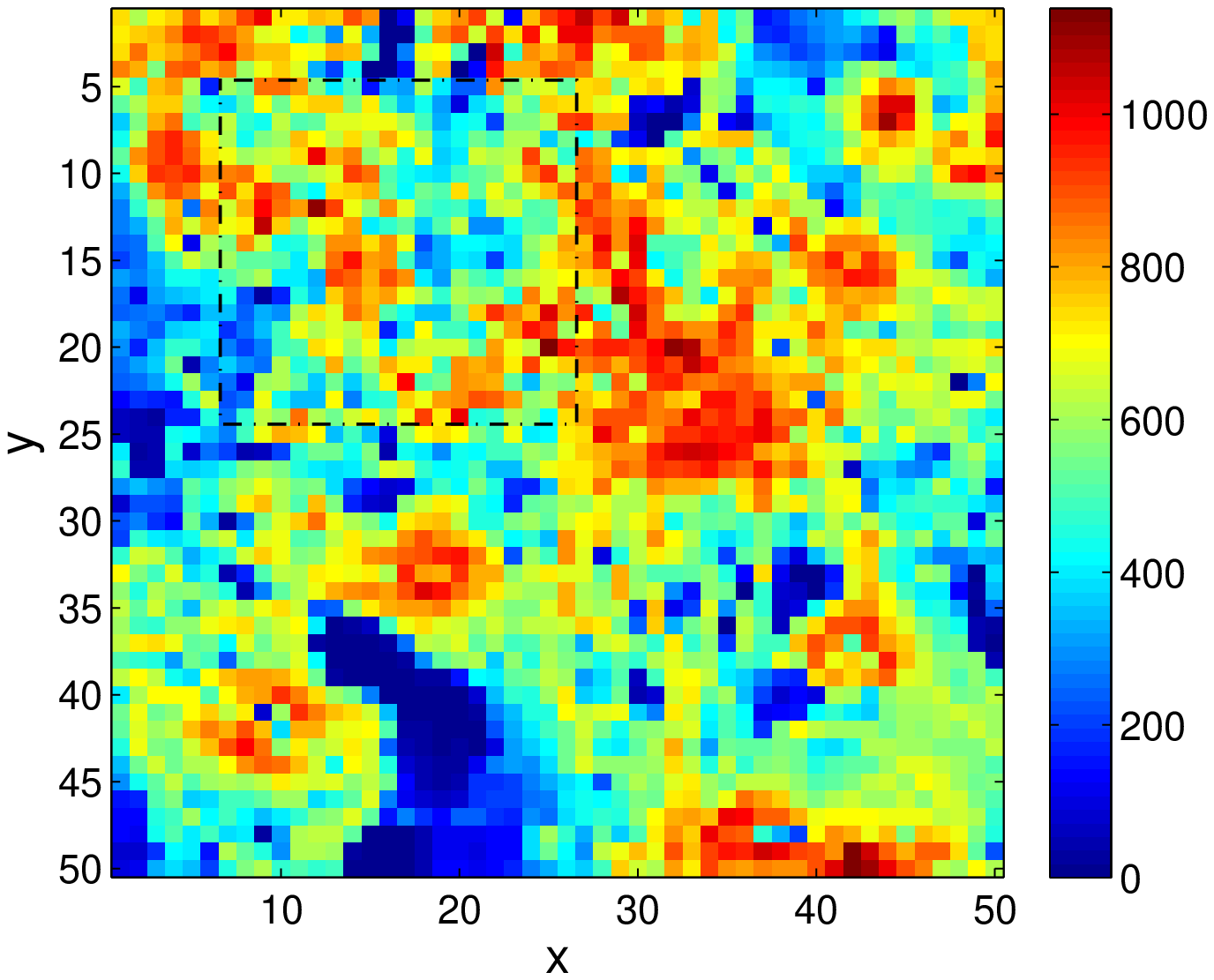}} \\
    \subfigure[Realization 2]{\label{fig:zr_XY_walker_blk20_real3}
    \includegraphics[scale=0.46,clip]{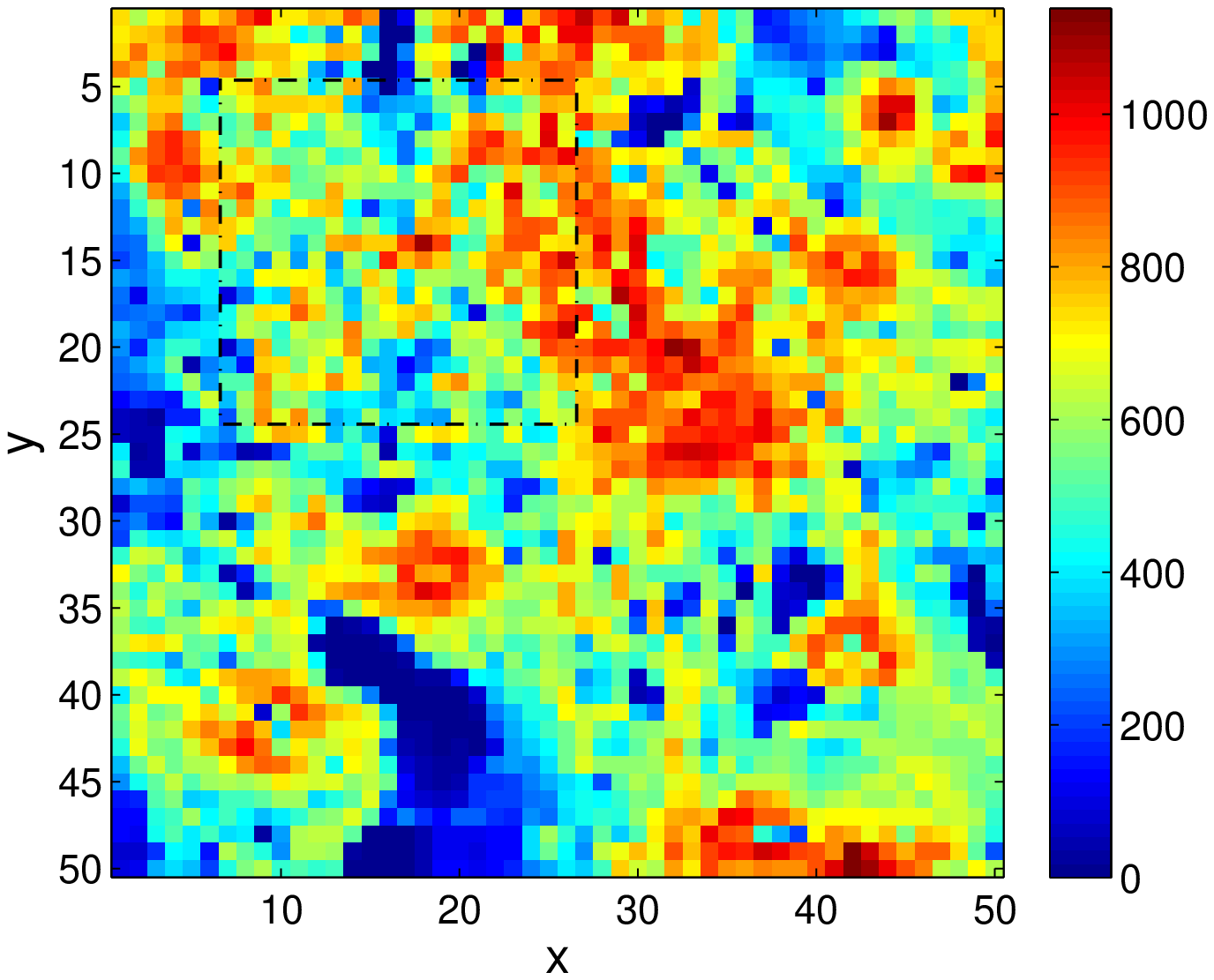}}
		\subfigure[Realization 3]{\label{fig:zr_XY_walker_blk20_real4}
    \includegraphics[scale=0.46,clip]{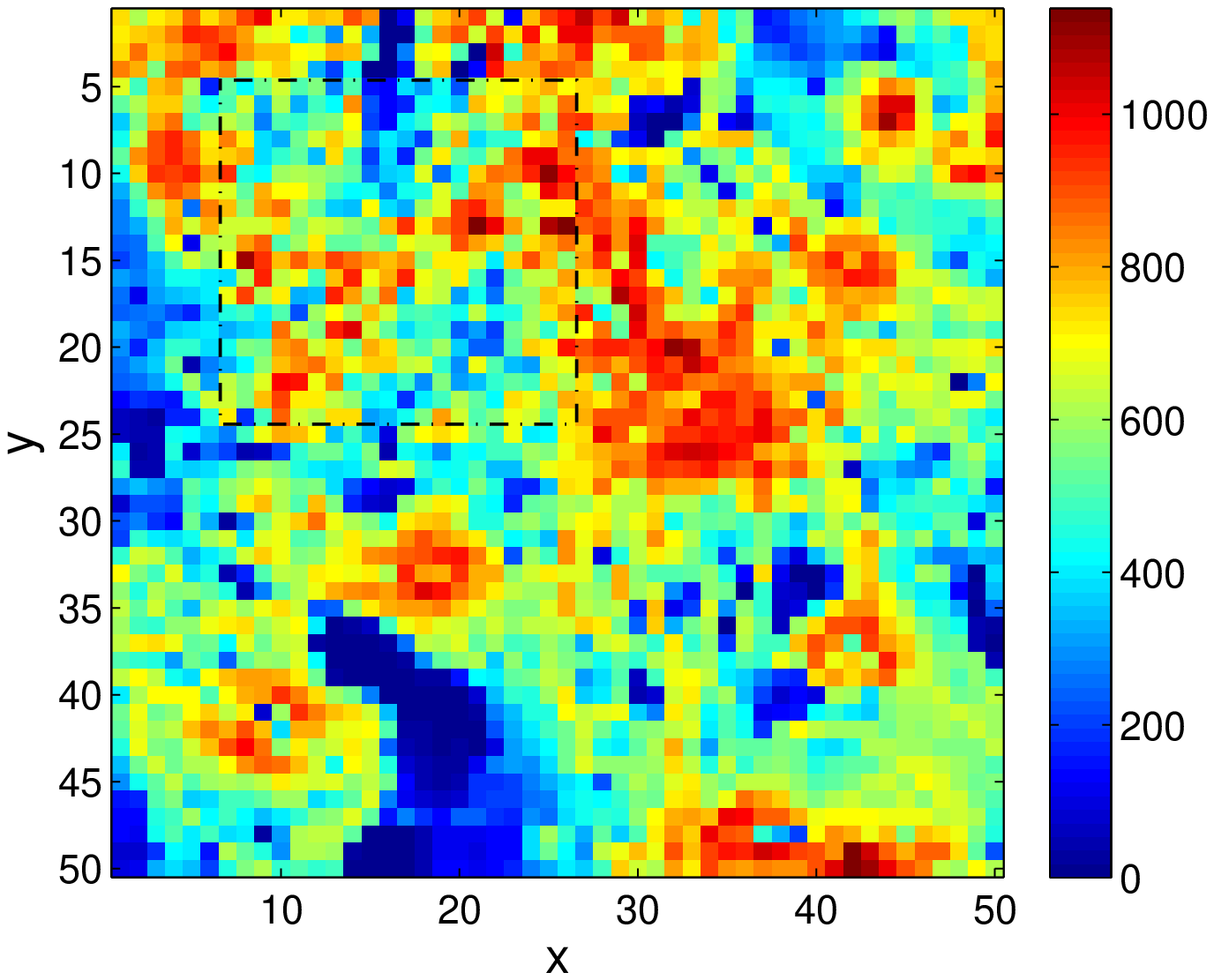}}
\end{center}
    \caption{Investigation of \mpr conditional simulations of Walker lake data with a missing square block: (a) Standard deviation of block values based on $M=100$ equilibrium realizations. (b)-(d) Three selected realizations (the area of the missing block in the upper left corner is indicated by means of the thin dashed line).}
  \label{fig:walker_fluct}
\end{figure}

The \mpr predictions represent conditional means over $M$ equilibrium realizations generated by means of conditional simulation. Hence, even though the \mpr predictions are relatively smooth, the  prediction variance is appreciable, as shown in Fig.~\ref{fig:std_XY_walker_blk20} for the Walker lake data. In addition, visual supervision of  individual \mpr-reconstructed realizations in Figs.~\ref{fig:zr_XY_walker_blk20_real1}-\ref{fig:zr_XY_walker_blk20_real4}
demonstrate that the intra-block spatial variability is well reconstructed and that the patterns look ``natural''.

\subsection{Comparison with the Gaussian model}
Given the resemblance of the \mpr energy function~\eqref{Hamiltonian_mod} with the quadratic form, one may wonder if the \mpr model has any advantages over the simpler Gaussian model. The latter is analytically solvable and has been applied in data reconstruction problems such as the image restoration~\citep{tana02,kuwa14,kata17}.
There are two approaches for implementing the latter. In the first, the energy is expressed as a quadratic function of the original field values $z_{i}$, and in the second as a quadratic function of the respective spin angles.  Since the first approach strongly penalizes large deviations from the mean, we opted for the second.

In the following, we compare the gap-filling performance of the \mpr model, which is defined by the pair interactions $H_{ij}=\cos[q\Delta\phi_{ij}]$ with the GMRF, which is defined by the pair interactions $H_{ij}=-\Delta\phi_{ij}^2/2\pi^2+1$. In both cases, the energy is given by ${\mathcal H} = -\sum_{\langle i, j\rangle}H_{ij}$, while the \emph{spin angle contrast} takes values  $\Delta\phi_{ij} = \phi_i-\phi_j \in [-2\pi,2\pi]$, and the pair interactions $H_{ij} \in [-1,1]$.

Based on the comparison between  the SEM method  and \emph{maximum likelihood estimation (MLE)} that give practically identical results for the one-dimensional \mpr model (see Section~\ref{ssec:sem-mle} below), we estimated parameters for both the \mpr and the GMRF using SEM. The \mpr sample specific energy for temperature inference is given by~\eqref{eq:mpr-sse}, while the
corresponding GMRF  sample specific energy is given by
\[
e_s = \frac{1}{N_{SP}}\sum_{i = 1}^{N}\sum_{j \in nn(i)}(\phi_i-\phi_j)^2/2\pi^2-1.
\]

As evidenced in the upper part of Table~\ref{tab:lognormal}, for  synthetic Gaussian data with various WM covariance parameters, there are no significant differences between the gap filling performance of the two models.
This is not surprising, since for symmetric Gaussian data the two models behave quite similarly.

However, the differences between the two models can be substantial for data with non-Gaussian distributions.
In the lower part of Table~\ref{tab:lognormal} we present results for synthetic data
that follow the \emph{lognormal distribution}, i.e.,  $\log Z \sim N(m = 5, \sigma = 1)$ and different  
parametrization of the WM covariance. Hence, the lognormal random field that generates the data has a median  $z_{0.50}= \exp(m) \approx 148.41$ and a respective standard deviation of  $\sigma_{Z}= \left[ \exp(\sigma^{2})-1 \right]^{1/2} \exp(\mu+ \sigma^{2}/2)\approx  320.75$. The resulting probability distribution thus has a right tail that extends to large positive values. 

For all the cases examined, the \mpr model validation measures are clearly superior to the GMRF. In addition, the CPU time is practically the same for both models, in spite of the  higher computational cost of the cosine compared to the quadratic function. The reason is that the evaluation of the energy function represents a relatively small fraction of the total CPU time. Nevertheless, as mentioned above, the GMRF admits an explicit solution that does not require MC simulations~\citep{tana02,kuwa14,kata17}.

\begin{figure}[]
\begin{center}
\vspace*{-10 mm}
    \subfigure{\label{fig:MAE_L32_M5_S1_R2_NU_025}
    \includegraphics[scale=0.35,clip]{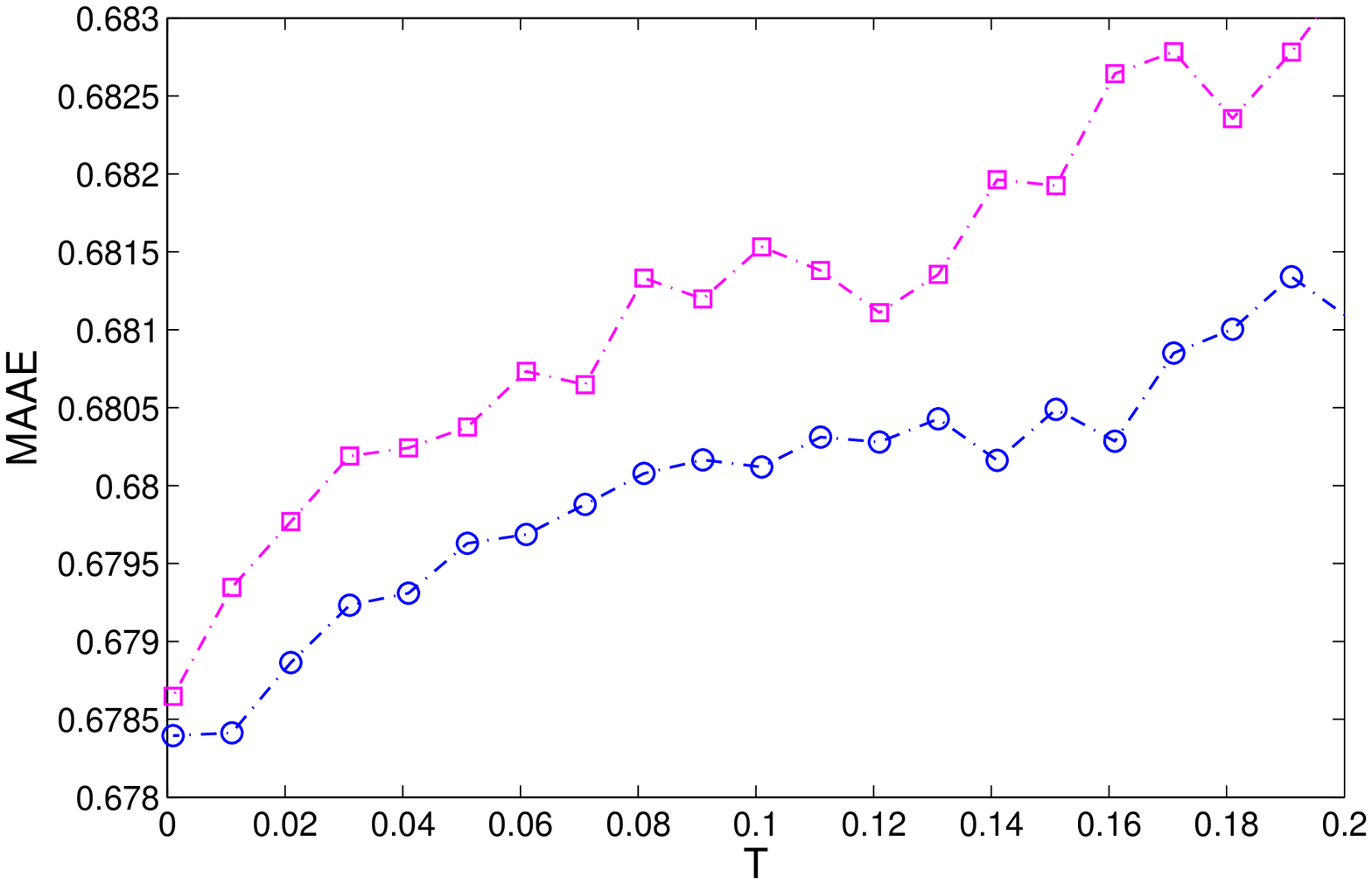}}
    \subfigure{\label{fig:MAE_L32_M5_S1_R2_NU_025_LOGN}
    \includegraphics[scale=0.35,clip]{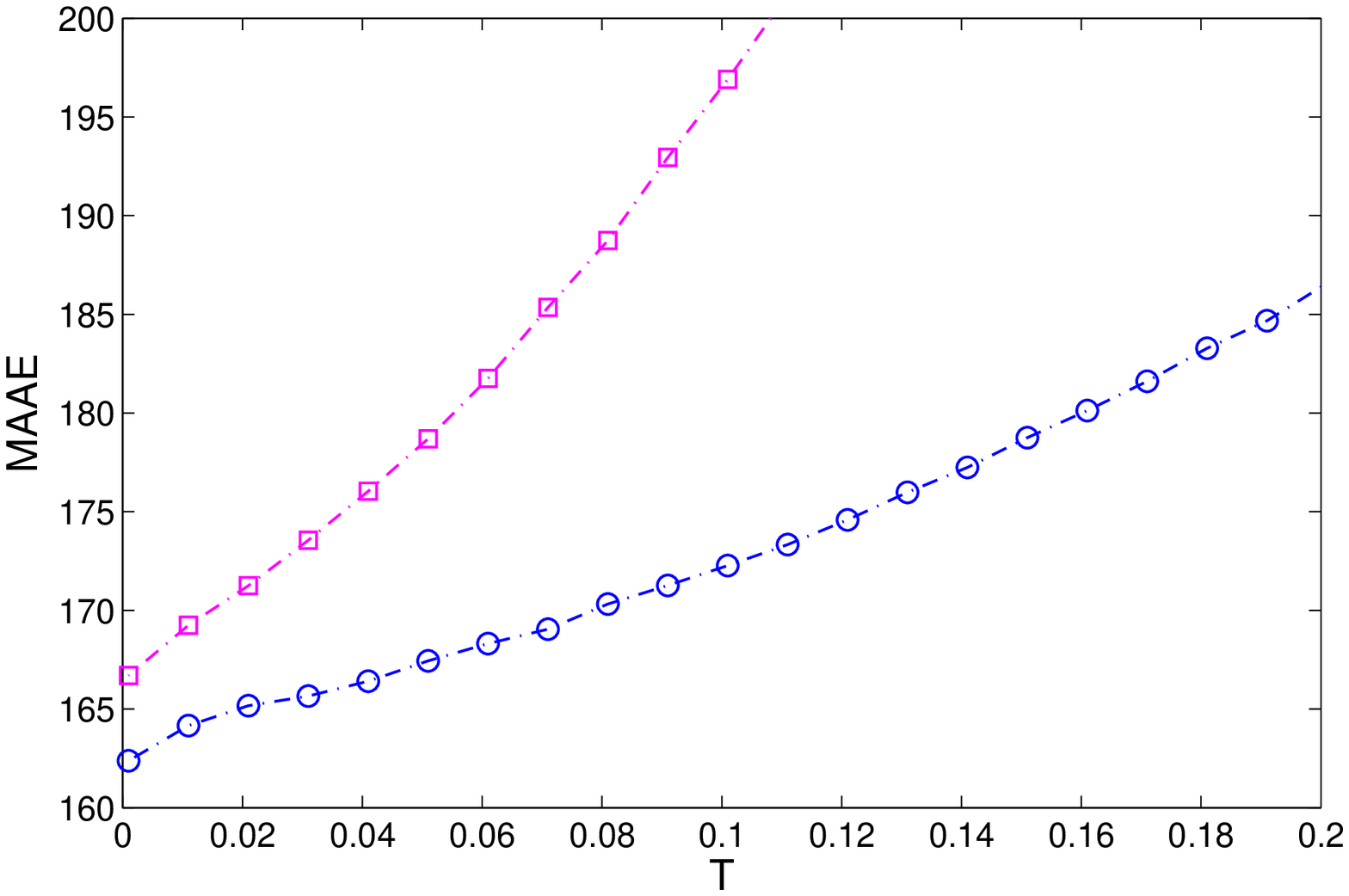}}\vspace*{-5 mm}\\
		\subfigure{\label{fig:MRE_L32_M5_S1_R2_NU_025}
    \includegraphics[scale=0.35,clip]{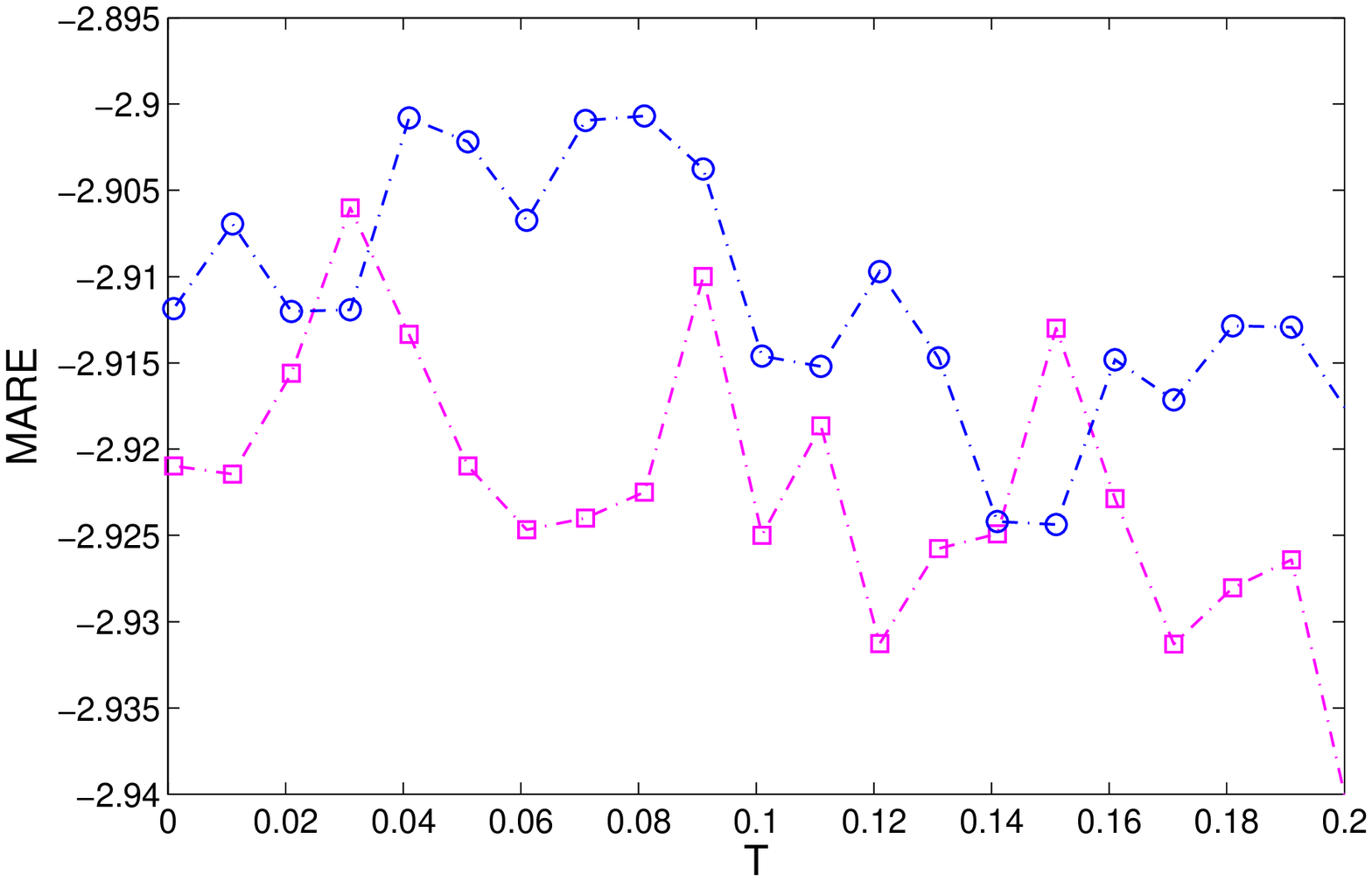}}
    \subfigure{\label{fig:MRE_L32_M5_S1_R2_NU_025_LOGN}
    \includegraphics[scale=0.35,clip]{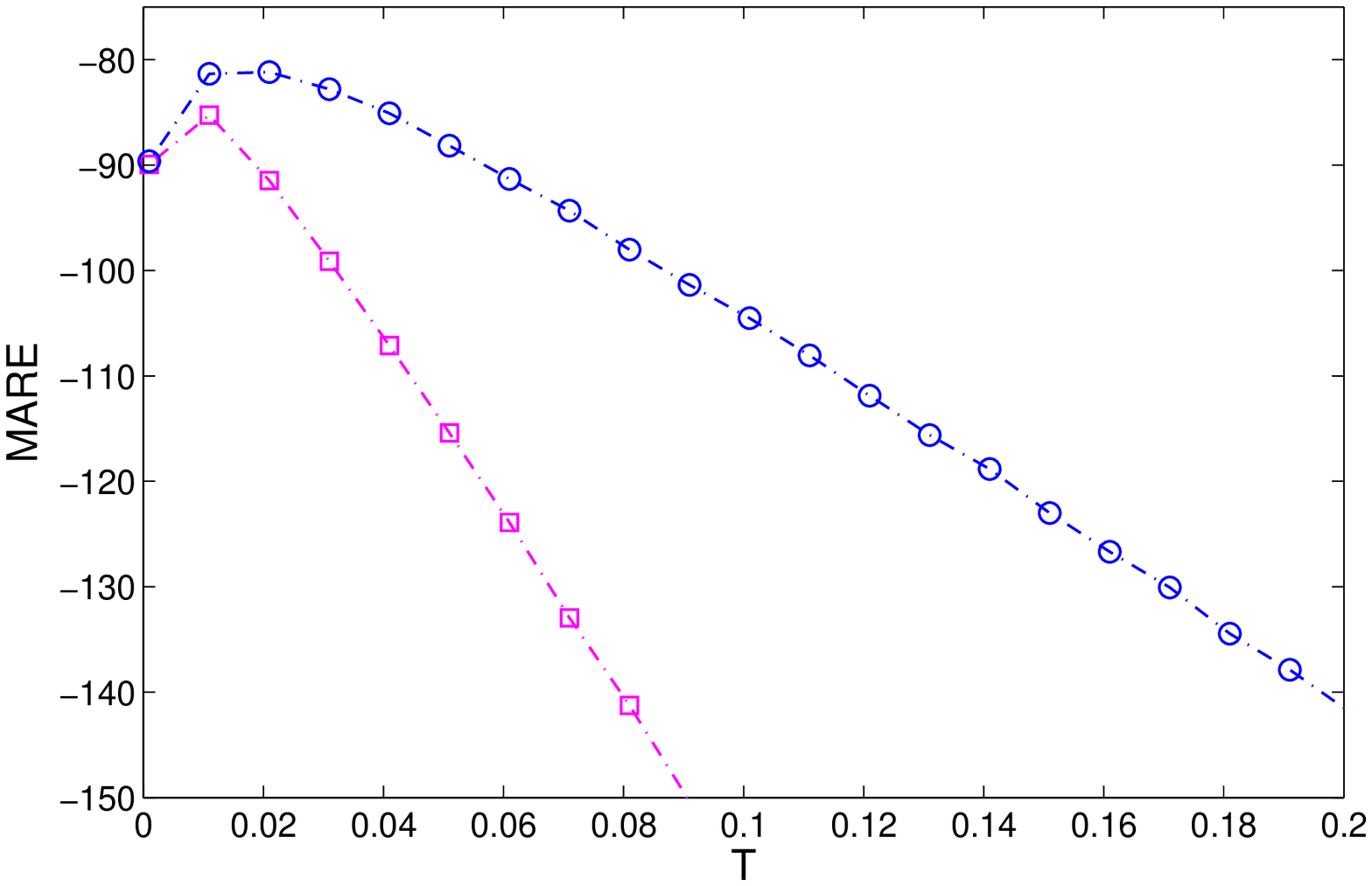}}\vspace*{-5 mm}\\
		\subfigure{\label{fig:MARE_L32_M5_S1_R2_NU_025}
    \includegraphics[scale=0.35,clip]{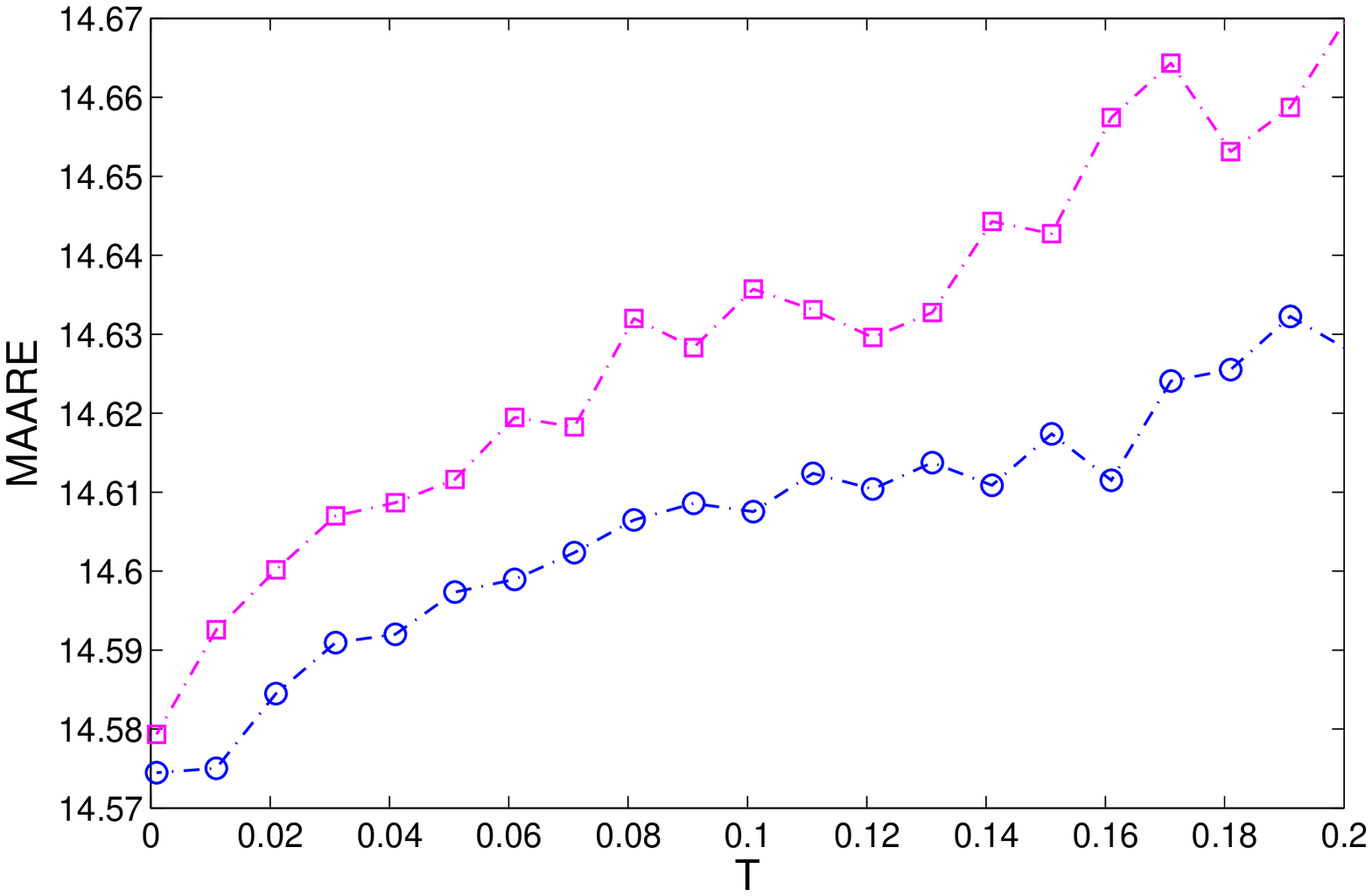}}
    \subfigure{\label{fig:MARE_L32_M5_S1_R2_NU_025_LOGN}
    \includegraphics[scale=0.35,clip]{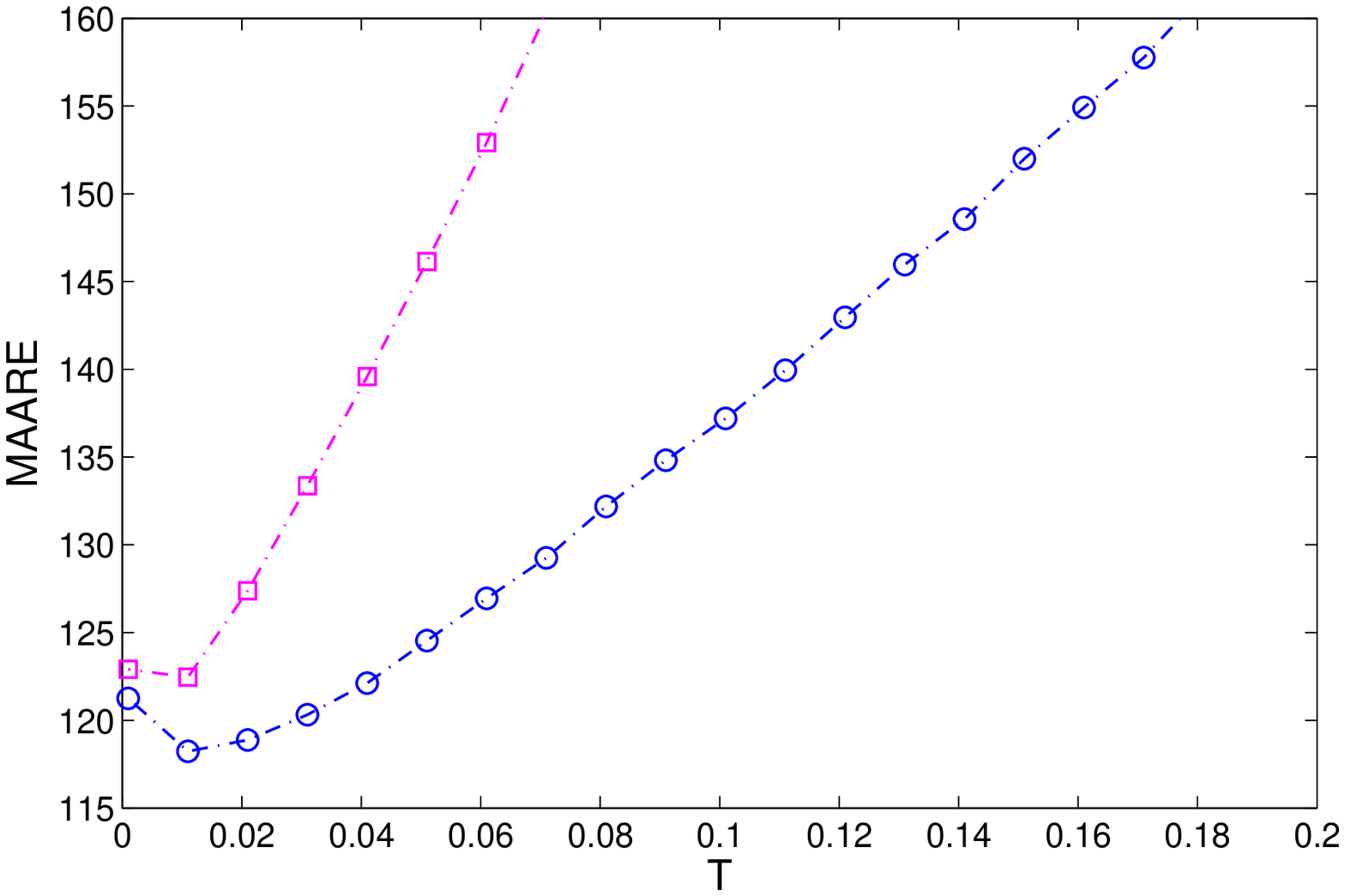}}\vspace*{-5 mm}\\
		\subfigure{\label{fig:RMSE_L32_M5_S1_R2_NU_025}
    \includegraphics[scale=0.35,clip]{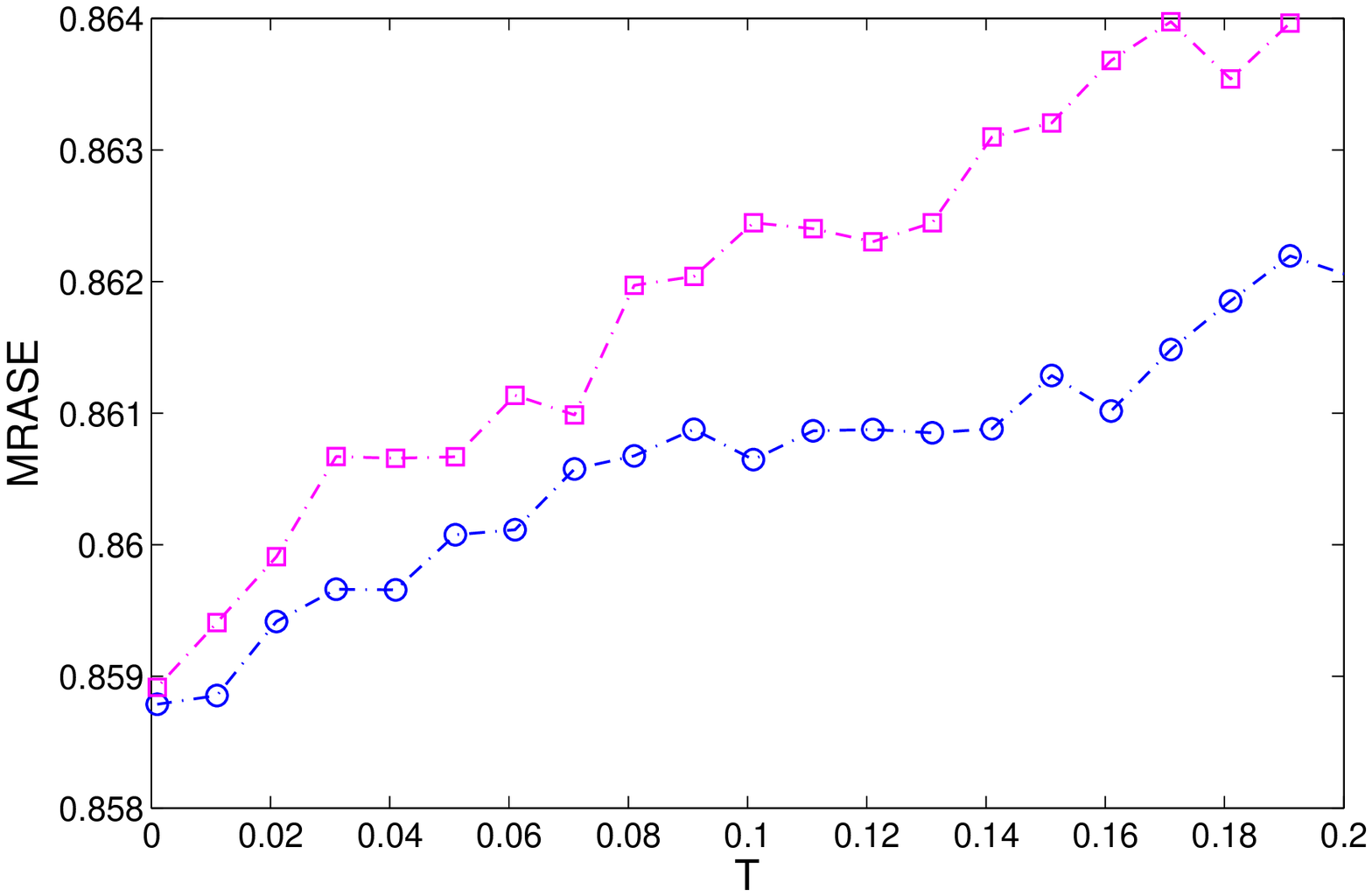}}
    \subfigure{\label{fig:RMSE_L32_M5_S1_R2_NU_025_LOGN}
    \includegraphics[scale=0.35,clip]{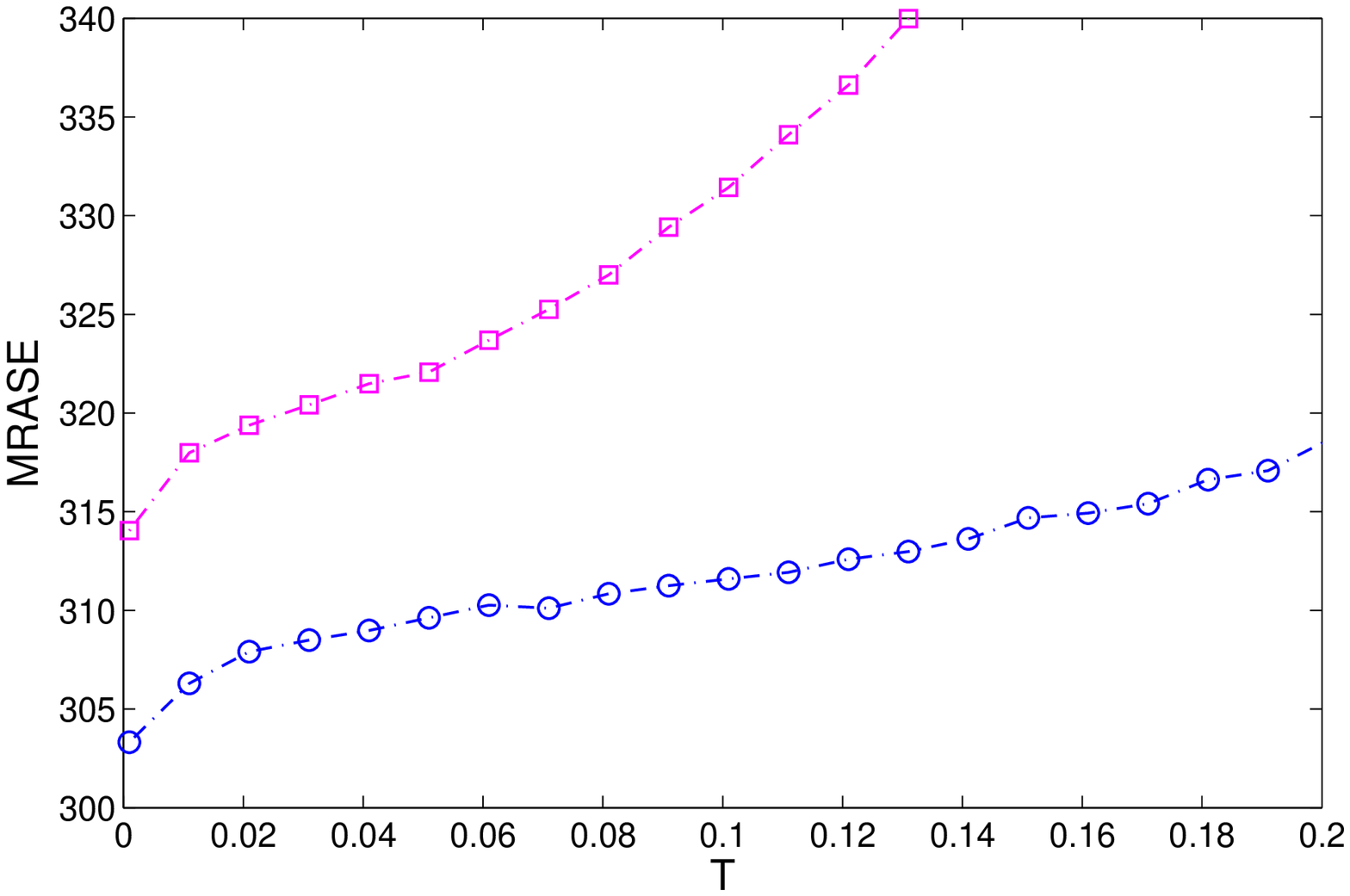}}\vspace*{-5 mm}\\
		\subfigure{\label{fig:COR_L32_M5_S1_R2_NU_025}
    \includegraphics[scale=0.35,clip]{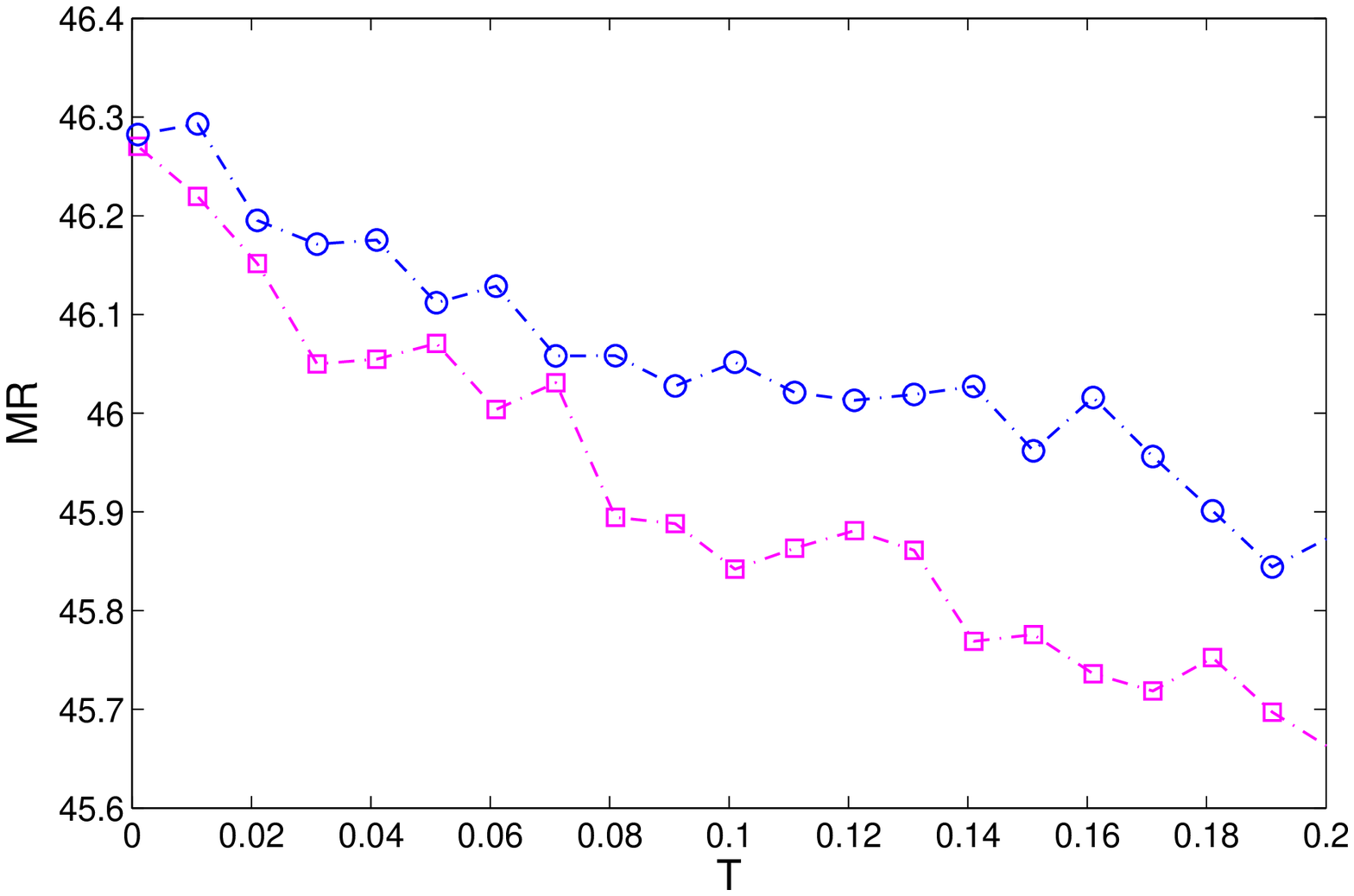}}
    \subfigure{\label{fig:COR_L32_M5_S1_R2_NU_025_LOGN}
    \includegraphics[scale=0.35,clip]{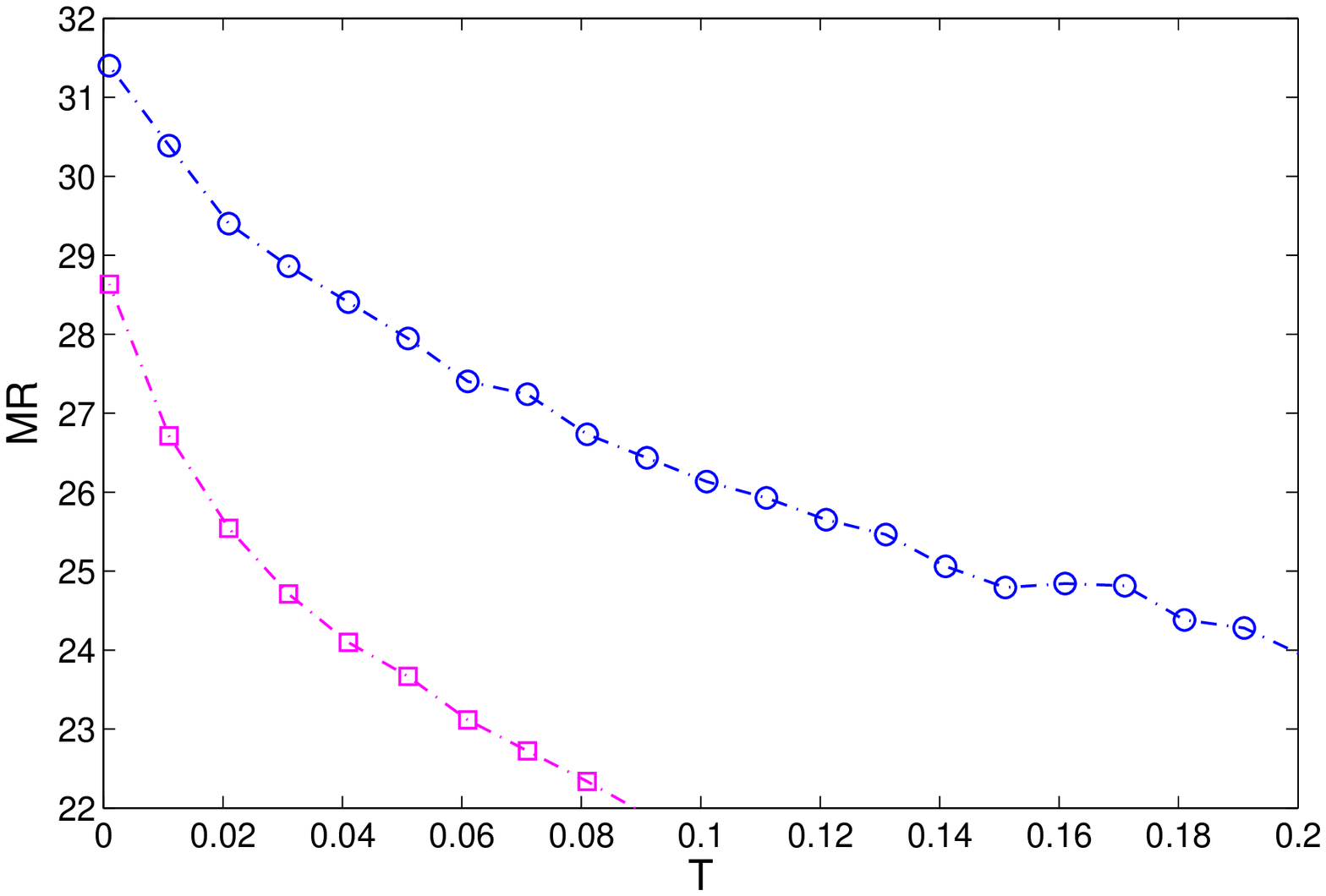}}\vspace*{-5 mm}
\end{center}
    \caption{Validation measures versus temperature for the GMRF (magenta squares) and \mpr (blue circles) energy functions, using samples from the Gaussian distribution, $Z \sim N(m=5,\sigma=1,\kappa=0.5,\nu=0.25)$ (left column), and  the lognormal distribution, i.e., $\log Z \sim N(m = 5, \sigma = 1)$ (right column) on a square grid of size $L=32$ with $p=33\%$ random thinning.}
  \label{fig:error_curves}
\end{figure}

As stated above, in 1D systems the SEM and ML parameter estimation methods yield similar results. Nonetheless, in order to eliminate any potential impact of parameter inference in the 2D system on  prediction performance, in Fig.~\ref{fig:error_curves} we plot the validation measures  for both the \mpr and the GMRF models as functions of the temperature.
For the Gaussian data (left column) the validation measures and their optimal values are similar for both models (notice the scale of the vertical axes). On the other hand, for the lognormal data (right column) all the \mpr validation measures are superior to their GMRF counterparts over the entire temperature range.



\begin{table}[t]
\addtolength{\tabcolsep}{0pt} \caption{Comparison of the interpolation validation measures for the \mpr method with the \mpr $H_{ij}=\cos[q(\phi_i-\phi_j)]$ and GMRF $H_{ij}=-(\phi_i-\phi_j)^2/2\pi^2+1$ pair interaction functions. $S=100$ samples are generated from the Gaussian $Z \sim N(m = 50, \sigma = 10)$ and lognormal $\log Z \sim N(m = 5, \sigma = 1)$ random fields, on  a square grid with side length $L=32$. Four covariance models, WM($\kappa,\nu$), with $\kappa=0.2,0.5$ and $\nu = 0.25,0.5$ are used. Missing data are generated by (a) $p=33\%$ (b) $p=66\%$ random thinning and (c) random removal of a square data block  with side length $L_B=20$.} \vspace{3pt} \label{tab:lognormal}
\begin{scriptsize}
\resizebox{1\textwidth}{!}{
\begin{tabular}{|c|c|c|c|ccc|ccc|ccc|ccc|ccc|ccc|}
\hline
& & & & \multicolumn{3}{c|}{MAAE}  & \multicolumn{3}{c|}{MARE [\%]} &
 \multicolumn{3}{c|}{MAARE [\%]} & \multicolumn{3}{c|}{MRASE} &
 \multicolumn{3}{c|}{MR [\%]}  & \multicolumn{3}{c|}{$ \langle t_{\mathrm{cpu}} \rangle $}   \\
Distr. & $\nu$ & $\kappa$ & $H_{ij}$ &(a) & (b) & (c) & (a) & (b) & (c) & (a) & (b) & (c) & (a) & (b) & (c) & (a) & (b) & (c) & (a) & (b) & (c) \\
\hline
\parbox[t]{2mm}{\multirow{8}{*}{\rotatebox[origin=c]{90}{Gaussian}}} &0.5 &$0.5$ & \mpr &5.23&5.77& 7.34&$-$2.06&$-$2.47&$-$4.16&11.14&12.34&15.98&6.63&7.29&9.23&70.54&62.42&24.47&0.03&0.04&0.03   \\
& & & Gauss &5.22&5.78&7.32&$-$2.08&$-$2.47&$-$3.99&11.14&12.37&15.93&6.63&7.31&9.20&70.53&62.26&25.17&0.03&0.04&0.03   \\
& &$0.2$ & \mpr &3.48&4.03&7.05&$-$1.08&$-$1.46&$-$5.60&7.41&8.70&16.49&4.36&5.10&9.10& 90.66&87.12&45.95&0.03&0.04&0.03   \\
& & & Gauss &3.49&4.05&7.10&$-$1.09&$-$1.46&$-$5.46&7.43&8.73&16.56&4.37&5.12&9.17&90.59&87.00&45.10&0.03&0.04&0.03  \\
& 0.25 &$0.5$ & \mpr &6.83&7.11&7.74&$-$2.95&$-$3.41&$-$8.77&14.67&15.35&17.66&8.64&9.07&9.74&45.59& 37.30&11.64&0.03&0.04&0.03   \\
& & & Gauss &6.83&7.11&7.74&$-$2.96&$-$3.42&$-$9.07&14.67&15.35&17.70&8.64&9.07&9.73&45.68&37.23&12.17&0.03&0.04&0.03   \\
& &$0.2$ & \mpr &5.50&5.81&6.72&$-$1.74&$-$1.90&$-$2.20&11.54&12.19&13.88&6.86&7.32&8.47&65.93&59.84&27.53&0.03&0.04&0.03   \\
& & & Gauss &5.50&5.82&6.62&$-$1.77&$-$1.92&$-$2.38&11.55&12.21&13.72&6.87&7.33&8.35&65.84&59.72&29.52&0.03&0.04&0.043  \\
\hline
\parbox[t]{2mm}{\multirow{8}{*}{\rotatebox[origin=c]{90}{lognormal}}} &0.5 &$0.5$ & \mpr &129.27&159.53&385.47&$-$75.82&$-$137.23&$-$477.32&101.91&157.80&482.10&204.73&226.69&430.15&58.55&48.92&9.67&0.03&0.04&0.03  \\
& & & Gauss &135.63&175.26&464.64&$-$91.84&$-$167.22&$-$571.52&114.82&184.28&574.94&208.11&237.41&509.93&57.49&47.10&7.05&0.03&0.04&0.03  \\
& &$0.2$ & \mpr &94.29&122.05&304.22&$-$45.68&$-$102.72&$-$468.40&66.46&120.65&474.34&152.03&178.21&345.64&83.48&77.53&29.67&0.03&0.04&0.03   \\
& & & Gauss &98.39&133.20&348.69&$-$58.12&$-$128.42&$-$540.95&76.84&144.03&545.85&153.72&185.62&388.15&83.19&76.31&25.66&0.03&0.04&0.03  \\
& 0.25 &$0.5$ & \mpr &174.32&214.51&641.86&$-$111.63&$-$203.28&$-$826.90&142.83&224.25&829.20&311.83&336.86&708.91&25.58&15.30&-2.11&0.03&0.04&0.03  \\
& & & Gauss &188.23&251.07&833.25&$-$141.66&$-$264.21&$-$1051.75&167.57&280.22&1053.09&325.58&365.79&900.42&22.04&13.07&2.62&0.03&0.04&0.03  \\
& &$0.2$ & \mpr &134.54&166.64&414.54&$-$69.55&$-$131.57&$-$442.22&96.93&151.55&442.22&220.83&244.32&462.01&50.44&39.63&10.08&0.03&0.04&0.03   \\
& & & Gauss &142.19&188.23&521.53&$-$87.84&$-$168.33&$-$548.43&111.19&184.30&551.04&225.85&259.71&569.52&48.61&37.24&8.41&0.03&0.04&0.03  \\
\hline
\hline
\end{tabular}
}
\end{scriptsize}
\end{table}

The advantage of the \mpr over the GMRF is due to the fact that the former has higher probability for larger spin angle contrasts, i.e., larger differences between neighboring values of the spin angles.  Since skewed data with heavier than normal right tail (e.g., following the lognormal distribution) can lead to spatial configurations with larger contrasts, the \mpr model is  more suitable than the GMRF.

\begin{figure}[t!]
\centering
\subfigure{\includegraphics[scale=0.5,clip]{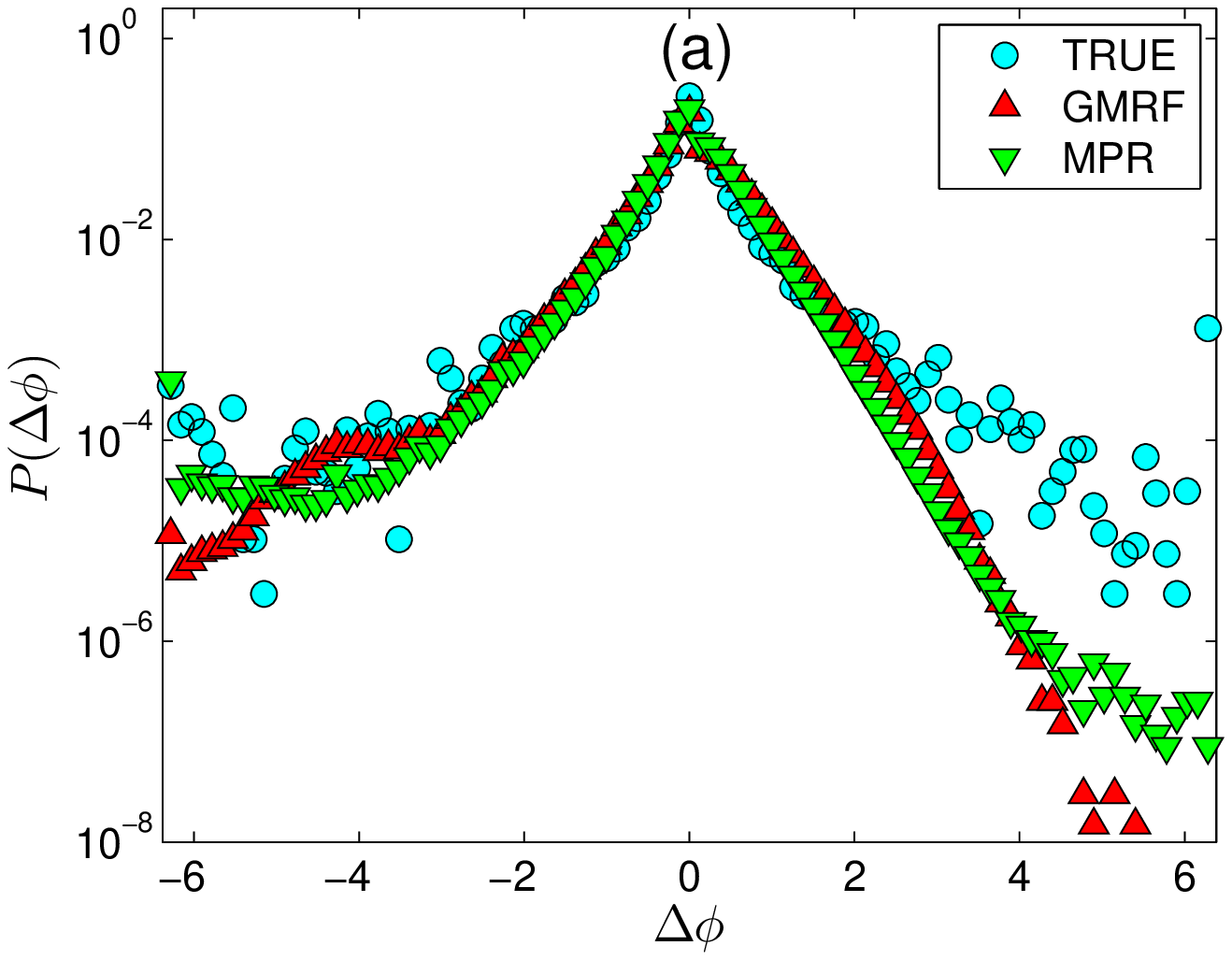}\label{fig:hist_delta_phi}}
\subfigure{\includegraphics[scale=0.5,clip]{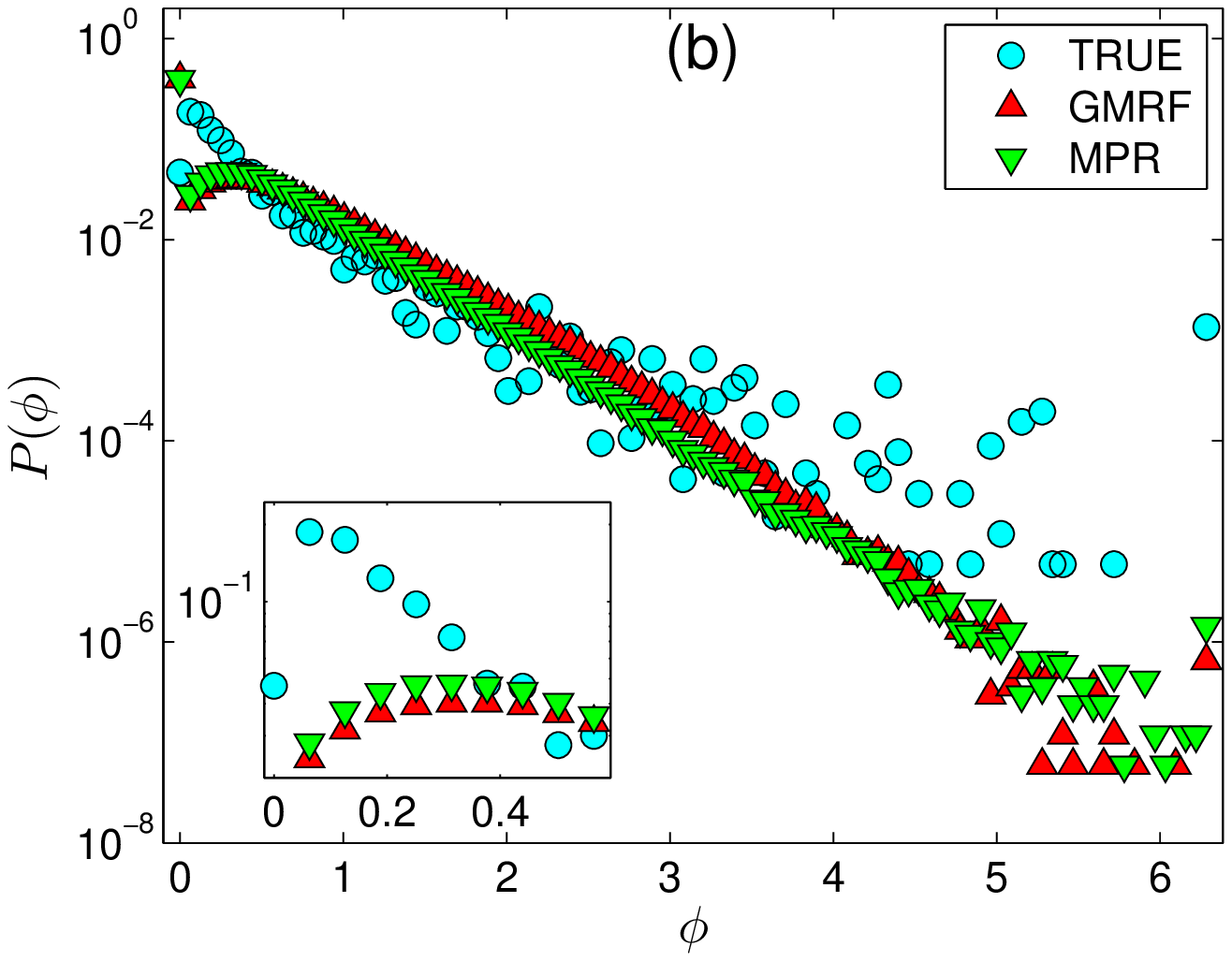}\label{fig:hist_phi}}
\caption{Histograms of (a) spin-angle contrast, $\Delta \phi$, and (b) spin angle, $\phi$, obtained by simulations of the GMRF and \mpr energy functions as well as true values, for lognormal data. The inset in (b) shows a magnified view at small $\phi$. The histograms are based on results obtained from 500 missing data configurations that are generated from the same field realization by means of 33\% random thinning.}
\label{fig:H_phi}
\end{figure}

We conducted a number of numerical experiments to confirm and investigate the above hypothesis. In particular, we generated $S=500$ spatial configurations with $p=33\%$ missing data from the same lognormal random field realization with WM correlations determined by $\kappa=0.5$ and $\nu=0.25$. We sampled  the spin angle contrasts $\Delta \phi$ at all the prediction sites in the equilibrium regime of the simulations. We then constructed the spin angle contrast histogram based on the contrast values sampled in  the equilibrium regime.
The histogram frequencies are obtained by dividing the cumulative frequency of occurrence with the number of the prediction sites, the number of nearest neighbors (four) per site, and the number of MC sweeps in the equilibrium regimes. The resulting histograms approximate the probability density function of the nearest-neighbor spin-angle contrast $\Delta \phi$. We also compare in Fig.~\ref{fig:H_phi} the histograms obtained from the GMRF and \mpr model predictions with histograms of the true values at the prediction sites. The latter are obtained based on the  500 data sets that are removed from the full realization to generate the missing data configurations.

As it is evident in Fig.~\ref{fig:hist_delta_phi} both the GMRF and \mpr histograms overestimate smaller contrasts and underestimate the larger ones. Nevertheless, it is apparent that extremely large contrasts of about $|\Delta\phi| \gtrsim 5$ are  better reproduced using the \mpr model. Inspection of the spin angle ($\phi$) histograms, shown in Fig.~\ref{fig:hist_phi}, reveals slightly fatter tails in the \mpr histogram, which better approximate those of the true values. Note that the histogram of the true values exhibit considerable fluctuations. 
This is due to the significant variance of the simulated lognormal distribution and the finite probability for very large (extreme) values as discussed above.

\begin{figure}[t!]
\centering
\includegraphics[scale=0.7,clip]{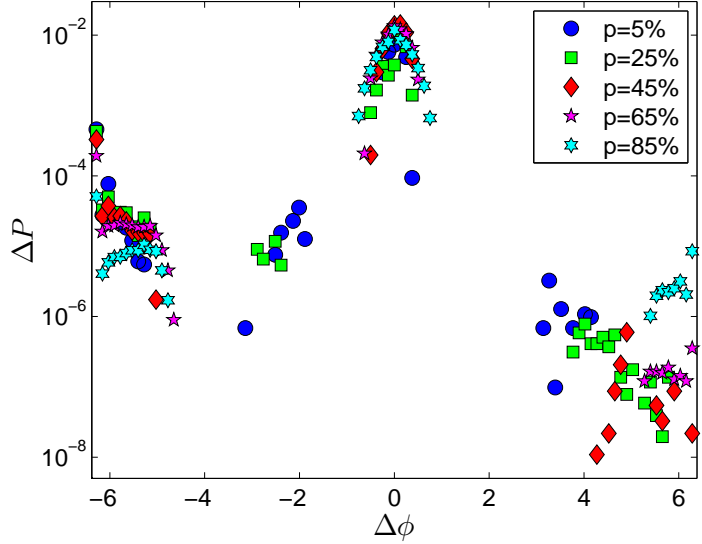}
\caption{Difference of the spin-angle contrast histograms, $\Delta P=P_{\rm MPR}(\Delta\phi)-P_{\rm GMRF}(\Delta\phi)$, for lognormal data obtained using the \mpr and GMRF models for datasets with different degrees of sparsity (percentage of missing points). The histograms are based on results from 500 missing data configurations that are generated from the same field realization.}
\label{fig:delta_hist_delta_phi}
\end{figure}

Differences between the \mpr and GMRF spin-angle contrast histograms also appear at other contrasts and persist even when the data sparsity changes. Fig.~\ref{fig:delta_hist_delta_phi} shows the difference $\Delta P=P_{\rm MPR}(\Delta\phi)-P_{\rm GMRF}(\Delta\phi)$ between the \mpr and GMRF histograms for the cases when $P_{\rm MPR}(\Delta\phi) > P_{\rm GMRF}(\Delta\phi)$~\footnote{Missing values at extremely large contrasts for $p=5\%$ and 25\% are due to the absence of such contrasts in the generated histograms and do not imply that $P_{\rm MPR}(\Delta\phi) < P_{\rm GMRF}(\Delta\phi)$.} and for different thinning degrees $p$. It is evident that the thicker tails in the \mpr histogram observed for $p=33\%$ (cf. Fig.~\ref{fig:hist_delta_phi}) persist for all the studied sparsity values. Small contrast values of $|\Delta\phi|\lesssim 0.65$ are also more frequent in the \mpr realizations, while the GMRF model has higher frequency at certain intermediate values.

%

Based on the evidence examined above, we conclude that the \mpr model has the advantage over the GMRF with respect to filling gaps in skewed, non-Gaussian  spatial data. The \mpr's  performance is due to a combination of factors that include the probability distribution of the dataset as well as the properties of the spatial correlation function (the latter has not been investigated).

In future research it is possible to generalize the \mpr model by introducing additional parameters to control non-linearity, e.g., by including higher-order interactions \citep{mz-gk17}, and to  capture other common features of spatial data, such as geometric anisotropy and non-stationarity.



\section{Discussion}
\label{sec:discuss}

\subsection{Model parameter inference}
The  reduced temperature is the only parameter of the \mpr model that needs to be inferred from the data. In the case of spin models, standard statistical inference procedures, e.g., maximum likelihood estimation, are not easy to apply. The problem is the calculation of the partition function (normalizing factor), which is  intractable even for moderately large systems. Consequently, one has to resort to tractable approximations. However, some approximate solutions, such as the maximum pseudolikelihood approach or Markov chain Monte Carlo techniques, can be inaccurate or/and prohibitively slow for large data sets.
As described in Section~\ref{ssec:temp-infer}, we use the SEM principle to estimate the temperature, ${\hat T}$, used in the \mpr conditional simulation.

\subsubsection{Performance of the SEM approach}
To test the performance of the SEM temperature estimator, we compare ${\hat T}$ inferred from various samples with the ``optimal'' temperature $T_{\textrm{opt}}$. For each sample, $T_{\textrm{opt}}$ is defined by means of
\begin{equation}
\label{T_opt}
T_{\textrm{opt}}=\sum_{i}w_iT_{\textrm{opt},i},
\end{equation}
where $T_{\textrm{opt},i}$ is the temperature that optimizes the i-th validation measure, ${\rm VM}_{\textrm{opt},i}$, and  VM= \{ AAE, ARE, AARE, RASE, R \}.
Hence, the lowest values are optimal for AAE, ARE, AARE, and RASE, while the highest value is optimal for $R$. The coefficients $w_i$ $(i=1, \ldots, 5)$ represent  weights defined as follows
\begin{equation}
\label{weights}
w_i={\Big |}({\rm VM}_{\textsc{SEM},i}-{\rm VM}_{\textrm{opt},i})/{\rm VM}_{\textsc{SEM},i}{\Big |}{\Big /}\sum_{i}{\Big |}({\rm VM}_{\textsc{SEM},i}-{\rm VM}_{\textrm{opt},i})/{\rm VM}_{\textsc{SEM},i}{\Big |},
\end{equation}
where ${\rm VM}_{\textsc{SEM},i}$ is the validation measure at the temperature ${\hat T}$ inferred by SEM.

\begin{figure}[t!]
\centering
\includegraphics[scale=0.7,clip]{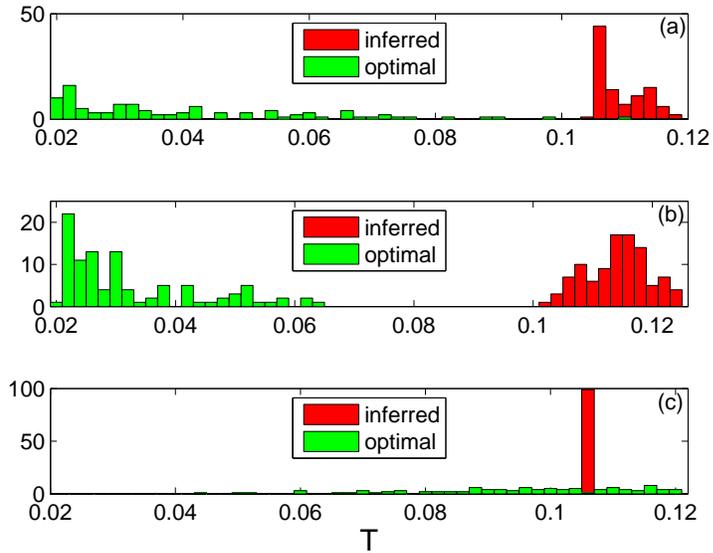}
\caption{Histograms of temperature values ${\hat T}$, inferred from the energy, vs. the optimal values ${T}_{\textrm{opt}}$, giving overall the best validation measures, based on $S=100$ samples of WM($\kappa=0.2,\nu = 0.5$) data with $L=256$, generated by (a) $p=33\%$, (b) $p=66\%$ random thinning and (c) random removal of a solid block of data of the side length $L_B=20$.}\label{fig:par_inf_r5_nu05}
\end{figure}

As evidenced in the results for the synthetic WM($\kappa=0.2,\nu = 0.5$) data that are presented in Fig.~\ref{fig:par_inf_r5_nu05}, there is considerable variation between the
 inferred temperatures using SEM  and the optimal values $T_{\textrm{opt}}$. Namely, SEM  tends to overestimate $T_{\textrm{opt}}$, especially in the case of  randomly thinned data.

Next, we investigate the impact on prediction performance of using  optimal temperatures instead of the SEM estimates by repeating the \mpr simulations at  temperatures $T_{\textrm{opt}}$ and analyzing the validation measures thus obtained. Table~\ref{tab:MPR_comp} lists the \emph{relative validation measures} ${\rm VM^*}={\rm VM}_{\textrm{opt}}/{\rm VM}_{\textsc{SEM}}$ for the synthetic data with $L=256$. As expected, the overall prediction performance improves by using $T_{\textrm{opt}}$. Nevertheless, considering the large differences between the ${\hat T}$ inferred from SEM and $T_{\textrm{opt}}$, the relative differences between the respective validation measures are surprisingly small, typically  $\approx 0.1\%$. These results support the robustness of the \mpr method against  fluctuations of $\hat{T}$ that might result from the presence of noise and outliers, or from limited  inference precision due to small sample size or data sparsity.

\begin{table}[t!]
\addtolength{\tabcolsep}{0pt} \caption{Relative validation measures ${\rm VM^*}={\rm VM}_{\textrm{opt}}/{\rm VM}_{\textsc{SEM}}$ obtained as the ratio of the validation measure based on the optimal temperature $T_{\textrm{opt}}$ over the same measure based on the temperature ${\hat T}$ estimated by means of specific energy matching.
$S=100$ samples are generated from a Gaussian random field with mean equal to 50 on  a square grid with side length $L=256$. Two covariance models, WM($\kappa=0.2,\nu = 0.5$) and WM($\kappa=0.2,\nu = 0.25$) are used. Missing data are generated by (a) $p=33\%$ (b) $p=66\%$ random thinning and (c) random removal of square data block  with side length $L_B=20$ (same data as in Table~\ref{tab:synt_int_MPR} for $L=256$). Boldfaced values mark cases for which the validation measure obtained at $T_{\textrm{opt}}$ is inferior to that obtained at ${\hat T}$.} \vspace{3pt} \label{tab:MPR_comp}
\begin{scriptsize}
\resizebox{1\textwidth}{!}{
\begin{tabular}{|c|ccc|ccc|ccc|ccc|ccc|}
\hline
 & \multicolumn{3}{c|}{${\rm MAAE^*}$}  & \multicolumn{3}{c|}{${\rm MARE^*}$} &
 \multicolumn{3}{c|}{${\rm MAARE^*}$} & \multicolumn{3}{c|}{${\rm MRASE^*}$} &
 \multicolumn{3}{c|}{MR$^*$}   \\
$\nu$ & (a) & (b) & (c) & (a) & (b) & (c) & (a) & (b) & (c) & (a) & (b) & (c) & (a) & (b) & (c) \\
\hline
$0.25$ &1.00&1.00&1.00&{\bf 1.00}&1.00&0.94&1.00&1.00&0.99&1.00&1.00&1.00&1.00&1.00&1.02 \\
$0.5$ &1.00&1.00&1.00&1.00&1.00&{\bf 1.02}&1.00&1.00&1.00&1.00&1.00&1.00&1.00&1.00&1.01 \\
\hline
\end{tabular}
}
\end{scriptsize}
\end{table}

\subsubsection{Comparison of SEM and MLE approaches in 1D}
\label{ssec:sem-mle}
As stated above, the SEM-based procedure seems to overestimate the temperature with respect to the ``optimal'' value, at least in configurations involving randomly missing data.  This effect diminishes in denser data sets. However, the ``optimality criterion''~\eqref{T_opt} is based on an \textit{ad hoc} linear combination of various validation measures, since the standard MLE procedure cannot be applied to 2D data.

\begin{figure}[t]
\begin{center}
    \includegraphics[scale=0.8,clip]{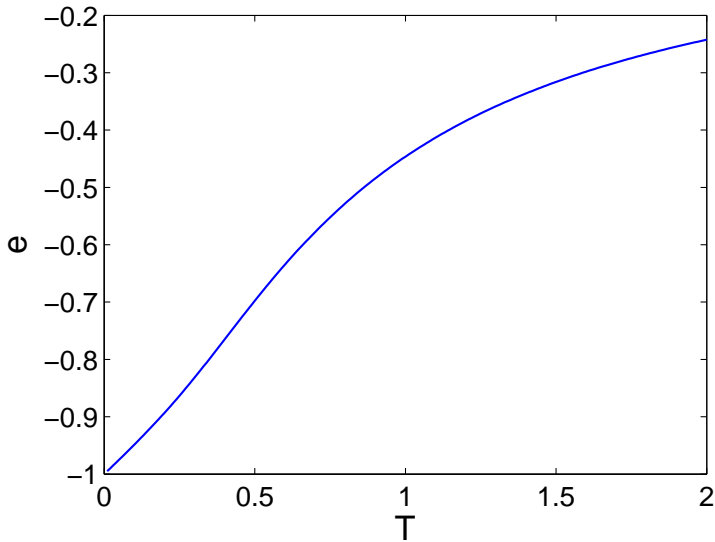}
\end{center}
    \caption{Specific energy of the 1D \mpr model as a function of temperature based on~\eqref{ene_1D}.}
  \label{fig:e-T}
\end{figure}

To test the reliability of SEM parameter inference, we compare it below with  MLE for the one-dimensional (1D) \mpr model. The partition function of the \mpr chain with an open boundary condition admits a closed-form expression~\citep{tane94} as
\begin{equation}
\label{part_fun}
Z(\beta) = I_0(\beta)^{N-1},
\end{equation}
where $\beta=1/T$ is the inverse temperature, $N$ is the chain length, and $I_0$ is the modified Bessel function of the first kind, which leads to the following log-likelihood function
\begin{equation}
\label{part}
\log L(\beta,\Phi_{s}) = -\beta E_s - N_{SP}\log I_0(\beta).
\end{equation}

\noindent In the above, $\Phi_{s}$ represent the sample data, $E_s=-\sum_{i=1}^{N}\cos[q(\phi_{i}-\phi_{i+1})]$ is the total sample energy calculated from the nearest-neighbor sample values and $N_{SP}$ is the number of the nearest-neighbor sample pairs.

The MLE estimates $\hat{T}=1/\hat{\beta}$ are obtained by minimizing numerically $-\log L(\beta,\Phi_{s})$, i.e., the negative log-likelihood (NLL). We perform the optimization with the gradient-free Nelder-Mead simplex search algorithm. The  termination criteria are that both $\beta$ and the NLL cost function  change less than $\epsilon=10^{-6}$ between consecutive steps. The initial guess for the inverse temperature is $\beta^{(0)}=1$. The algorithm was implemented using the Matlab\circledR\ function \verb+fminsearch+.

For the SEM method we need the specific (internal) energy. Knowing the partition function, the latter can be obtained in closed form as follows
\begin{equation}
\label{ene_1D}
e = -\frac{1}{N-1}\frac{\partial}{\partial \beta}\log Z(\beta) = -I_1(\beta)/I_0(\beta).
\end{equation}
The temperature dependence of the 1D-\mpr specific energy is plotted in Fig.~\ref{fig:e-T}. The SEM temperature for a given sample is obtained as the value corresponding to the sample's specific energy $e_s=E_s/N_{SP}$, i.e., by means of $\hat{T}=e^{-1}(e_s)$.

\begin{figure}[t]
\begin{center}
    \subfigure{\label{fig:T_MLE_vs_SEM}
    \includegraphics[scale=0.55,clip]{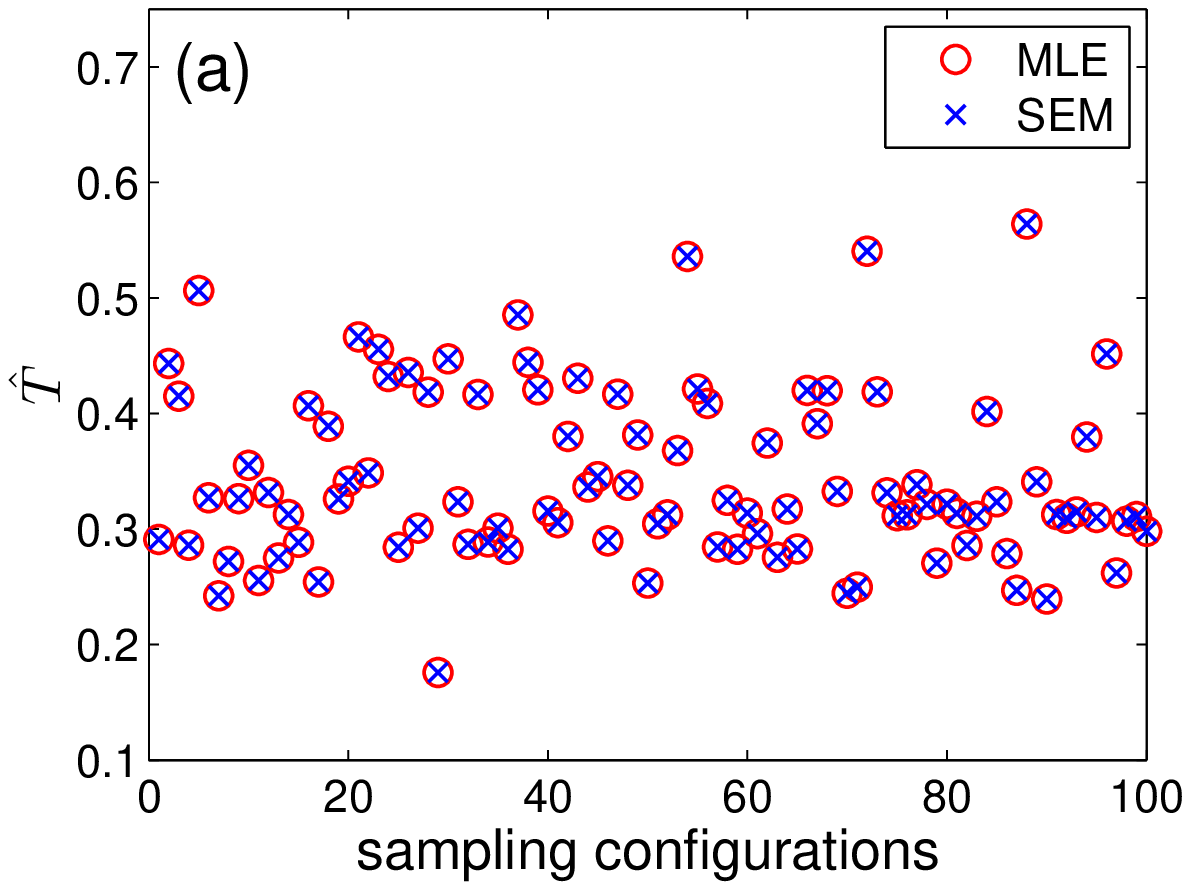}}
    \subfigure{\label{fig:MT_MLE-T_SEM}
    \includegraphics[scale=0.55,clip]{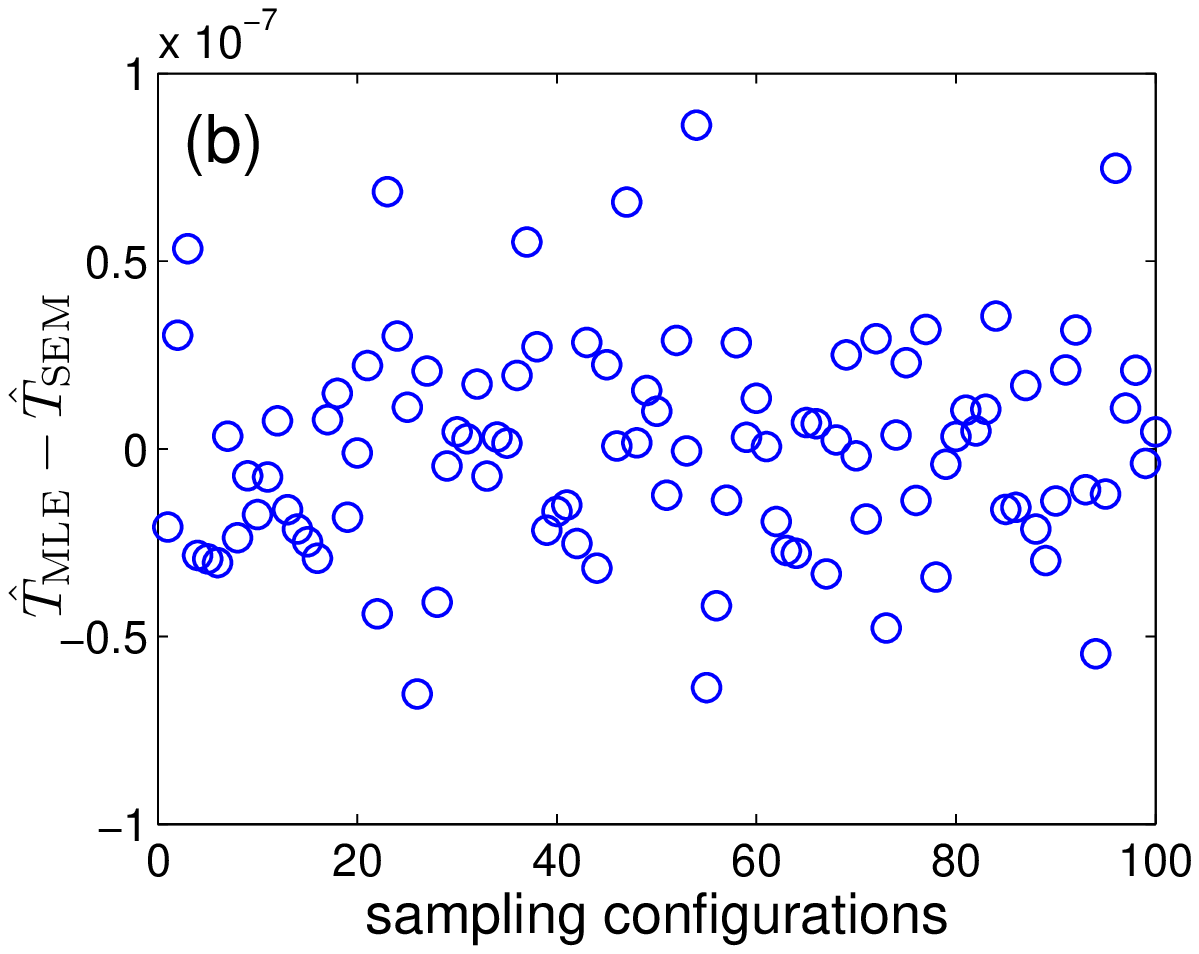}}
\end{center}
    \caption{(a) Temperature estimates for the 1D \mpr model obtained  by means of MLE (red circles) and SEM (blue crosses).  (b) Difference of the respective SEM and MLE estimates.}
  \label{fig:T_inf}
\end{figure}

To compare the MLE and SEM temperature estimates, we performed tests on synthetic data mirroring those used for the 2D case. Namely, we first generated a 1D data (time series) of length $N=100$ from the Gaussian distribution $Z \sim N(m = 50, \sigma = 10)$ with  WM covariance parameters $\kappa=0.5,\nu=0.25$. Then we randomly removed $p=33\%$ of the data to generate $S=100$ different sampling configurations. As shown in Fig.~\ref{fig:T_MLE_vs_SEM} both MLE and SEM lead to practically identical estimates. Fig.~\ref{fig:MT_MLE-T_SEM} displays  the difference between the respective SEM and MLE estimates, the values are smaller in magnitude than the tolerance $\epsilon=10^{-6}$ used for MLE optimization. These results demonstrate that the SEM estimates are as reliable as the MLE ones, at least in the 1D case.

\subsection{Computational efficiency}

\begin{figure}[t!]
\centering
\includegraphics[scale=0.7,clip]{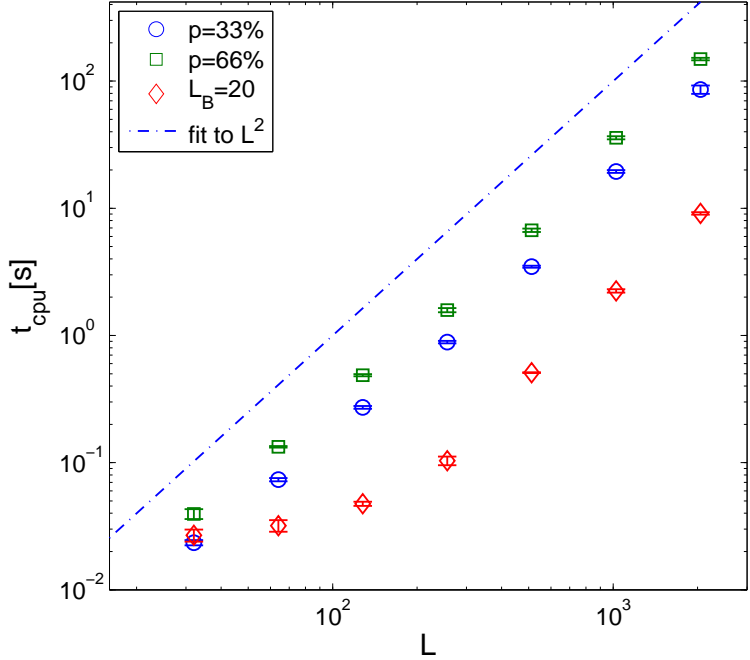}
\caption{CPU time scaling versus the grid size $L$ based on $S=100$ samples of WM($\kappa=0.2,\nu = 0.5$) data. Three plots generated by  $p=33\%$ random thinning (circles), $p=66\%$ random thinning (squares), and random removal of a solid block with side length $L_B=20$ (diamonds) are shown. The dash-dot line is a guide to the eye for linear dependence.}\label{fig:cpu_time_r5_nu05_new}
\end{figure}

The computational efficiency of the \mpr approach crucially depends on  an \emph{efficient updating scheme} that can bring the system to thermodynamic equilibrium as fast as possible. After equilibrium is established, a predefined number of realizations can be sampled  to derive predictive means. The hybrid algorithm that combines   restricted Metropolis and  over-relaxation dynamics provides such an updating scheme. The resulting relaxation time in terms of MC sweeps is of the order of tens of hybrid sweeps even for the largest grid sizes considered, and it seems to plateau at this level. Additionally, the short-range nature of the interaction between the spin variables enables vectorization  by means of the checkerboard algorithm, so that each sweep can be completed in just two steps. Naturally, the physical CPU time per sweep, and thus also the total CPU time $t_{\mathrm{cpu}}$ (including both the relaxation and sampling time), is expected to increase with data size. In Fig.~\ref{fig:cpu_time_r5_nu05_new} we plot the total CPU time as a function of the data size obtained based on $S=100$ simulations of Gaussian data with WM($\kappa=0.2,\nu = 0.5$) and different patterns of missing values. The log-log plots indicate that, at least on grids with side length up to $L=2048$, the CPU time does grow \emph{at most linearly} with the data size.

\section{Conclusions and Further Research}
\label{sec:conclusions}
We have introduced a novel \emph{Gibbs Markov random field} based on the modified planar rotator (\mpr) model.
Unlike the well-known non-Gaussian Ising model that is suitable for binary-valued
data, the \mpr model is applicable to continuous data that take values in a closed subset of the real numbers.
The \mpr is amenable to computationally efficient conditional simulation  suitable for the reconstruction of missing data on regular spatial grids. Hence, it is useful for the imputation of missing data in remote sensing datasets (e.g., satellite and airborne lidar data). The computational efficiency derives from the \emph{local nature} of the spin interactions and the use of a \emph{hybrid Monte Carlo} simulation algorithm.

Using empirical tests on both randomly missing and contiguous missing block data, we have demonstrated the competitiveness of \mpr with respect to several interpolation methods used for gap filling.
The \mpr model is  promising for automated processing of partially sampled data sets due to its simplicity, computational efficiency, and  dependence on a single tunable parameter (temperature). The latter can be estimated without user supervision, thus making the \mpr model suitable for the automated prediction of missing data.

Another important feature is the potential of the \mpr algorithm to process big data in near real-time.  This goal requires further gains in efficiency and  memory use optimization that can be achieved through parallelization of the algorithm. Such parallelization is feasible due to the short-range (nearest-neighbor) nature of the interactions between the \mpr variables (spins). Recent developments in spin model simulations~\citep{weig12}, as well as our preliminary tests, have demonstrated that much larger data sizes can be handled, and significant speed-ups by factors of up to $1000$ can be achieved by using a highly parallel architecture of graphics processing units (GPUs).

One may wonder whether the advantage of the \mpr model that derives from its dependence on a single parameter  limits the scope of its applications.
The flexibility of the model can be enhanced at some computational cost. One possibility is to allow the model to automatically select the optimal value of $q$ in the interval $[0, 1/2]$ by means of a cross-validation procedure ---instead of arbitrarily setting it equal to $1/2$. In order to better capture additional spatial features, such as geometric anisotropy or non-stationarity, potentially necessary for huge Earth observation data sets over extended spatial domains,  additional coupling parameters can be introduced. Possible extensions in this direction include the generalization of the \mpr Hamiltonian by incorporating (i) exchange interaction anisotropy (ii) an external ``magnetic'' field that can generate spatial trends and (iii) interactions beyond nearest neighbors.  Furthermore, the double-checkerboard decomposition that enables processing data in several non-overlapping windows can provide computational benefits for the modeling of nonstationary and/or anisotropic data~\citep{weig12}.

The \mpr model could also be extended to irregularly spaced data by means of kernel functions, in the spirit of stochastic local interaction models~\citep{dth15}. However, in the case of irregularly spaced data some of the computational efficiency that derives from the lattice geometry will be sacrificed. Another appealing direction is the extension of the present approach to three dimensions, where efficient methods for modeling large spatio-temporal data sets are still lacking~\citep{wang12}.

\section*{Acknowledgments}
This work was supported by the Scientific Grant Agency of Ministry of Education of Slovak Republic (Grant Nos. 1/0474/16 and 1/0331/15). We also acknowledge support for a short visit by M.~\v{Z}. at the Technical University of Crete from the Hellenic Ministry of Education - Department of Inter-University Relations, the State Scholarships Foundation of Greece and the Slovak Republic's Ministry of Education through the Bilateral Programme of Educational Exchanges between Greece and Slovakia. We would like to thank Yusuke Tomita for useful comments on the computational implementation. Finally, we thank the two anonymous reviewers who offered useful and
constructive suggestions which greatly enhanced the clarity of this manuscript.


\end{document}